\documentclass[aps,pre,twocolumn,superscriptaddress,floats,floatfix]{revtex4-2}
\usepackage{amssymb,amsmath,amsthm}
\usepackage{graphicx}
\usepackage{epstopdf}
\usepackage{color}
\usepackage{subfigure}
\usepackage{epsfig}
\usepackage{rotating}
\usepackage{mwe}
\usepackage{float}
\usepackage{dcolumn,bm}
\usepackage{verbatim}
\usepackage{hyperref}
\usepackage{pgf,tikz}
 \usetikzlibrary{calc}
\usetikzlibrary{arrows.meta, angles, quotes}
\usepackage{makecell}
\usetikzlibrary{3d}
\usepackage[normalem]{ulem}
\usepackage{xcolor}
\definecolor{Darkred}{RGB}{180, 0, 0}
\definecolor{Darkblue}{RGB}{26,68, 171}
\hypersetup{
  colorlinks = true,     %Colours links instead of ugly boxes
  urlcolor   = Darkblue, %Colour for external hyperlinks
  linkcolor  = Darkblue, %Colour of internal links
  citecolor  = Darkred   %Colour of citations
}

\begin{document}
\begin{abstract}
We develop a theoretical framework that allows us to explore the coupled motion of neutron-superfluid vortices and proton-superconductor flux tubes in a gravitationally collapsed condensate, which describe neutron stars that form pulsars. Our framework uses the 3D Gross-Pitaevskii-Poisson-Equation (GPPE) for neutron Cooper pairs, the Real-Time-Ginzburg-Landau equation (RTGLE) for proton Cooper pairs, the Maxwell equations for the vector potential ${\bf A}$, and Newtonian gravity and interactions, both direct and induced by the Poisson equation, between the neutron and proton subsystems. For a pulsar we include a crust potential, characterized by an angle $\theta$, and frictional drag. By carrying out extensive direct numerical simulations of this model, we obtain a variety of interesting results. We show that a rotating proton superconductor generates a uniform London magnetic field and the field distribution around flux tubes changes. In the absence of any direct interaction between the two species, they interact through the gravitational Poisson equation. 
The inclusion of the current-current interaction and the complete Maxwell equations allows us to quantify the entrainment effect that leads to induced magnetization of neutron vortices. We demonstrate that, with a strong external magnetic field ${\bf B}_{\rm ext}$, proton flux tubes are anchored to the crust, whereas neutron vortices leave the condensate and lead to abrupt changes of the crust angular momentum ${\rm J}_c$. The frictional term in the dynamical equation for $\theta$ yields stick-slip dynamics that leads, in turn, to glitches in the time series of ${\rm J}_c$. By calculating various statistical properties of this time series, we demonstrate that they display self-organised criticality (SOC) that has been found in observations for several pulsars. We compare our results with those of earlier explorations of pulsar-glitch statistics in GPE-based minimal models for pulsars.
\end{abstract}

\title{Neutron-superfluid vortices and proton-superconductor flux tubes: Development of a minimal model for pulsar glitches}
\author{Sanjay Shukla }%$^1$}
\email{shuklasanjay771@gmail.com}
\affiliation{Centre for Condensed Matter Theory, Department of Physics, Indian Institute of Science, Bangalore 560012, India}
\author{Marc E. Brachet }
\email{brachet@phys.ens.fr}
\affiliation{Laboratoire de Physique de l’École Normale Supérieure, ENS, Université PSL,
CNRS, Sorbonne Université Université de Paris, 24 Rue Lhomond, 75005 Paris, France}

\author{Rahul Pandit $^1$}
\email{rahul@iisc.ac.in}

\maketitle

\section{Introduction}
Recent advances in the Gross-Pitaevskii-Poisson (GPP) modelling of bosonic~\cite{Akhilesh_2016} and axionic~\cite{shukla_2023} stars have led to an elegant, minimal model for pulsars~\cite{AK_verma_2022}, which includes a crust potential, and leads naturally to pulsar glitches~\cite{Radhakrishnan_1969,Boynton_1969,manchester_2017}. These GPP models have, so far, accounted only for bosons, e.g., the neutron superfluid in a neutron star~\cite{Gordon_1969}. Within the outer core of a neutron star, characterized by a density ranging from $\rho \simeq 5 \times 10^{13}$ to $10^{15} \ \rm g \ cm^{-3}$~\cite{Gordon_1969}, lies a region consisting predominantly of neutrons ($95\%$ of the total mass) and some protons ($5\%$ of the total mass), and sufficient electrons to maintain charge neutrality; this region exhibits extraordinary properties: the neutron Cooper pairs form a superfluid~\cite{MIGDAL_1959} and the proton Cooper pairs a superconductor~\cite{Gordon_1969,Chamel_2008}. Pulsars are rapidly rotating and highly magnetized neutron stars~\cite{Haskell_2015,Basu_2018} with magnetic fields $\simeq 10^{12} \ \rm G$. Our goal is to generalise the GPP modelling framework for pulsars~\cite{AK_verma_2022} to include protons, which are in a superconducting state that is affected strongly by the magnetic field.

Neutrons in a neutron-star interiors are in a superfluid state, so when the star rotates with a sufficiently large angular velocity, quantized vortices are formed; each vortex has an angular momentum that is an integer multiple of the quantum of circulation $\mathcal{K}=\frac{h}{m_n^*}$, where $m_n^*$, the mass of a neutron Cooper pair,  is twice the mass of a neutron. By contrast, the protons in a neutron star form an Abrikosov phase of a Type II superconductor~\cite{Mendell_1991,Graber_2015}, in which the external magnetic field leads to an array of flux tubes, each carrying a magnetic flux quantum $\Phi_0=\frac{hc}{q}$, with $q=2e$ being the charge of a proton Cooper pair. Early studies by Ginzburg and Kirzhnits~\cite{Ginzburg_1964}, Wolf~\cite{Wolf_1966}, and Baym, Pethick, and Pines~\cite{Gordon_1969} laid the foundations for our understanding of neutron superfluidity and proton superconductivity in neutron stars. Also, see the recent review on superfluidity and superconductivity in neutron stars~\cite{Haskell2018} and references therein.

Pulsars exhibit sudden increases, known as glitches, in their rotational frequencies~\cite{Radhakrishnan_1969,Boynton_1969}. The interaction of the pulsar crust with the neutron superfluid may provide an explanation for these glitches, as first suggested by Baym, Pethick, and Pines~\cite{Gordon_1969}, and as explored recently by our group in Ref.~\cite{AK_verma_2022}. Pulsar-glitch observations~\cite{Radhakrishnan_1969,Boynton_1969,manchester_2017} suggest that there is a connection of glitches with the transfer of angular momentum, stored in the quantized vortices of the neutron superfluid, to the solid crust of a pulsar. Various models, such as those based on vortex avalanches~\cite{khomenko_haskell_2018,Howitt_2022} or superfluid vortex-crust interaction~\cite{Watanabe_2017,AK_verma_2022,Poli_2023}, have been proposed to study the glitching phenomenon.

As we have noted above, a neutron superfluid dominates the outer core of a pulsars.  Therefore, Refs.~\cite{Warszawski_2011,Warszawski_2012} have utilized a simple model for a pulsar in which the Gross-Pitaevskii equation (GPE) is used for the neutron superfluid together with a pinning potential for the crust and a rotating container that is defined by a quadratic confining potential. In Ref.~\cite{AK_verma_2022}, our group has removed the confining potential but introduced Newtonian gravity, which leads to a gravitationally collapsed bosonic condensate that displays glitches whose statistical properties are akin to those seen in several pulsars.

Even though protons constitute only about $\simeq 5 \%$ of the mass of a pulsar, they play a crucial role in its rich dynamics because the strong magnetic field leads to the formation of an array of flux tubes. The flux tubes in this lattice can interact with the vortices in the neutron superfluid~\cite{Sauls_1989}. References~\cite{Drummond_2017} and \cite{Drummond_2017_b} have examined such interactions, both in equilibrium and out-of-equilibrium conditions, in the 2D and 3D GPE systems, but with (a) a \textit{static Ansatz} for the proton flux tubes and (b) \textit{harmonic confinement}. We go beyond approximation (a) and replace the harmonic potential in (b) by Newtonian gravity that leads to a gravitationally collapsed condensate. In particular, we develop a theoretical framework by combining the Maxwell equations for the electromagnetic fields with the 3D Gross-Pitaevskii-Poisson-Equation (GPPE) for the neutron superfluid and the Real-Time-Ginzburg-Landau equation (RTGLE) for the proton-superconductor system. Moreover, we include (i) density-density and (ii) current-current direct interaction between neutron and proton Cooper pairs.

Before going into the details of our calculations, we present the principal results of our study of the coupled GPPE, RTGLE, and Maxwell systems:

\begin{itemize}
    \item {We show that the rotation of the proton superconductor leads to a London moment~\cite{Verheijen_1990}, i.e., inside the superconductor there is a uniform magnetic field, whose magnitude depends on the rotation frequency.}

    \item {We evolve the magnetic field of the proton flux tubes by using the Maxwell equations; this leads to a precise characterization of the entrainment of protons around neutron vortices, by virtue of which these vortices also become magnetised. This is the first demonstration of such entrainment in the GPPE context.}
 
    \item {If $\Theta > 0$ is the initial angle between the rotation axis and the external magnetic field, then, eventually, the proton-superconductor flux tubes tend to align themselves along the rotation axis. We demonstrate this alignment by calculating the magnetic moment of the proton Cooper pairs.}

    \item {We follow the real-time dynamics of the GPPE and RTGLE together with the crust potential, for the illustrative case  $\Theta=0$ and with the neutron and proton Cooper pairs interacting only via the gravitational potential. This gives rise to a collapsed condensate, with a crust angular momentum that displays glitches with signatures of self-organized criticality (SOC)~\cite{Bak_1987,Jensen_1998,Donald_Turcotte_1999,Melatos_2008,AK_verma_2022}.}
\end{itemize}

The remainder of this paper is organised as follows: In Section~\ref{sec:the_model}, we provide a comprehensive description of the GPPE and RTGLE models. Section~\ref{sec:units_dimension} outlines the units and dimensionless forms of GPPE and RTGLE, accompanied by an elucidation of the pseudospectral method employed for our study. Our results are presented in Section~\ref{sec:results}, followed by a discussion of conclusions in Section~\ref{sec:conclusions}.

\section{The Model}
\label{sec:the_model}
The total Lagrangian $\mathcal{L}$ governing the dynamics of neutron and proton Cooper pairs within the system is composed of distinct Lagrangians. Section~\ref{sec:neutron_lag} delves into $\mathcal{L}_{\rm n}$, which encapsulates the dynamics of neutron Cooper pairs. Similarly, Section~\ref{sec:proton_lag} focuses on $\mathcal{L}_{\rm p}$ describing the proton Cooper pairs, while the electromagnetic field is described by $\mathcal{L}_{\rm EM}$ (Section~\ref{sec:proton_lag}). The interactions between neutron and proton Cooper pairs are addressed through $\mathcal{L}_{\rm np}$ in Section~\ref{sec:neutron_proton_int_lag}. Finally, we present the governing equations of motion in Section~\ref{sec:total_lag} by using the total Lagrangian $\mathcal{L}$.

\subsection{Neutron Superfluid}
\label{sec:neutron_lag}
In a pulsar, neutrons form Cooper pairs that lead to a superfluid~\cite{MIGDAL_1959,Rezzolla:2018jee}. At temperatures below the transition temperature $T_{\lambda}$, these Cooper pairs lead to a Bose-Einstein condensate (BEC), characterized by a macroscopic complex wavefunction $\psi_n$. The Lagrangian of a weakly interacting rotating BEC in a self-gravitating potential $\Phi$ is given by~\cite{AK_verma_2022}:
\begin{eqnarray}
\mathcal{L}_{\rm n} &=& \frac{i\hbar}{2} \left( \psi_n^* \frac{\partial \psi_n}{\partial t}- \psi_n \frac{\partial \psi_n^*}{\partial t}\right) -\frac{\hbar^2}{2m_n}|\nabla \psi_n|^2  \nonumber\\
&-&\frac{g}{2} \bigg(|\psi_n|^2-\frac{\mu_n}{g}\bigg)^2 -m_n\Phi|\psi_n|^2- \frac{1}{8\pi G} (\nabla \Phi)^2 \nonumber \\
&-& V_{\theta} |\psi_n|^2+\frac{i\hbar}{2}({\bf\Omega} \times {\bf r}) \cdot (\psi_n\nabla \psi_n^* - \psi_n^*\nabla \psi_n)\,,
\label{eq:Lag_n}
\end{eqnarray}
where $g=4\pi a \hbar^2/m_n$ is the interaction strength between neutron Cooper pairs, $a$ is the s-wave scattering length, $m_n$ and $\mu_n$ are, respectively,
the mass and chemical potential of neutron Cooper pairs, $G$ is Newton's gravitational constant, and $\bf\Omega$ is the rotational velocity. Here $V_{\theta}$ represents the crust potential, which is located just above the outer core [see Fig.~\ref{fig:neutron_star_cartoon}]. This crust potential contains a lattice of atomic nuclei, with each lattice point acting as a pinning center where neutron vortices can be pinned, so they corotate with the crust [details about the crust potential are given in Section~\ref{sec:crust_pot}].

\subsection{Proton Superconductor}
\label{sec:proton_lag}
 Proton Cooper pairs, which also form in a pulsar, yield a Type II superconductor with an Abrikosov flux lattice~\cite{Gordon_1969}. This superconducting system can be described by the complex wavefunction $\psi_p$, coupled to a vector potential $\bf A$ and self-gravitating potential $\Phi$, resulting in the following Lagrangian, in which we include the rotational velocity $\bf\Omega$:
\begin{equation}
\begin{aligned}
\mathcal{L}_{\rm p} &= \frac{i\hbar}{2} \left( \psi_p^* \frac{\partial \psi_p}{\partial t}- \psi_p \frac{\partial \psi_p^*}{\partial t}\right) -\frac{1}{2m_p}|D_{\bf A}\psi_p|^2 \\
&- \frac{\alpha_s}{2}\bigg(|\psi_p|^2-\frac{\mu_p}{\alpha_s}\bigg)^2 -q \phi |\psi_p|^2 -m_p\Phi|\psi_p|^2 - V_{\theta} |\psi_p|^2\\
&+\frac{1}{2}({\bf \Omega} \times {\bf r}) \cdot (\psi_pD_{\bf A} \psi_p^* + \psi_p^*D_{\bf A} \psi_p)\,,
\end{aligned}
\label{eq:Lag_p}
\end{equation}
where $D_{\bf A} \equiv [\frac{\hbar}{i}\nabla -q {\bf A}$] is the magnetic gradient operator, $m_p$ is the mass of a proton Cooper pair, $\alpha_s$ is the interaction strength between the proton Cooper pairs, $\mu_p$ is the proton chemical potential, $q=2e$ is the charge of a proton Cooper pair, and $\phi$ is the electric scalar potential.

The evolution of the vector potential ${\bf A}$ follows the Maxwell equations, which can be obtained from the electromagnetic Lagrangian~\cite{Kosyakov2007}:
\begin{equation}
\begin{aligned}
\mathcal{L}_{\rm EM} &= \epsilon_0 \bigg[ -\frac{1}{2} [{\bf E}^2 - c^2({\bf B}-{\bf B}_{ext})^2] \\
+ {\bf E} &\cdot \bigg(-\nabla \phi - \frac{\partial {\bf A}}{\partial t }\bigg) -c^2({\bf B}-{\bf B}_{\rm ext}) \cdot (\nabla \times {\bf A})\bigg]\,,
\label{eq:lagrangian_EM}
\end{aligned}
\end{equation}
where $\bf E$ and $\bf B$ are the electric and magnetic fields, respectively, ${\bf B}_{\text{ext}}$ is the external magnetic field, and $c$ is the speed of light. In the context of neutron stars, ${\bf B}_{\text{ext}}$ is the mean internal magnetic field, which reorganizes into flux tubes when the proton subsystem condenses into a superconducting state. This mean magnetic field is subtracted in Eq.~\eqref{eq:lagrangian_EM} so that it only appears in terms of the curl of a uniform field in the Maxwell equation (see Eq.\eqref{eq:vector_pot}). If $\langle{\bf A}\rangle$ is periodic then $\langle{\bf B}\rangle=\langle\nabla \times {\bf A}\rangle=0$, where $\langle .\rangle$ denotes the spatial average. So, in our calculations using periodic boundary conditions, we use ${\bf A}$ that is solid-rotation-like at the position of the star and thus controls the mean magnetic field of the star ${\bf B}$, which we call ${\bf B}_{\rm ext}$ [see Eq.~\eqref{eq:vec_pot_initial}].

\subsection{Interaction between Neutron and Proton Cooper Pairs}
\label{sec:neutron_proton_int_lag}
We consider (i) the density-density and (ii) the current-current interactions between the neutron and proton subsystems~\cite{Drummond_2017, Eysden_2011} and use the Lagrangian
\begin{equation}
\begin{aligned}
\mathcal{L}_{\rm np} &= \gamma \bigg\{ g_{np} |\psi_n|^2 |\psi_p|^2\\
&-\frac{\hbar}{4i} ( \psi_n \nabla \psi_n^* - \psi_n^* \nabla \psi_n)\cdot \left[ \psi_p D_{\bf A} \psi_p^* + \psi_p^* D_{\bf A} \psi_p\right] \bigg\}\,,
\label{eq:interaction_lag}
\end{aligned}
\end{equation}
where $\gamma$ is the overall interaction strength, $g_{np}$ is the density-density coupling constant~\cite{Bhattacharya_1991}. The current-current interaction, given in the second row, causes neutron-superfluid vortices to drag proton-superconductor flux tubes. 

\subsection{Total Lagrangian: equations of motion}
\label{sec:total_lag}
By combining the Lagrangians~(\ref{eq:Lag_n})-(\ref{eq:interaction_lag}) we obtain the total Lagrangian
\begin{eqnarray}
    \mathcal{L} = \mathcal{L_{\rm n}}+\mathcal{L_{\rm p}}+\mathcal{L_{\rm EM}}+\mathcal{L_{\rm np}}\,.
\end{eqnarray}

The Euler-Lagrange equations for $\mathcal{L}$ yield the following:
\begin{itemize}
  \item the Gross-Pitaevskii-Poisson equation (GPPE) for neutron Cooper pairs (variation with respect to $\psi_n^*$):
\begin{eqnarray}
i\hbar\frac{\partial \psi_n}{\partial t} &=& -\frac{\hbar^2}{2m_n}\nabla^2 \psi_n-\mu_n \psi_n  + g|\psi_n|^2 \psi_n+m_n\Phi\psi_n\nonumber\\ 
 &+& i\hbar ({\bf \Omega} \times {\bf r}) \cdot \nabla\psi_n+ V_{\theta} \psi_n+\gamma g_{np}|\psi_p|^2 \psi_n\nonumber\\
 &-& \frac{\gamma\hbar}{2i} \left[ \nabla \psi_n \cdot {\bf J}_p + \nabla \cdot \left( \psi_n {\bf J}_p\right)\right]\,;
\label{eq:GPE_neutron}
\end{eqnarray}
here, ${\bf J}_p$ is the proton current density:
\begin{eqnarray}
    {\bf J}_p &=& \frac{\hbar}{2i}(\psi_p^*\nabla\psi_p - \psi_p\nabla \psi_p^*) - q {\bf A}_{\rm eff}|\psi_p|^2\,;
\label{eq:JnJp}
\end{eqnarray}  
\item  the real-time Ginzburg-Landau-Poisson equation (RTGLPE) for proton Cooper pairs (variation with respect to $\psi_p^*$):
\begin{equation}
\begin{aligned}
    i\hbar \left( \frac{\partial }{\partial t} + \frac{i}{\hbar}q\phi_{\rm eff} \right) \psi_p &= \frac{1}{2m_p}\bigg(\frac{\hbar}{i}\nabla- q{\bf A}_{\rm eff}\bigg)^2 \psi_p-\mu_p\psi_p \\
    &+\alpha_s |\psi_p|^2\psi_p+m_p \Phi \psi_p+ V_{\theta}\psi_p \\
    &+ \gamma g_{np} |\psi_n|^2 \psi_p+\gamma q({\bf J}_n\cdot {\bf A})\psi_p \\
    &- \frac{\gamma\hbar}{2i}[{\bf J}_n.\nabla \psi_p + \nabla \cdot({\bf J}_n \psi_p)]\,;
\label{eq:GLE_proton}
\end{aligned}
\end{equation}
here,
\begin{eqnarray}
    \phi_{\rm eff} &=& \phi-\frac{m_p}{2q}\Omega^2 r^2\,;\nonumber \\
    {\bf A}_{\rm eff} &=& {\bf A}+\frac{m_p}{q}({\bf \Omega} \times {\bf r})\,; \nonumber \\
    {\bf J}_n &=& \frac{\hbar}{2i}(\psi_n^*\nabla\psi_n - \psi_n\nabla \psi_n^*)\,;
    \label{eq:eff}
\end{eqnarray}    

\item  the Maxwell equation for the vector potential 
${\bf A}$ and Poisson equations for the gravitational potential $\Phi$ and the electric scalar potential $\phi$:
\begin{equation}
\begin{aligned}
    \frac{1}{c^2}\frac{\partial^2 {\bf A}}{\partial t^2} -\nabla^2 {\bf A}-\nabla \times {\bf B}_{\rm ext}&= \mathbb{P}\bigg[\frac{q}{m_pc^2\epsilon_0} {\bf J}_{p}\\
    &-\frac{\gamma q}{c^2\epsilon_0} {\bf J}_n|\psi_p|^2\bigg]\,;
\end{aligned}
\label{eq:vector_pot}
\end{equation}
\begin{equation}
    \begin{aligned}
        \nabla^2 \Phi &= 4\pi G \bigg(m_n|\psi_n|^2 +m_p|\psi_p|^2 -\rho_{\rm bg}\bigg)\,;
    \end{aligned}
\label{eq:grav_pot}
\end{equation}
\begin{eqnarray}
    \rho_{bg} = m_n\langle|\psi_n|^2\rangle+m_p\langle|\psi_p|^2\rangle\,;
\label{eq:back}
\end{eqnarray}
\begin{eqnarray}
    \nabla^2\phi = -\frac{1}{\epsilon_0} (q|\psi_p|^2-qn_p)\,.
\label{eq:scalar_pot}
\end{eqnarray}
\end{itemize}

With $\gamma=0$ and without crust potential ($V_{\theta}=0$), Eq.~(\ref{eq:GPE_neutron}) has been used extensively in Refs.~\cite{Chavanis_2011,Chavanis_2016,Chavanis_2018} to study self-gravitating BECs at temperature $T=0$ (by using a Gaussian Ansatz for $|\psi_n|^2$). References~\cite{Eric_2016,Guzman_2006} have performed numerical simulations without rotation (${\bf \Omega}=0$). Furthermore, Ref.~\cite{Madarassy_2015} has included rotation in the GPPE to study the dynamical properties of BEC dark matter (but still with $\gamma=0$ and $V_{\theta}=0$). In our previous studies, we have used the GPPE to study the formation of compact bosonic objects at finite temperatures~\cite{Akhilesh_2016} and their axionic counterparts, by including a quintic nonlinearity~\cite{shukla_2023}. The imaginary time ($t\to -it$) version of Eq.~(\ref{eq:GLE_proton}) with $\gamma=0$, $V_{\theta}=0$, and $\Phi=0$ is the well-known time-dependent Ginzburg-Landau equation~\cite{tinkham2004introduction} whose solutions give Type I and Type II superconductors. In conventional calculations for superconductors, the term $\nabla\times {\bf B}_{\rm ext}$ in Eq.~\eqref{eq:vector_pot} vanishes because of the uniformity of ${\bf B}_{\rm ext}$; and it appears only as a boundary condition. In our calculations, which use periodic boundary conditions, we consider ${\bf B}_{\rm ext}$ such that it is periodic in the domain; and it reorganises itself in the form of flux tubes as time progresses. In the context of neutron stars, this corresponds to the mean internal magnetic field.

In writing Eq.~(\ref{eq:vector_pot}), we have used the Coulomb gauge $\nabla \cdot {\bf A}=0$. The Helmholtz projector, which has components $\mathbb{P}_{ij}:= \delta_{ij}-\mathcal{F}^{-1} \frac{k_ik_j}{k^2}\mathcal{F}$, with $\mathcal{F}$ the Fourier-transform operator, projects a field onto its divergence-free part; its application in Eq.~(\ref{eq:vector_pot}) maintains the Coulomb gauge. In Eq.~(\ref{eq:grav_pot}) for $\Phi$, the gravitational potential, the subtraction of the background mean density $\rho_{\rm bg}$ [often called the Jeans Swindle~\cite{Falco_2013}]  can be justified either by defining a Newtonian cosmological constant~\cite{KIESSLING_2003} or by accounting for cosmological expansion~\cite{Peebles_1980,Falco_2013}. Furthermore, in Eq.~(\ref{eq:scalar_pot}) for the scalar potential $\phi$, we subtract the mean charge density, $qn_p$, to maintain charge neutrality in the system. In the context of a neutron star, $qn_p$ corresponds to the background charge coming from electrons. 

The important term considering the interaction between neutron and proton Cooper pairs is the last term in Eq.~(\ref{eq:vector_pot}), which is of the form $\frac{\gamma q}{c^2\epsilon_0} {\bf J}_n|\psi_p|^2$. This term, not considered hitherto in the GPPE and RTGLPE, causes neutron-superfluid vortices to drag proton-superconductor flux tubes, generate an entrained-proton current, because of which the neutron-superfluid vortices become magnetized, as we show below.
\begin{figure}[!htb]
    \centering 
    \includegraphics[scale=0.16]{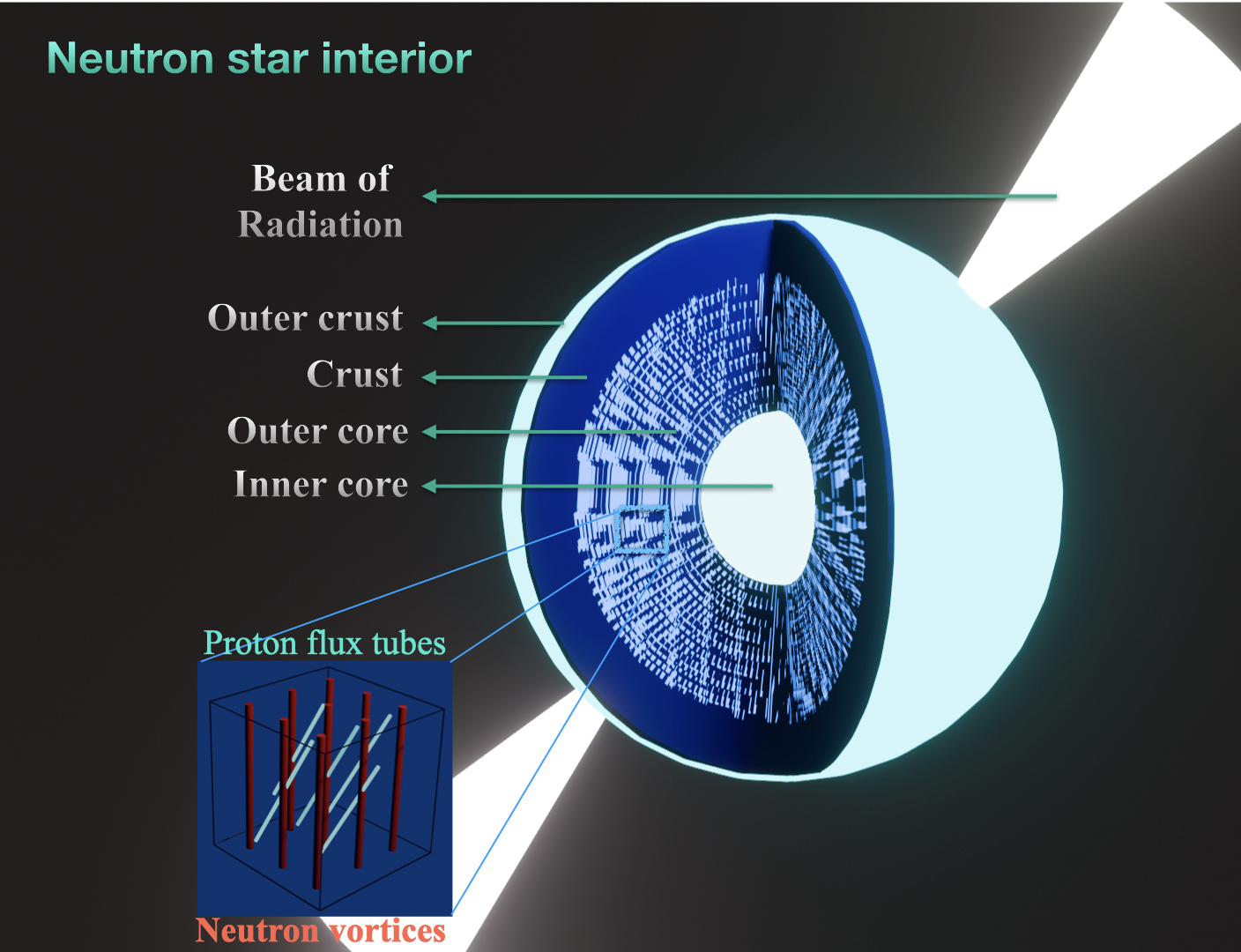}
    \caption{A schematic diagram [cf. Refs.~\cite{lattimer2004physics,Gibney_nature_2017}] of the interior of a pulsar (magnetized neutron star). The light-blue luminous central region represents the \textit{inner core}, characterized by ultra-dense matter where neutrons and protons break down into quarks and gluons. Surrounding this core lies the \textit{outer core} (shaded blue-white), composed of a neutron superfluid and proton superconductor, with neutron-superfluid vortices and proton-superconductor flux tubes, respectively [magnified view in the bottom-left inset]. The dark-blue crust has a crystalline lattice structure (not shown) and consists of heavy atomic nuclei and free neutrons and free electrons. The neutrons in the crust exist in the form of a superfluid that is threaded by vortices. The glowing white region, often called the outer crust, comprises atomic nuclei and free electrons. The white conical regions show radiation beams emerging from the poles of the pulsar. }
    \label{fig:neutron_star_cartoon}
\end{figure}

\subsection{Crust Potential ($V_{\theta}$)}
\label{sec:crust_pot}

The region just above the outer core of a neutron star, known as the crust, contains heavy nuclei arranged in a lattice structure. This crust plays a crucial role in models of pulsar glitches. Neutron vortices become pinned to the lattice sites and corotate with the crust. As the crust spins down, the superfluid within remains unaffected because of its zero viscosity. This differential rotation causes vortices to unpin from their pinning sites, thus transferring momentum to the crust and resulting in glitches. Flux tubes are also anchored to the crust by the strong magnetic field, and there is a depletion of the proton cooper pairs. In our model, spherically collapsed neutron and proton condensates contain vortices and flux tubes that are located away from the cubic domain boundary. The crust potential lies just above the condensate, with the magnetic field inside the flux tubes passing through the crust anchoring them. In this Section, we model the crust using a Gaussian potential $V_{\theta}$ modulated by equally spaced pinning centers. In the absence of the crust potential $V_{\theta}$, Eqs.~(\ref{eq:GPE_neutron})-(\ref{eq:scalar_pot}) govern the interplay between neutron-superfluid vortices and proton-superconductor flux tubes in the outer-core region shown in the schematic diagram of Fig.~\ref{fig:neutron_star_cartoon}. 
%The potential $V_{\theta}$ comes into play in the crust, which lies between the envelope and the outer core of a neutron star and provides pinning sites for vortices and flux tubes (see Fig.~\ref{fig:neutron_star_cartoon} and Ref.~\cite{lattimer2004physics}). 
At the level of a minimal model for pulsars~\cite{AK_verma_2022} the dynamics of this crust is characterised by a single polar angle $\theta$~\cite{AK_verma_2022} that evolves as follows:
\begin{equation}
\begin{aligned}
I_c\frac{d^2\theta}{dt^2} &= \frac{1}{N_n}\bigg(\int d^3x\partial_{\theta}V_{\theta}|\psi_n|^2 +\frac{n_n}{n_p}\int d^3x\partial_{\theta}V_{\theta}|\psi_p|^2\bigg)\\
&-\delta \frac{d\theta}{dt}\,;\\
V_{\theta} ({\bf r}_p) &= V_0\exp \left[-\frac{(|{\bf r}_p|-r_{\rm crust})^2}{(\Delta r_{\rm crust})^2}\right] \tilde{V}(x_{\theta},y _{\theta})\,; 
 \end{aligned}
 \label{eq:crust_pot_eq}
\end{equation}
here, $I_c$ is the moment of inertia of the crust, the angle $\theta$ represents the angular rotation of the neutron star's crust about the rotation axis, $N_n = \int |\psi_n|^2 d^3 x$ is the total number of neutron Cooper pairs, $n_n/n_p$ is the ratio of the number densities of neutrons and protons, and the slowing down of the crust is controlled by the friction coefficient $\delta$. The first two terms on the right-hand side of upper Eq.~\eqref{eq:crust_pot_eq} couple the crust to the superfluid and superconductor, respectively. These terms ensure that the superfluid and superconductor act on the crust. The last term on the right-hand side of upper Eq.~\eqref{eq:crust_pot_eq} represents the friction, which slows down the crust and creates a differential rotation between it and the superfluid. The evolution Eq.~\eqref{eq:crust_pot_eq} for the crust potential can be written in the compact form $I_c\ddot{\theta}={\rm F}_s -\delta \dot{\theta}$, where ${\rm F}_s$, the force of the superfluid on the crust, is given by the term in parentheses in the first line of Eq.~\eqref{eq:crust_pot_eq}. 

We choose $\tilde{V} (x_{\theta},y_{\theta})=3+\cos(n_{\rm crust}x_{\theta})+\cos(n_{\rm crust}y_{\theta})$, with $x_{\theta}=\cos(\theta)x_p+\sin(\theta)y_p$ and $y_{\theta}=-\sin(\theta)x_p+\cos(\theta)y_p$ as in Ref.~\cite{AK_verma_2022}; here, $n_{\rm crust}$ determines the number of pinning sites in the crust, $r_{\rm crust}$ is the radius at which $V_{\theta}$ assumes its maximum value, and $\Delta r_{\rm crust}$ is the thickness of the crust. We use a $2\pi$-periodic version of the coordinates, namely, ${\bf r}_p= (x_p,y_p,z_p)$, which is $\pi$-centered,  with $z_p=\pi$,
\begin{eqnarray}
    x_p&=&-\sum_{n=1}^{10}\exp(-\frac{16}{100} n^2)(-1)^n \frac{\sin(n(x-\pi))}{n}\,,\nonumber\\
    {\rm{and}}\,\,y_p&=&-\sum_{n=1}^{10}\exp(-\frac{16}{100} n^2)(-1)^n \frac{\sin(n(y-\pi))}{n}\,. \nonumber\\
    \label{eq:x_y_periodic}
\end{eqnarray}

For the proton superconductor, in the absence of rotation, we have ${\bf A}_{\rm eff}\to{\bf A}$ and $\phi_{\rm eff}\to \phi$ [Eq.~\ref{eq:eff}]. A superconductor that is subjected to rotation (${\bf \Omega} \neq 0$) and which is in a uniform external magnetic field displays a captivating interplay of quantum phenomena. Consider first a non-rotating Type-II superconductor in an external magnetic field; it can display a vortex-lattice phase in which flux tubes are arranged in the form of an Abrikosov lattice~\cite{Abrikosov_2004}; the quantized magnetic flux $\Phi_B=\int {\bf A} \cdot d{\bf l} $ passes through each vortex. Each of these magnetised vortices contributes a discrete quantum of magnetic flux to the net magnetic field inside the superconductor, which is zero outside the vortices. Next consider a rotating superconductor without an external magnetic field; this displays a uniform magnetic field, known as the London moment~\cite{Verheijen_1990,Hipkins_1996}. [This field is uniform away from the boundary, for distances larger than $\lambda_p$, the London penetration depth.] The London moment follows from  ${\bf A}_{\rm eff}$ [Eq.~\ref{eq:eff}] because the RTGLPE [Eq.~\ref{eq:GLE_proton}] has an additional term with a vector potential (the subscript $L$ stands for London)
\begin{equation}
    {\bf A}_{\rm L} = -\frac{m_p}{q}({\bf \Omega }\times {\bf r}) \,,
    \label{eq:london_vecpot}
\end{equation}
so that, in the absence of flux tubes,  ${\bf A}_{\rm eff}=0$ inside the superconductor. For a rotating Type II superconductor~\cite{Egor_2014}, in a uniform magnetic field, there is a critical rotational speed ${\rm \Omega}_c^p$ beyond which vortices (here, proton-superconductor flux tubes) enter the system. The critical ${\rm \Omega}_c^p$, which follows by minimizing the energy in the rotating frame $E'\equiv E - {\bf \Omega} \cdot {\bf L}_z$, with ${\bf L}_z$ the angular momentum along the rotation axis, is
\begin{eqnarray}
    {\rm \Omega}_c^p = \frac{\hbar }{m_p\lambda_p^2}\ln \bigg(\frac{\lambda_p}{\xi_p}\bigg)\,,
    \label{eq:omega_c_super}
\end{eqnarray}
where $\xi_p$, and $\lambda_p$ are, respectively, the superconducting coherence length and the London penetration depth.
In the Abrikosov-lattice phase, a London moment ($\propto {\bf \Omega }$) is present inside the superconducting together with flux-tube lattice. 

\section{Units and Numerical Method}

\subsection{Non-dimensionalisation}
\label{sec:units_dimension}
We use the dimensionless forms of Eqs.~(\ref{eq:GPE_neutron})-(\ref{eq:scalar_pot}), which we obtain by using the general reference length $L_{\rm ref}$ and speed $V_{\rm ref}$. The scaled position ${\bf x}$, time $t$, vector potential ${\bf A}$, and scalar potential $\phi$ are
\begin{eqnarray}
    {\bf x} &=& L_{\rm ref}{\bf x}'\,, \nonumber\\
    t &=& \frac{L_{\rm ref}}{V_{\rm ref}} t'\,, \nonumber\\
    {\bf \Omega} &=& \frac{L_{\rm ref}}{V_{\rm ref}} {\bf \Omega}'\,,\nonumber\\
    {\bf A} &=&  \frac{H_{\rm c2}L_{\rm ref}}{\kappa} {\bf A}^{'}\nonumber\,, \\
    {\rm {and}} \,\, \phi &=& \frac{ L^2_{\rm ref}}{\tau} \frac{H_{\rm c2}}{\kappa}  \phi^{'}\,,
    \label{eq:non_dimensional}
\end{eqnarray}
 where $H_{\rm c2}$ is the (zero-temperature) upper critical magnetic field of the superconductor, $\kappa = \frac{\lambda}{\xi_p}$ is the London ratio, and $\tau = \frac{L_{\rm ref}}{V_{\rm ref}}$. In Table~\ref{tab:paramters}, we provide all the parameters and dimensionless ratios that follow from our non-dimensionalization. The wavefunctions are normalized as $\psi_n=\sqrt{n_n}\psi_n'$ and $\psi_p=\sqrt{n_p}\psi_p'$, so the non-dimensionalised neutron GPPE, proton RTGLPE, and vector, gravitational, and scalar potential equations are, respectively, (for notational simplicity we now drop the 
 primes that come from non-dimensionalization):
\begin{equation}
\begin{aligned}
i\frac{\partial \psi_n}{\partial t} &= -\alpha \nabla^2 \psi_n+ \beta (|\psi_n|^2-1) \psi_n  + \mathfrak{G} \Phi\psi_n  \\ 
&+i ({\bf \Omega} \times {\bf r}) \cdot \nabla\psi_n+ V_{\theta} \psi_n+ \gamma_p \mathfrak{g} |\psi_p|^2 \psi_n \\
&-\frac{\gamma_p\alpha}{i}  \left[ \nabla \psi_n \cdot {\bf J}_p + \nabla \cdot \left( \psi_n {\bf J}_p\right)\right]\,;
\end{aligned}
  \label{eq:GPE_neutron_ndim}
\end{equation}
\begin{equation}
\begin{aligned}
i \left( \frac{\partial }{\partial t} + i\frac{L_{ref}^2}{\kappa \xi_p^2} \phi_{\rm eff}\right) \psi_p &= \alpha \bigg(\frac{\nabla}{i}-\frac{L_{\rm ref}^2}{\xi_p^2\kappa} {\bf A}_{\rm eff}\bigg) \psi_p +\mathfrak{G}\Phi \psi_p\\
&+\beta \frac{\xi_n^2}{\xi_p^2}(|\psi_p|^2 -1)\psi_p +V_{\theta}\psi_p  \\
&+ \gamma_n \mathfrak{g} |\psi_n|^2 \psi_p\\
&-  \gamma_n \alpha (2{\bf J}_n \cdot D_{\bf A}+i  \psi_p \nabla \cdot {\bf J}_n)\,;
\end{aligned}
\label{eq:GLE_proton_ndim}
\end{equation}
\begin{equation}
\begin{aligned}
\frac{V_{ref}^2}{c^2}\frac{\partial^2 {\bf A}}{\partial t^2}-\nabla^2 {\bf A} -\nabla \times {\bf B}_{\rm ext}&=& \mathbb{P}\bigg[\frac{1}{\kappa} {\bf J}_{p}-\frac{\gamma_n}{\kappa} {\bf J}_n|\psi_p|^2\bigg]\,;
\label{eq:vector_pot_ndim}
\end{aligned}
\end{equation}
\begin{eqnarray}
    \nabla^2 \Phi &=&|\psi_n|^2 +\frac{n_p}{n_n}|\psi_p|^2 -n_{\rm bg}\,;
\label{eq:gravity_ndim}
\end{eqnarray}
\begin{eqnarray}
\nabla^2\phi &=& -\frac{\beta}{\kappa} \bigg(\frac{c}{c_s}\bigg)^2 (|\psi_p|^2-1)\,.
\label{eq:scalar_pot_ndim}    
\end{eqnarray}
The dimensionless current densities and effective vector and scalar potentials are, respectively: 
\begin{eqnarray}
    {\bf J}_n &=& \frac{1}{2i} (\psi_n^*\nabla \psi_n-\psi_n\nabla\psi_n^*)\,; \nonumber\\
{\bf J}_p &=& \frac{1}{2i} (\psi_p^*\nabla \psi_p-\psi_p\nabla\psi_p^* ) - \frac{L_{ref}^2}{\xi_p^2\kappa} {\bf A}_{\rm eff}|\psi_p|^2\,;\nonumber\\
    {\bf A}_{\rm eff}&=&{\bf A}+\frac{\xi_p^2}{L_{\rm ref}^2} \frac{\kappa}{2\alpha} ({\bf \Omega }\times {\bf r})\,;\nonumber\\
    { \phi}_{\rm eff}&=&{\phi}-\frac{\xi_p^2}{L_{\rm ref}^2} \frac{\kappa}{4\alpha} ({ \Omega^2 }{r^2})\,.
\end{eqnarray}
For the convenience of the reader, we define all the parameters and dimensionless ratios in Table~(\ref{tab:paramters}). 
\begin{table}[!hbt]
    \centering
    \begin{tabular}{|c|c|c|}
    \hline
    \thead{Parameters \\in Eqs.(\ref{eq:GPE_neutron_ndim})-(\ref{eq:scalar_pot_ndim})} & Description \\
    \hline
    $\alpha = \frac{c_s\xi_n}{\sqrt{2}L_{ref}V_{ref}}$  & \thead{Coefficient of the kinetic term \\in  Eqs.(\ref{eq:GPE_neutron_ndim}) and (\ref{eq:GLE_proton_ndim})}\\
    \hline
    $\beta = \frac{c_sL_{ref}}{\sqrt{2}\xi_nV_{ref}}$   & \thead{Coefficient of the nonlinear term \\in  Eqs.(\ref{eq:GPE_neutron_ndim}) and (\ref{eq:GLE_proton_ndim})}  \\
    \hline
    $\thead{\mathfrak{G}=\frac{L_{ref}^3 2\sqrt{2}\pi Gm_n n_n}{V_{ref}c_s\xi_n}}$ & Gravitational strength\\
    \hline
    $\thead{\mathfrak{g}=\frac{L_{ref}g_{np}}{\sqrt{2} \xi_n V_{ref}c_sm_nm_p}}$  & Density-Density coupling strength\\
    \hline
    $\thead{\frac{n_p}{n_n}}$ & \thead{Number-density ratio \\ (protons to neutrons)} \\
    \hline
    $\thead{\gamma}$  & Dimensional interaction strength \\
    \hline
    $\thead{\gamma_n=\gamma m_nn_n}$ & Interaction coefficient for protons\\
    \hline
    $\thead{\gamma_p=\gamma m_pn_p}$ & Interaction coefficient for neutrons\\
    \hline
    $\thead{\xi_n}=\frac{\hbar}{\sqrt{2m_n gn_n}}$ & \thead{Coherence length for \\neutron Cooper pairs}\\
    \hline
    $\thead{\xi_p}=\frac{\hbar}{\sqrt{2m_p\alpha_s n_p}}$   & \thead{Coherence length for \\ proton Cooper pairs}\\
    \hline
    $\thead{\lambda_p}$ & London penetration depth\\
    \hline
    $\thead{\frac{c}{c_s}}$ & \thead{Ratio of speed of light to \\ speed of sound}\\
    \hline
    $\thead{\kappa=\frac{\lambda_p}{\xi_p}}$ & London ratio\\
    \hline
    $\thead{ \Omega}$ & Dimensionless rotational speed\\
    \hline
    \end{tabular}
    \caption{Definitions of all the dimensionless parameters and ratios appearing in Eqs.~(\ref{eq:GPE_neutron_ndim})-(\ref{eq:scalar_pot_ndim}).}
    \label{tab:paramters}
\end{table}

We use pseudospectral direct numerical simulations (DNSs) to solve Eq.~(\ref{eq:crust_pot_eq}) and Eqs.~(\ref{eq:GPE_neutron_ndim})-(\ref{eq:scalar_pot_ndim}) in a cubic domain, with side $L=2\pi$ and $N^3$ collocation points, and periodic boundary conditions in all three directions. We employ the Fourier expansion for the function $\Psi\equiv (\psi_n,\psi_p,A_x,A_y,A_z,\Phi,\phi)$ as follows
\begin{eqnarray}
\Psi ({\bf x}) = \sum_{ {\bf k} } \hat{\Psi}_{{\bf k}} \exp (i {\bf k} \cdot {\bf x}) \,, 
\label{eq:Fourier}
\end{eqnarray}
and the $2/3$-rule for dealiasing, i.e., we truncate the Fourier modes by setting $\hat{\Psi}_{\bf k} \equiv 0$ for $ |{\bf k}| > k_{max}$~\cite{giorgio_2011,HOU_dealiasing}, with $k_{max} = [N/3]$. Given current computational resources, it is well-nigh impossible
to use such a DNS with astrophysically realistic values (say for a pulsar) for the parameters and ratios in Table~(\ref{tab:paramters}).
Nevertheless, as we show below, it is possible to obtain a large body of results that are qualitatively relevant for (a) interactions between proton-superconductor flux tubes and neutron-superfluid vortices and (b) a minimal model for pulsars and their glitches~\cite{AK_verma_2022}.

We will use the imaginary time versions of equations in the initial parts of the results, which can be obtained by using the substitution $t\to -it$ in Eqs.\eqref{eq:GPE_neutron_ndim}-\eqref{eq:GLE_proton_ndim}. The imaginary-time version of the Maxwell equation~\eqref{eq:vector_pot_ndim} is the following first-order partial differential equation [similar to the 
vector-potential equation used in the formulation of the time dependent Ginzburg-Landau model of superconductivity~\cite{Kato_PhysRevB}]:
\begin{eqnarray}
     \frac{V_{\rm ref}^2}{c^2}\frac{\partial {\bf A}}{\partial t} - \nabla^2 {\bf A} - \nabla \times {\bf B}_{\rm ext} = \mathbb{P}\bigg[\frac{1}{\kappa}{\bf J}_p-\frac{\gamma_n}{\kappa}{\bf J}_n |\psi_p|^2 \bigg]\,.\nonumber\\
     \label{eq:imaginary_time_vec_pot}
\end{eqnarray}

\subsection{Initial conditions}
\label{sec:initial_conditions}
To solve imaginary-time ($t \to -i t$) versions of the GPPE~\eqref{eq:GPE_neutron_ndim}, the RTGLE~\eqref{eq:GLE_proton_ndim}, and the Maxwell equation~\eqref{eq:imaginary_time_vec_pot}, we use the following initial conditions:
\begin{itemize}
    \item {\bf ICI1}: The imaginary-time ($t \to -i t$) version of GPPE with ${\bf \Omega}=0$ is first evolved by using a uniform density distribution and with small superimposed perturbations. This gives a spherically collapsed condensate. We now use this collapsed state as an initial condition in the same equation but with a small value of ${\bf \Omega}$. We keep increasing ${\bf \Omega}$ in small steps until we get a collapsed object threaded by vortices.
    \item {\bf ICI2}: {For the RTGLE, we follow the procedure used in {\bf ICI1}, but we insert vortices initially by choosing
    \begin{eqnarray}
        \psi_{pi} = \psi_{\rm uni}\times [\cos(kx)+i \cos(ky)]^n\,,
    \end{eqnarray}
    where $\psi_{\rm uni}$ is a uniform density distribution with small superimposed perturbations. Here, the integer $n$ denotes the multiplicity of a vortex; and $k$ is the number of vortices in the interval [$-\pi/n,\pi/n$].
    }
    \item {\bf ICI3}: {For the imaginary-time version of the Maxwell equations~\eqref{eq:imaginary_time_vec_pot}, we use the following initial condition: 
    \begin{eqnarray}
        A_x &=& -\frac{1}{2}\times y_p\times B_{\rm ext}\,,\nonumber\\
        A_y &=& \frac{1}{2}\times x_p\times B_{\rm ext}\,,
        \label{eq:vec_pot_initial}
    \end{eqnarray}
    where $x_p$ and $y_p$ are the periodic versions of the coordinates [Eqs.~\eqref{eq:x_y_periodic}] and $B_{\rm ext}$ is the uniform external magnetic field in the $z$-direction.
    }
\end{itemize}

\section{Results}
\label{sec:results}
Our results are presented in the following Subsections:

\begin{itemize}
    \item {Section~\ref{sec:imag_time_no_int_no_Vtheta_p}: we solve the imaginary-time versions of Eqs.~(\ref{eq:GPE_neutron_ndim})-(\ref{eq:scalar_pot_ndim}) without any interactions ( $\gamma=0$), no crust potential ($V_{\theta}=0$), and $\Theta=0$, where $\Theta$ is the angle between the rotation axis and external magnetic field. Note that, even if there is no direct interaction between the neutron-superfluid and the proton-superconductor (i.e., $\gamma=0$), they interact indirectly through the gravitational Poisson equation~(\ref{eq:gravity_ndim}). }
    \item {Section~\ref{sec:imag_time_int_no_Vtheta_p}: 
    we solve the imaginary-time  versions of Eqs.~(\ref{eq:GPE_neutron_ndim})-(\ref{eq:scalar_pot_ndim}), but with $\gamma \neq0$,  $V_{\theta}=0$, and $\Theta=0$.}
    \item {Section~\ref{sec:imag_time_non_zero_angle}: 
     we solve the imaginary-time  versions of Eqs.~(\ref{eq:GPE_neutron_ndim})-(\ref{eq:scalar_pot_ndim}), but with $V_{\theta}=0$ and $\Theta=30^{\circ}$ and 
    (i) $\gamma=0$ [Sec.~\ref{sec:imag_time_non_zero_angle_no_int_no_Vtheta}] and (ii) $\gamma\neq 0$ [Sec.~\ref{sec:imag_time_non_zero_angle_int_no_Vtheta}].}
    \item {Section~\ref{sec:real_time}: we solve the \textit{real-time} Eqs.~(\ref{eq:GPE_neutron_ndim})-(\ref{eq:scalar_pot_ndim}) with non-zero crust potential ($V_{\theta}\neq0$), no direct interactions ($\gamma=0$), and $\Theta=0$.} Note that the \textit{imaginary-time} evolution in Secs.~\ref{sec:imag_time_no_int_no_Vtheta_p}, \ref{sec:imag_time_int_no_Vtheta_p}, and \ref{sec:imag_time_non_zero_angle} has no dynamical significance; this evolution just provides us with a convenient way of obtaining the equilibrium configuration at very large imaginary time. 
\end{itemize}

\subsection{Imaginary-time study : $\gamma=0$, $V_{\theta}=0$, $\Theta=0$}
\label{sec:imag_time_no_int_no_Vtheta_p}
We solve imaginary-time ($t\to -i t$) versions of Eqs.~(\ref{eq:GPE_neutron_ndim})-(\ref{eq:vector_pot_ndim}) together with Eqs.~(\ref{eq:gravity_ndim})-(\ref{eq:scalar_pot_ndim}) with $\gamma=0$, $V_{\theta}=0$, and $\Theta=0$. The imaginary-time version of the Maxwell equation~\eqref{eq:vector_pot_ndim} is given in Eq.~\eqref{eq:imaginary_time_vec_pot}. For all the imaginary-time studies, we start with the initial conditions ${\bf ICI1}$, ${\bf ICI2}$, and ${\bf ICI3}$ given in Sec.~\ref{sec:initial_conditions}. The neutron star in our model rotates with an angular velocity ${\bf \Omega} =\Omega \hat{z}$; for specificity, we choose $\Omega=2.5$. Both the neutron-superfluid and proton-superconductor subsystems also rotate with this frequency; if we include an external magnetic field ${\bf B}_{\rm ext }$, the proton-superconductor responds directly to it. 
However, to isolate the effects of the rotation and the external magnetic field, it is useful to study the following three cases: (i) the neutron-superfluid rotates, but not the proton-superconductor, which is in an external magnetic field; (ii) both the neutron-superfluid and the proton-superconductor rotate, but there is no external magnetic field; and (iii) both the neutron-superfluid and the proton-superconductor rotate, and there is an external magnetic field. Clearly, only case (iii) is directly relevant to neutron stars. 

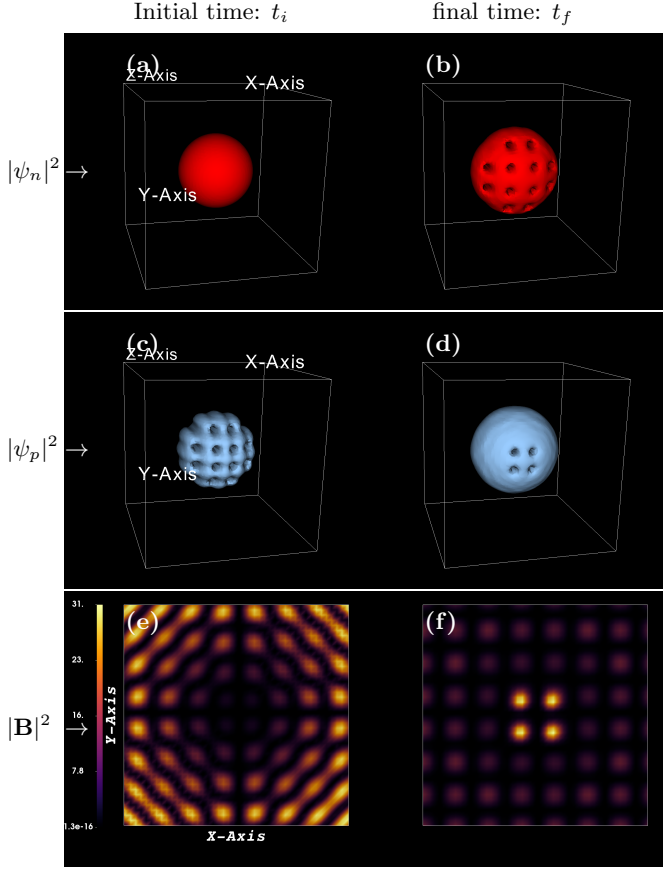
\begin{figure}[!htb]
    \centering
    \includegraphics[scale=0.11]{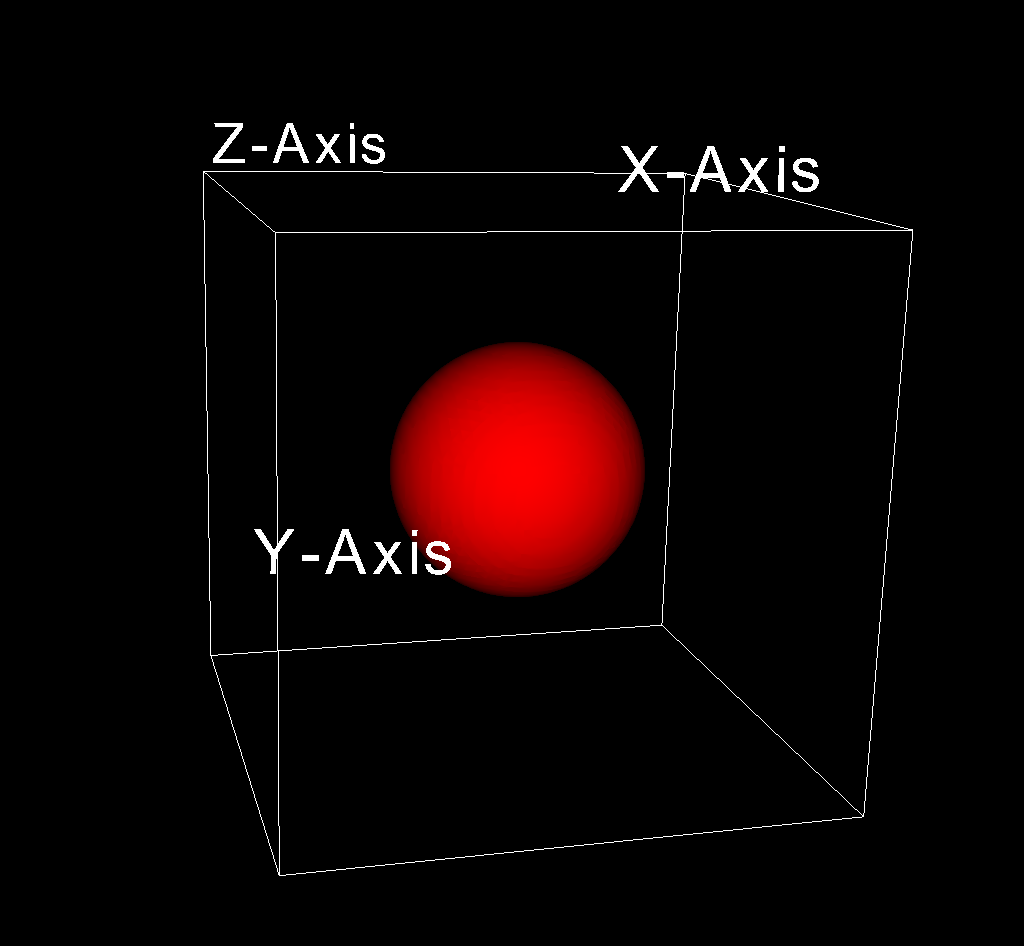}
    \put(-90,90){\color{white} \bf (a)}
    \put(-90,110){ Initial time: $t_i$}
    \put(-138,50){ $|\psi_n|^2 $}
    \put(-116,50){ \color{white} $\bf \rightarrow$}
    \includegraphics[scale=0.11]{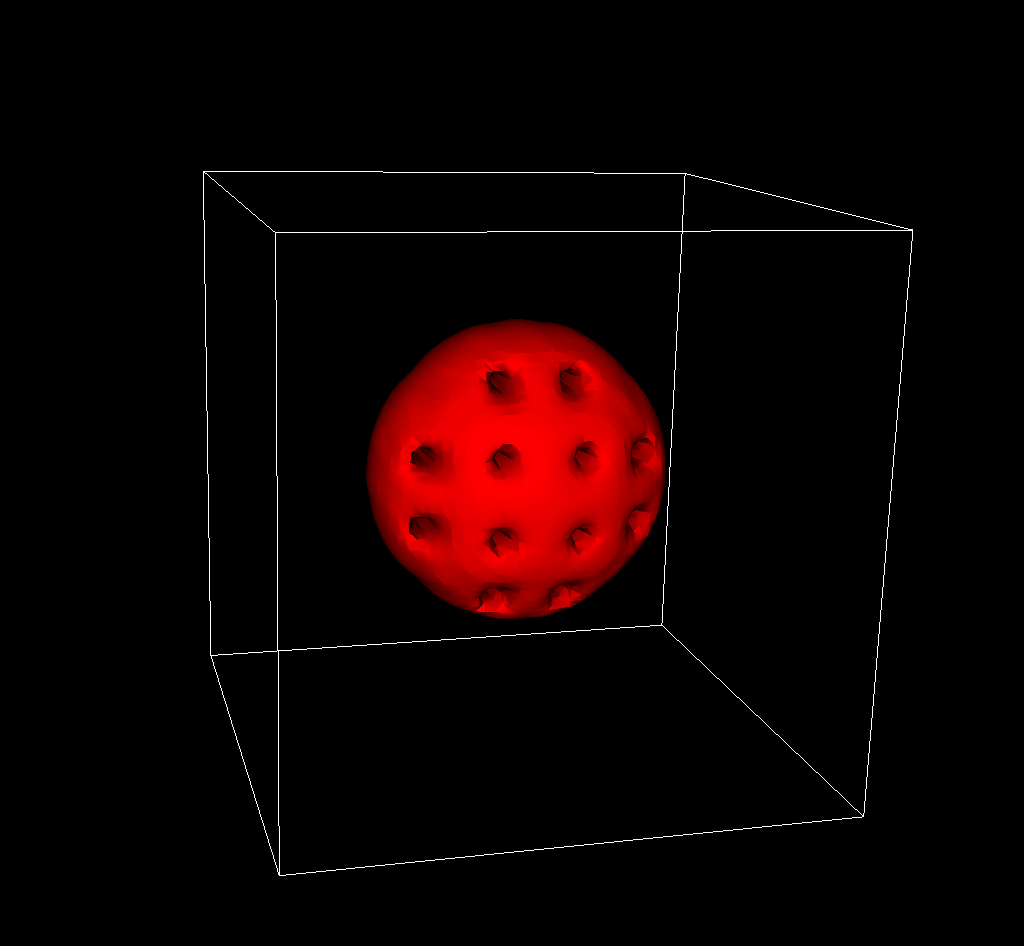}
    \put(-90,90){\color{white} \bf (b)}
    \put(-90,110){ final time: $t_f$}

    \includegraphics[scale=0.11]{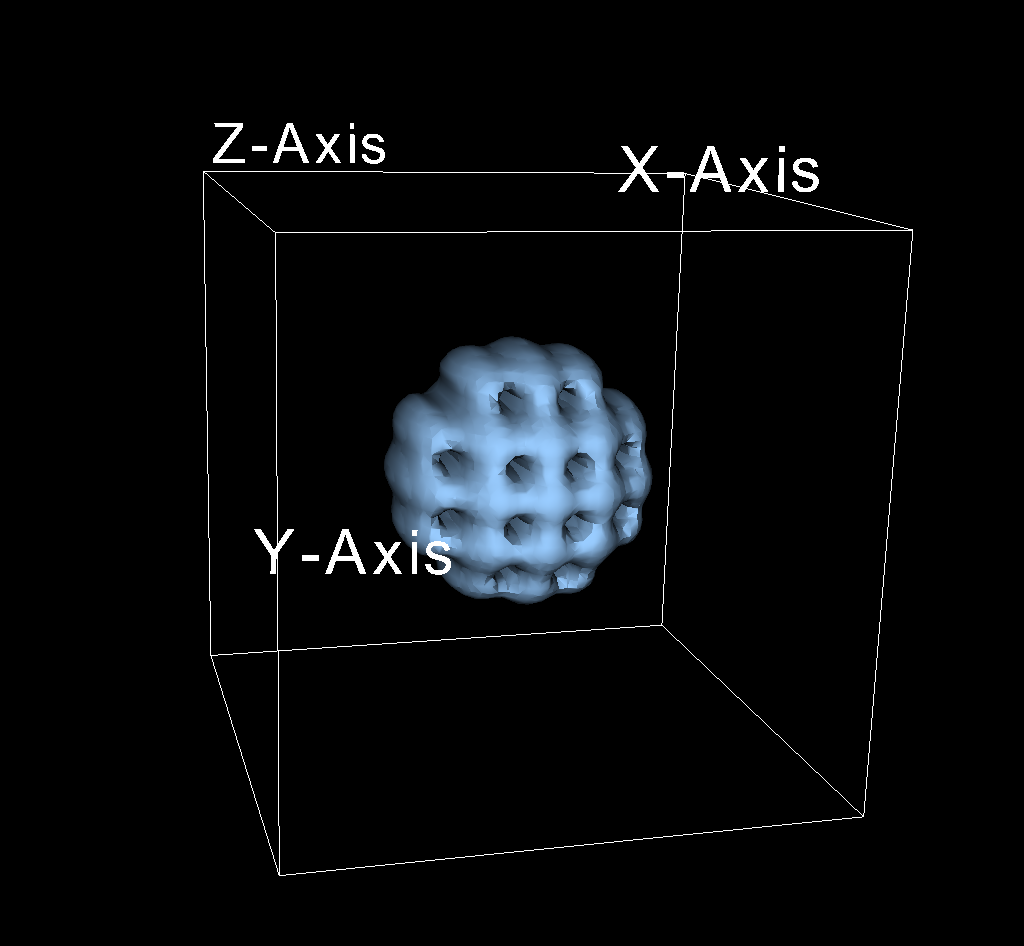}
    \put(-90,90){\color{white} \bf (c)}
    \put(-138,50){ $|\psi_p|^2 $}
    \put(-116,50){ \color{white} $\huge \to$}
    \includegraphics[scale=0.11]{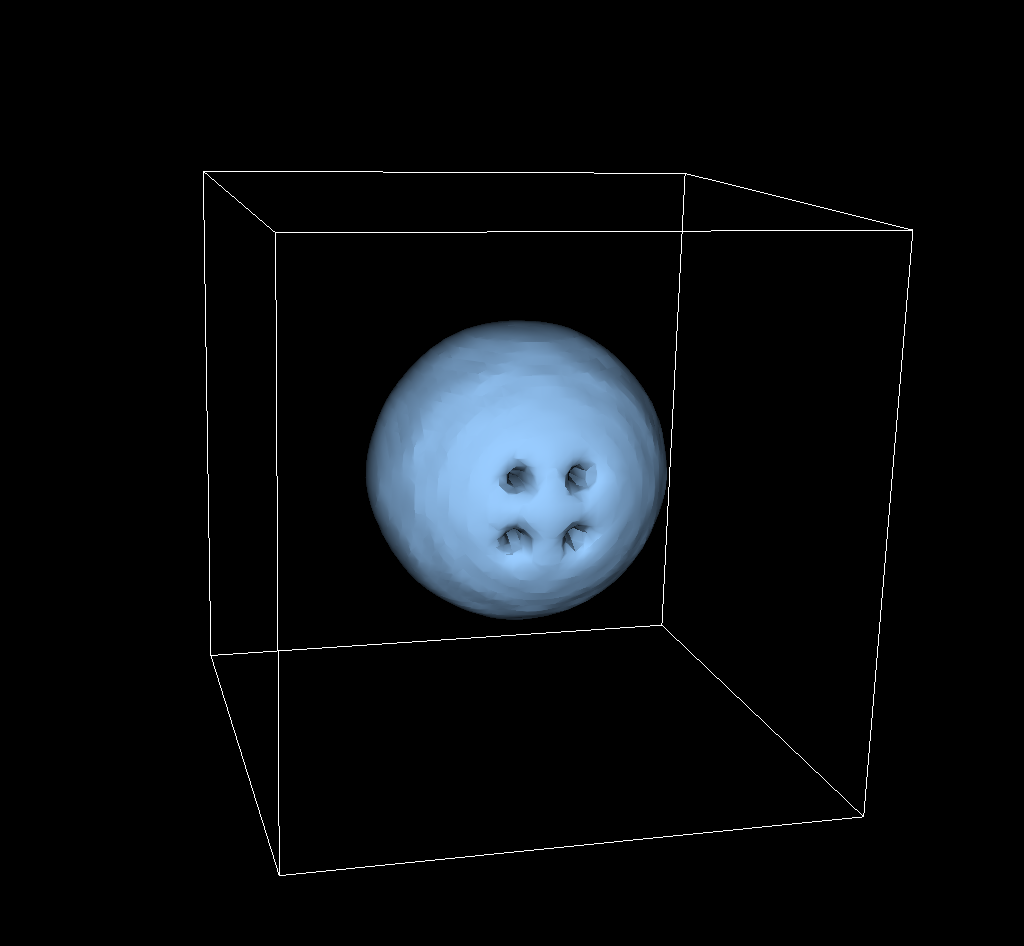}
    \put(-90,90){\color{white} \bf (d)}

    \includegraphics[scale=0.11]{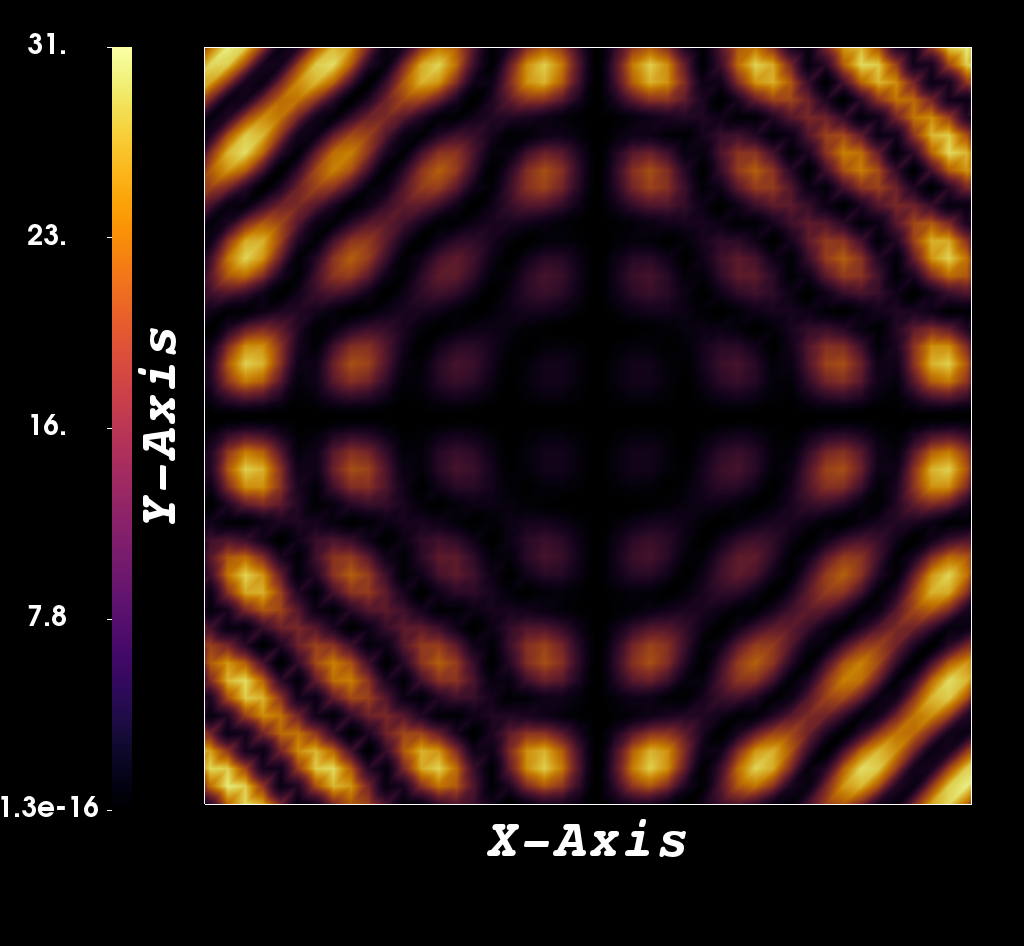}
    \put(-90,90){\color{white} \bf (e)}
    \put(-138,50){ $|{\bf B}|^2 $}
    \put(-116,50){ \color{white} $\huge \to$}
    \includegraphics[scale=0.11]{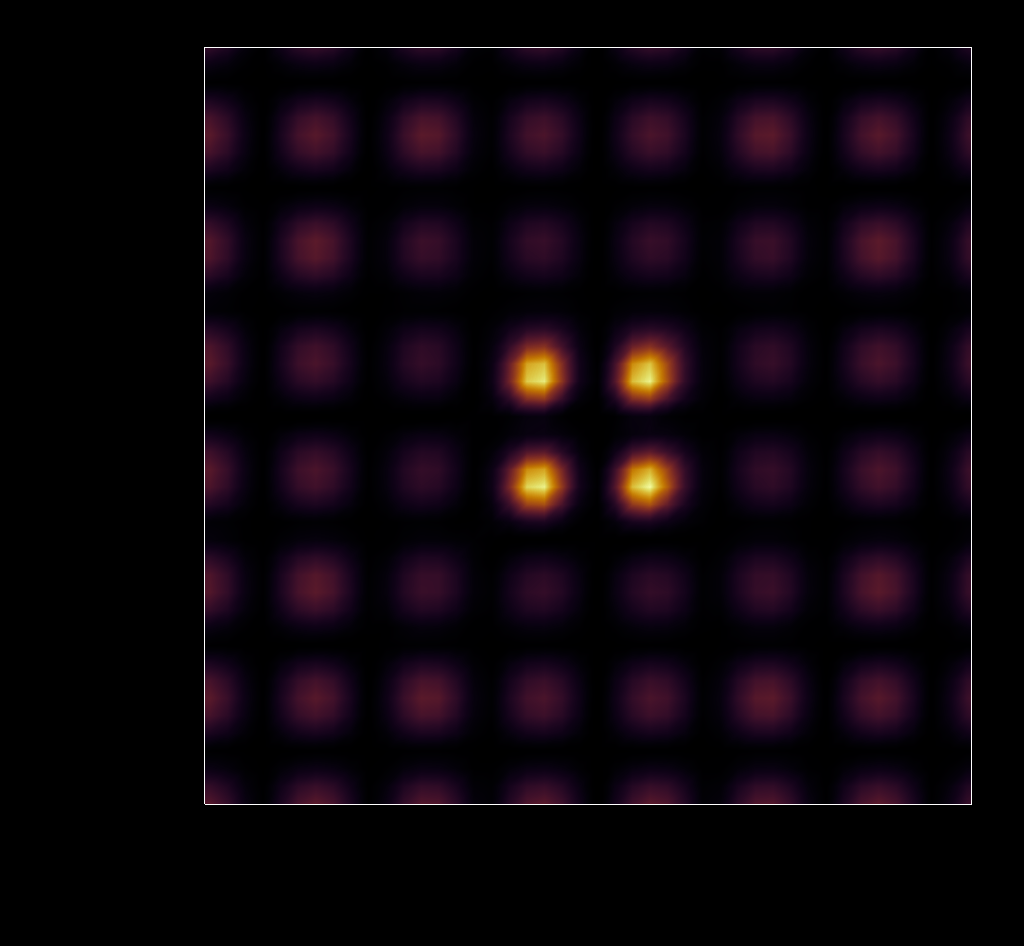}
    \put(-90,90){\color{white} \bf (f)} 
    \caption{One-level contour plots of (${\bf a}$)-(${\bf b}$) the neutron Cooper-pair density $|\psi_n|^2$ and (${\bf c}$)-(${\bf d}$) proton Cooper-pair density $|\psi_p|^2$, obtained by using the imaginary time ($t\to -it$) versions of the GPPE~(\ref{eq:GPE_neutron_ndim}) and RTGLPE~ (\ref{eq:GLE_proton_ndim}). Pseudocolor plots (${\bf e}$)-(${\bf f}$) of the magnetic field ${\bf B}=\nabla \times {\bf A}$ at the mid-plane $z=L/2$. 
    Here, the neutron condensate rotates with and angular velocity ${\bf \Omega} = \Omega \hat{z}$, with $\Omega=2.5$; the non-rotating proton condensate is placed in an external magnetic field ${\bf B}_{\rm ext}=B \hat{z}$, with $B=0.8$; neutron and proton Cooper pairs interact only through the gravitational potential (i.e., $\gamma=0$).
    }
    \label{fig:agle_rho_no_p_rotation}
\end{figure}

\textbf{Case(i):} The neutron condensate rotates [${\bf \Omega} = \Omega \hat{z}$] and the non-rotating proton condensate is in an external magnetic field ${\bf B}_{\rm ext}=B \hat{z}$. In Figs.~\ref{fig:agle_rho_no_p_rotation}(a)-(b) and Figs.~\ref{fig:agle_rho_no_p_rotation}(c)-(d) we show via contour plots of $|\psi_n|^2$  and  $|\psi_p|^2$, respectively, that the neutron condensate is threaded by vortices and the proton-superconductor displays an Abrikosov flux lattice. Each vortex in this lattice carries the quantum of magnetic flux $\Phi_B = \oint {\bf A} \cdot d{\bf l}$. Initially, the magnetic field is confined near the boundary [Fig.~\ref{fig:agle_rho_no_p_rotation}(e)]. Eventually these vortices penetrate the condensate [Fig.~\ref{fig:agle_rho_no_p_rotation}(f)] and each one of them contributes the unit magnetic flux $\Phi_B = \oint {\bf A} \cdot d{\bf l}$ to the overall magnetic field, which is confined within quantized flux tubes, solely inside the proton superconductor.  
Initially we insert flux tubes inside the proton condensate [Fig.~\ref{fig:agle_rho_no_p_rotation}(c) and {\bf ICI2} in Appendix~\ref{sec:initial_conditions}] because, in the absence of rotation, the system cannot generate flux tubes at $T=0$.

\begin{figure}[!htb]
    \centering
    \includegraphics[scale=0.1]{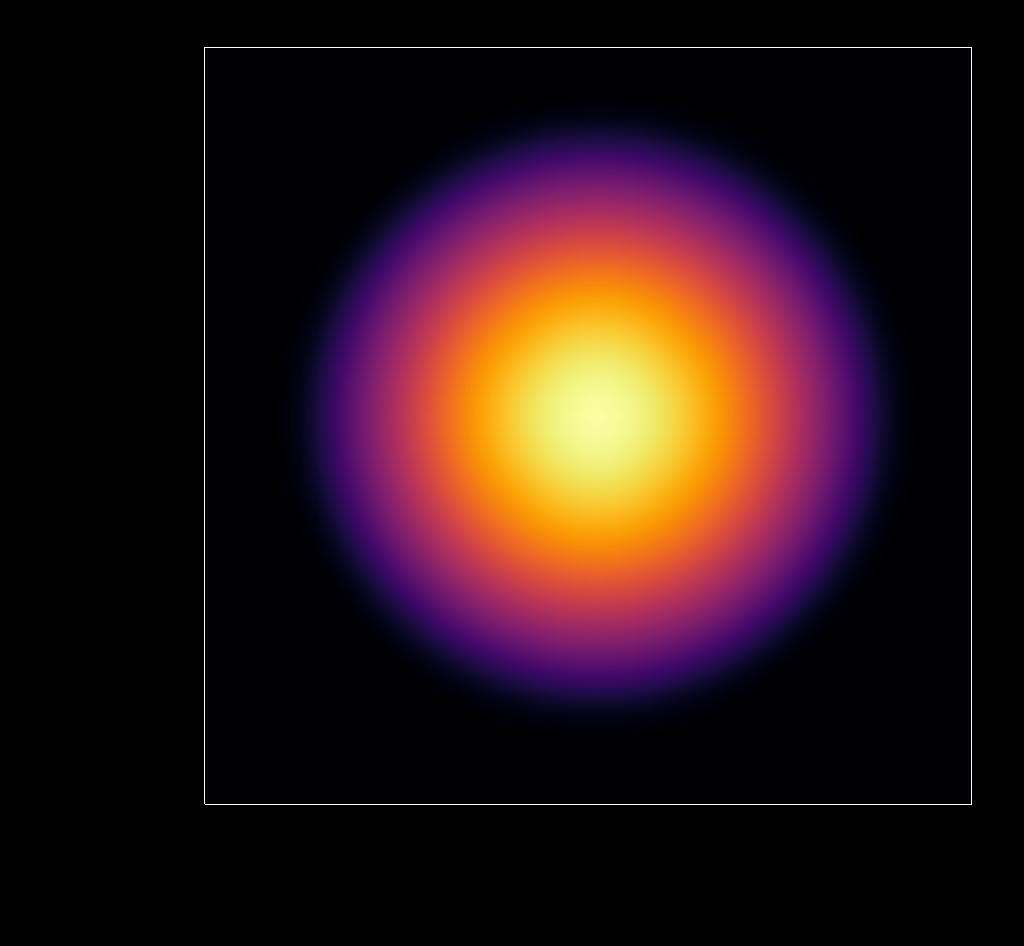}
    \put(-95,88){\color{white} \bf (a)}
    \put(-70,100){ $\Omega = 2.5 $}
    \put(-138,50){ $|\psi_p|^2 $}
    \put(-116,50){  $\bf \rightarrow$}
    \includegraphics[scale=0.1]{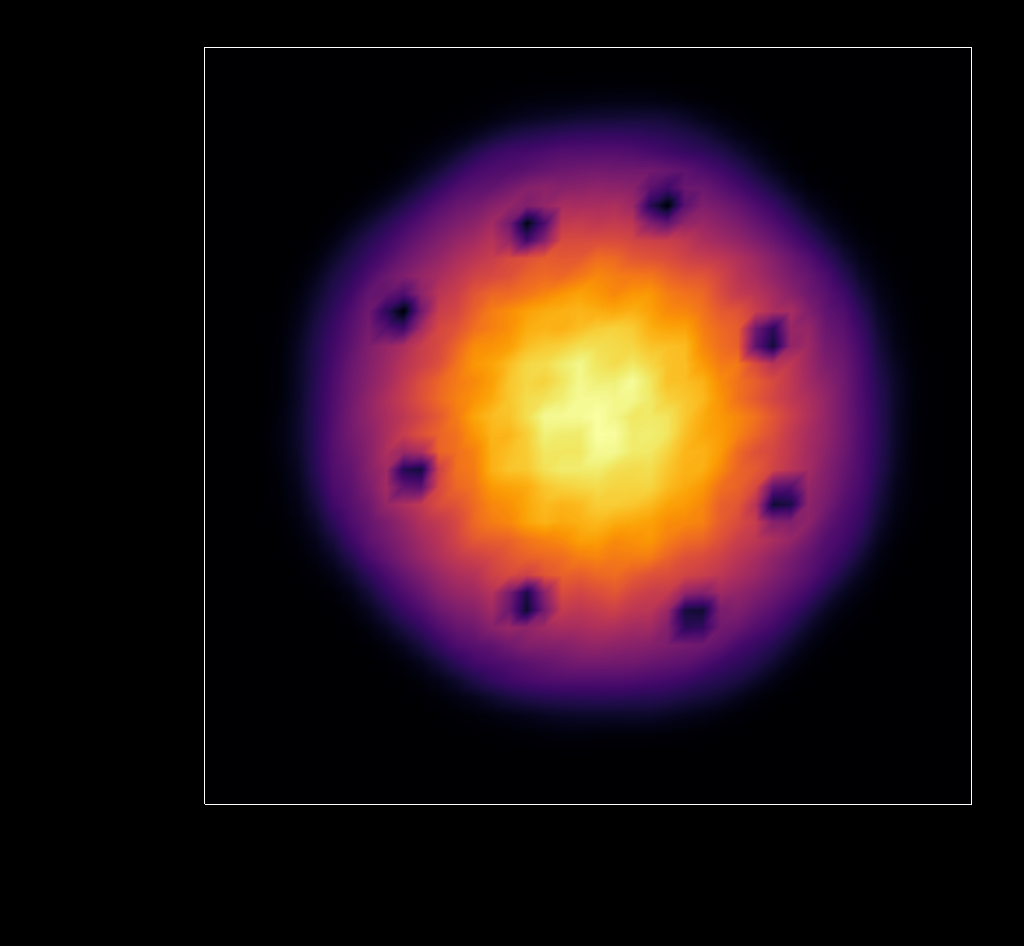}
    \put(-95,88){\color{white} \bf (b)}
    \put(-70,100){ $\Omega>\Omega_c=4.5$}

    \includegraphics[scale=0.10]{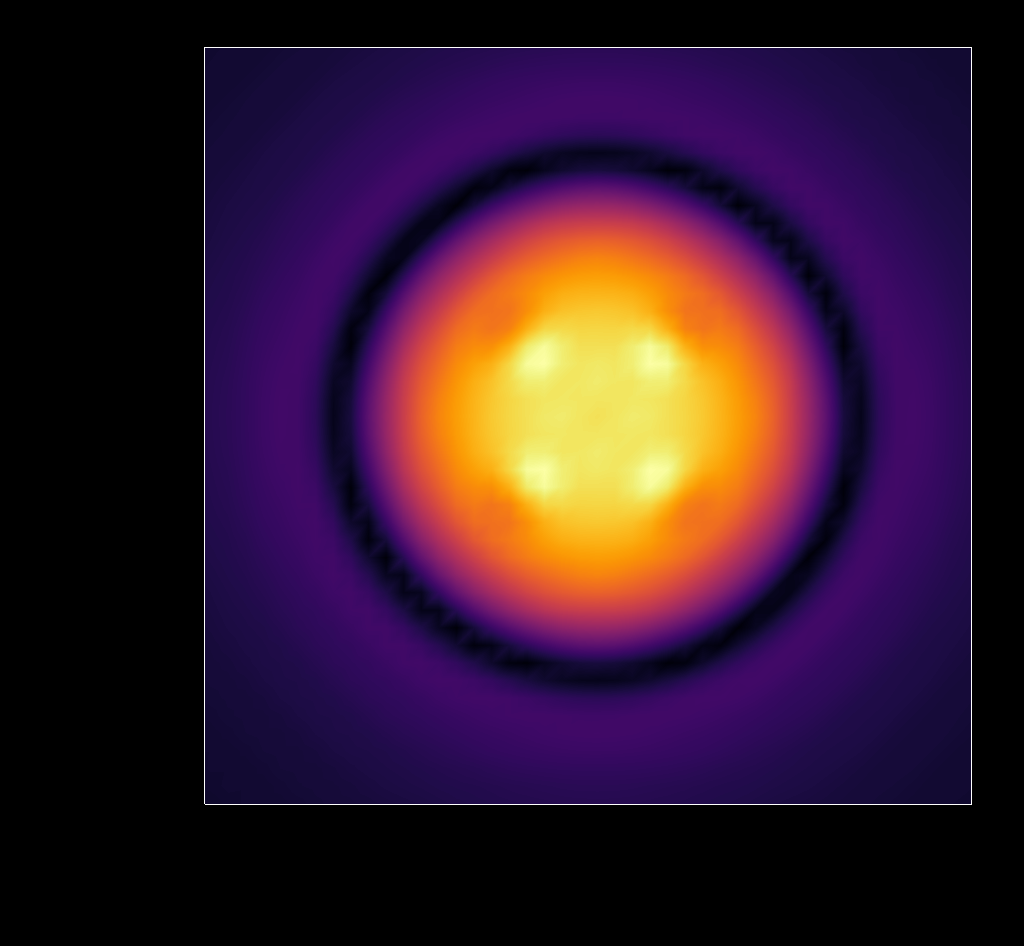}
    \put(-95,88){\color{white} \bf (c)}
    \put(-138,50){ $|{\bf B}|^2 $}
    \put(-116,50){ $\bf \rightarrow$}
    \includegraphics[scale=0.1]{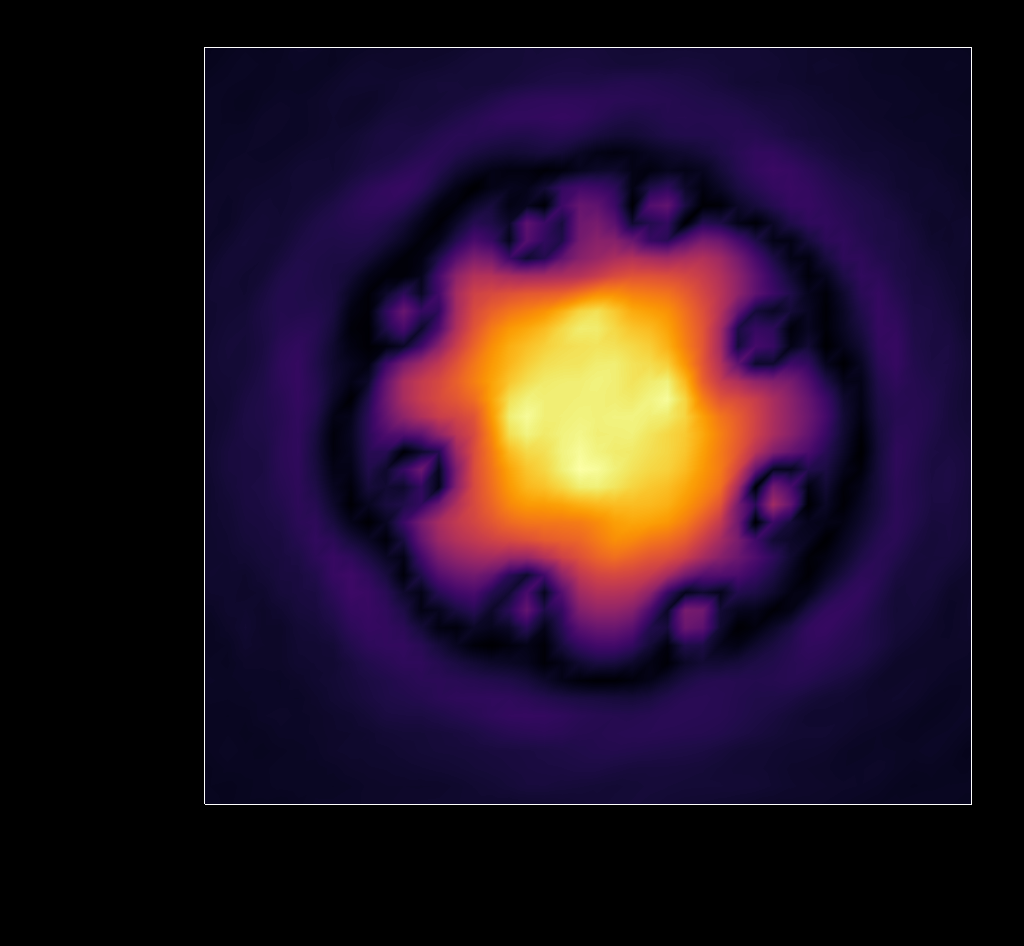}
    \put(-95,88){\color{white} \bf (d)}
    
    \caption{Pseudocolor plots of (${\bf a}$)-(${\bf b}$) the proton Cooper-pair density $|\psi_p|^2$ and (${\bf c}$)-(${\bf d}$) the magnetic field ${\bf B}=\nabla\times {\bf A}$  at the mid plane $z=L/2$. In columns 1 and 2, $\Omega=2.5$ and $\Omega>\Omega_c=4.5$, respectively. In these plots, we have ${\bf B}_{\rm ext}=0$. Neutron and proton Cooper pairs interact only indirectly through the gravitational potential (i.e., $\gamma=0$).}
    \label{fig:agle_rho_p_rotation}
\end{figure}

\begin{figure}[!htb]
    \centering
    \includegraphics[scale=0.21]{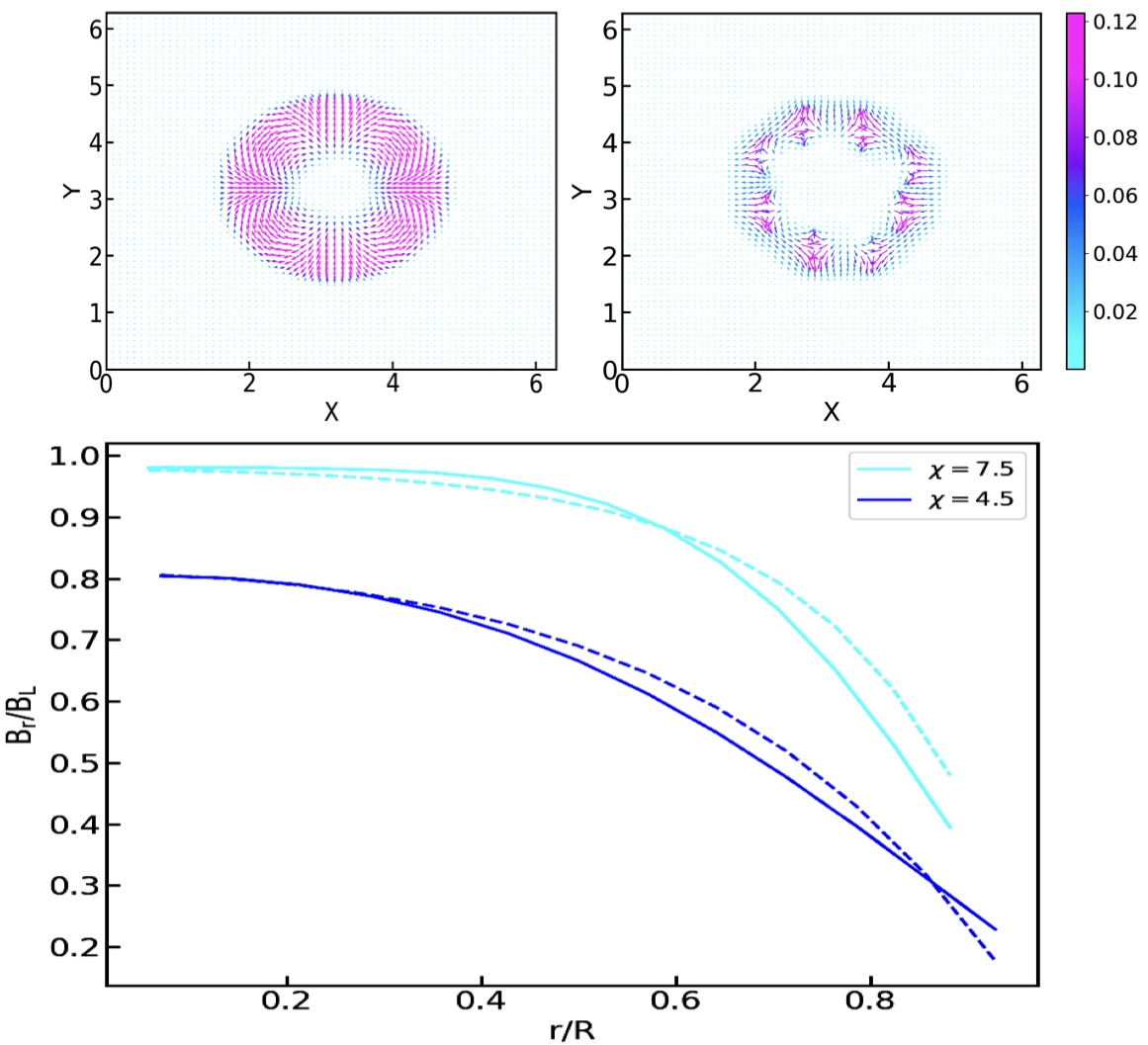}    
    \put(-220,210){ \bf (a)}
    \put(-110,210){ \bf (b)}    
    \put(-210,100){ \bf (c)}
    \caption{Two-dimensional (2D) vector plots of the proton Cooper-pair current densities $J_x$ and $J_y$ for (${\bf a}$) $\Omega=2.5$ and (${\bf b}$) $\Omega>\Omega_c=4.5$. ({\bf c}) The magnitude of the radial component of the magnetic field ${\bf B}=\nabla\times {\bf A}$, normalized by the London magnetic field $B_L$, plotted as a function of the distance $r/R$ from the centre of the superconductor, where $R$ is the radius of the spherical proton condensate. The solid curve is from the imaginary-time DNS of Eqs.~(\ref{eq:GPE_neutron_ndim})-(\ref{eq:scalar_pot_ndim}) and the dashed curve is the analytical relation~(\ref{eq:internal_field_analy}) for two values of $\chi = \frac{R}{\lambda_p}$, with $\lambda_p$ the London penetration depth of the superconductor.}
    \label{fig:agle_jx_jy_slow}
\end{figure}

\textbf{Case(ii):} both the neutron-superfluid and the
proton-superconductor rotate [with ${\bf \Omega} = \Omega \hat{z}$], but there is no external
magnetic field [${\bf B}_{\rm ext}=0$]. The neutron condensate is threaded by vortices [Figs.~\ref{fig:agle_rho_no_p_rotation}(a)-(b)], beyond a critical angular velocity, as in Case (i). The behaviour of the proton superconductor presents a compelling contrast. For slow rotation, less than a critical $\Omega_c^p$ (we use $\Omega=2.5$ here), the proton superconductor assumes a spherical shape devoid of vortices, as illustrated in Fig.~\ref{fig:agle_rho_p_rotation}(a). Furthermore, within the superconductor, a uniform London field emerges~(Fig.~\ref{fig:agle_rho_p_rotation}(c)), first from the superconductor's boundary at the characteristic length scale $\lambda_p$, which is the London penetration depth [see Section~\ref{sec:crust_pot}]. Any macroscopic rotation of a neutron superfluid results in the formation of quantized vortices; and the formation of flux tubes in a proton superconductor is driven by the magnetic field, not by macroscopic rotation. However, the rotation of a superconductor generates an additional magnetic field known as the London field. It is important to note that, considering realistic parameter values, the magnitude of the London magnetic field is very small for neutron stars [Ref.~\cite{Graber_2015}]. However, given the constrained parameter values in our simulations, the London field attains a reasonable finite value. In our simulation with Eq.~(\ref{eq:london_vecpot}), the magnitude of the dimensionless London field ${\bf B}_L = \nabla \times {\bf A}_L$ is
\begin{equation}
    {\rm B}_L=\frac{\xi_p^2}{L_{\rm ref}^2}\frac{\kappa}{\alpha}\Omega \simeq 1.12\ \Omega \,.
\end{equation}
The effect of rotation on the magnetic field distribution around flux tubes can be understood by comparing Figs.~\ref{fig:agle_rho_no_p_rotation}(f) and \ref{fig:agle_rho_p_rotation}(d). For a non-rotating proton superconductor in a magnetic field, the field is solely confined inside flux tubes [Fig.~\ref{fig:agle_rho_no_p_rotation}(f)]. For a rotating proton superconductor, flux tubes enter the system above a critical angular speed $\Omega_c$ [compare Figs.~\ref{fig:agle_rho_p_rotation}(c) and (d)], and the magnetic field passes through the centre of the flux tubes with a finite region around the center devoid of the magnetic field  [Fig.~\ref{fig:agle_rho_p_rotation}(d)]. Beyond this finite region, we observe a uniform distribution of the London field. The London magnetic field, a fundamental property of rotating superconductors, has been studied in a hydrodynamical model of a neutron-star interior in Ref.~\cite{Graber_2015}. The generation of this uniform magnetic field is facilitated by the macroscopic London current ${\bf J}_{\rm L}$, which is concentrated near the superconductor's boundary, as we show via 2D vector plots of $J_x$ and $J_y$  in Fig.~\ref{fig:agle_jx_jy_slow}(a). 


If ${\bf B}_{\rm ext}=0$, the rotation is so slow that there are no flux tubes [$\Omega < {\rm \Omega}_c^p$], and $\gamma=0$, then we can write Eq.~(\ref{eq:vector_pot}) in the steady state as
\begin{equation}
\begin{aligned}
    \nabla \times {\bf B}= \frac{q}{m_pc^2\epsilon_0}{\bf J}_s  - 
    \frac{qn_p}{c^2\epsilon_0} ({\bf \Omega \times {\bf r}})\,,
\end{aligned}
\end{equation}
where $n_p=|\psi_p|^2$. We now use the London equation 
\begin{equation}
\nabla \times {\bf J}_s = -\frac{n_pq}{m_p}{\bf B}\,,
\label{eq:London}
\end{equation}
with ${\bf J}_s=m_pn_p {\bf v}$, to obtain
\begin{equation}
\begin{aligned}
    \nabla \times \nabla \times {\bf v} = -\frac{1}{\lambda_p^2}({\bf v}- {\bf \Omega}\times {\bf r})\,,
    \label{eq:internal_vecpot}
\end{aligned}
\end{equation}
where $\lambda_p = \sqrt{\frac{m_pc^2 \epsilon_0}{qn_p^2}}$ is the London penetration depth. [A similar relation has been used in Ref.~\cite{Egor_2007} but for a multicomponent superconductor.] If we assume that the density distribution is spherically symmetric, then the proton superconductor has only the azimuthal component ${\bf v}=v{\bf e}_{\phi}$, so, by solving Eq.~(\ref{eq:internal_vecpot}), we get
\begin{equation}
\begin{aligned}
    {\bf v} = \bigg[ { \Omega }r+ \frac{C}{r^2} (\sinh\bigg(\frac{r}{\lambda_p}\bigg)-\frac{r}{\lambda_p} \cosh\bigg(\frac{r}{\lambda_p}\bigg))\bigg]{\bf e}_{\phi}\,,
    \label{eq:internal_vel}
\end{aligned}
\end{equation}
whence we obtain the radial component ${\rm B}_r$ of the magnetic field by using Eqs.~(\ref{eq:internal_vel}) and (\ref{eq:London}):
\begin{equation}
\begin{aligned}
    {\rm B}_r = \frac{m_p}{q}\bigg[ 2{ \Omega }+ \frac{2C}{r^3} \bigg(\sinh(\frac{r}{\lambda_p})-\frac{r}{\lambda_p} \cosh(\frac{r}{\lambda_p})\bigg)\bigg]\,.
    \label{eq:internal_mag}
\end{aligned}
\end{equation}
We determine the constant $C$ by demanding ${\rm B}_{r={\rm R}} =0$, where ${\rm R}$ is the radius of the spherical condensate, because ${\bf B}_{\rm ext}=0$. Finally, we get
\begin{eqnarray}
    \frac{{\rm B}_r}{{\rm B}_L}=1-\frac{1}{(r/\rm R)^3}\frac{\sinh(\frac{r}{\rm R}\chi)-\frac{r}{\rm R}\chi \cosh(\frac{r}{\rm R}\chi)}{\sinh(\chi)-\chi \cosh(\chi)}\,,
    \label{eq:internal_field_analy}
\end{eqnarray}
where $\chi = \frac{\rm R}{\lambda_p}$, which we show via dashed lines in  Fig.~\ref{fig:agle_jx_jy_slow}(c) for 
two representative values of $\chi$. The solid lines in 
Fig.~\ref{fig:agle_jx_jy_slow}(c) give the results of our
DNS, which agree well with the results of our analytical approximation [Eq.~(\ref{eq:internal_field_analy})]. We observe that, for large value of $\chi$ (small $\lambda_p$), the internal magnetic field is comparable to the London field ${\rm B}_L$; and it decreases as we decrease $\chi$.

Upon increasing the angular velocity ${ \Omega}$ beyond $\Omega_c^p \simeq 4.5$ [see Eq.~(\ref{eq:omega_c_super})], vortices begin to penetrate the proton superconductor, as we show in ~Fig.~\ref{fig:agle_rho_p_rotation}(b). Each of these vortices carries a quantum of magnetic flux $\Phi_{\Omega} = \oint ({\bf \Omega }\times {\bf r}) \cdot d{\bf l} $, which is the flux because of the uniform London field ${\rm B}_L$ within the superconductor  (Fig.~\ref{fig:agle_rho_p_rotation}(d)).
 Our DNS provides valuable insights into the dynamic interplay between rotation, vortices, and the London field in the proton superconductor. As vortices penetrate the condensate~\cite{Egor_2014}, they reduce the uniform magnetic field through the system~\footnote{This is in contrast to the behavior of the Abrikosov lattice, for a non-rotating superconductor in an external magnetic field~\cite{Egor_2014}.}. The distribution of currents, crucial for generating and sustaining this magnetic field, is revealed by the 2D  vector plots of $J_x$ and $J_y$ in Fig.~\ref{fig:agle_jx_jy_slow}(b). Note that these currents are concentrated near the superconductor's boundary and around the vortices. 

\begin{figure}[!htb]
    \centering
    \includegraphics[scale=0.11]{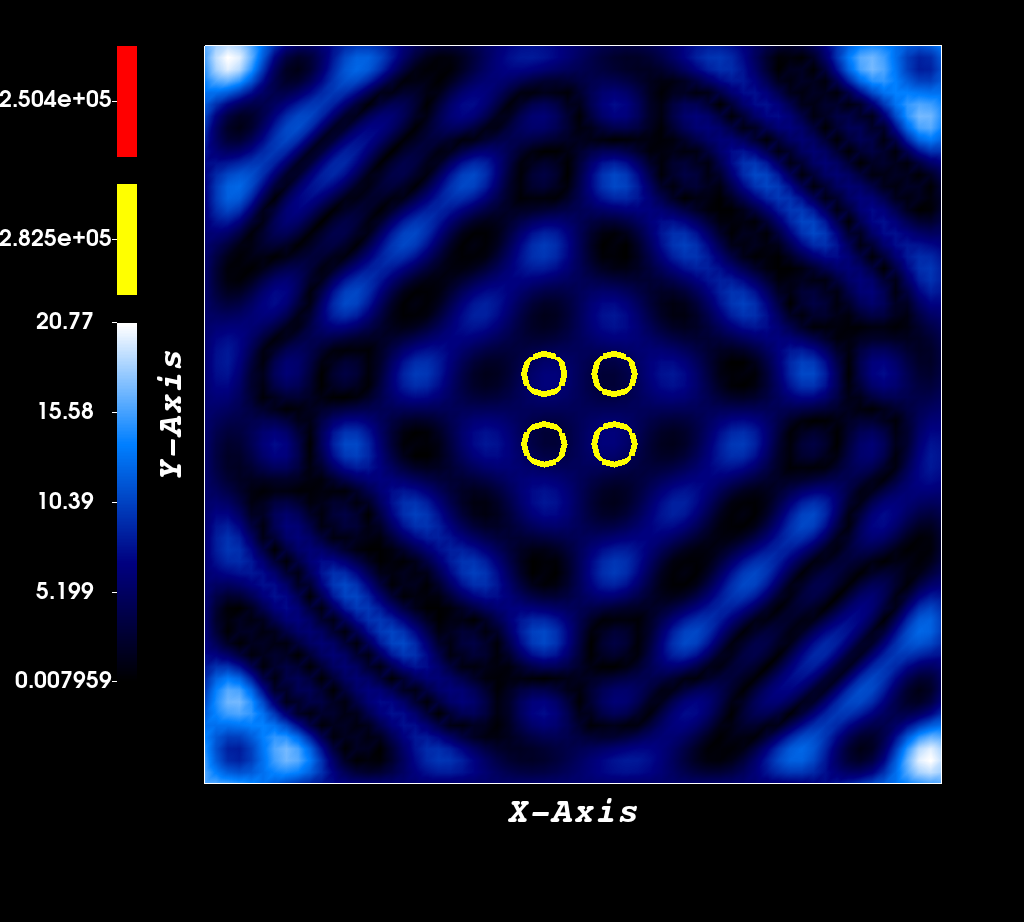}
    \put(-55,110){  $ t_i$}
    \put(-90,88){ \bf \color{white}(a)}
    \includegraphics[scale=0.11]{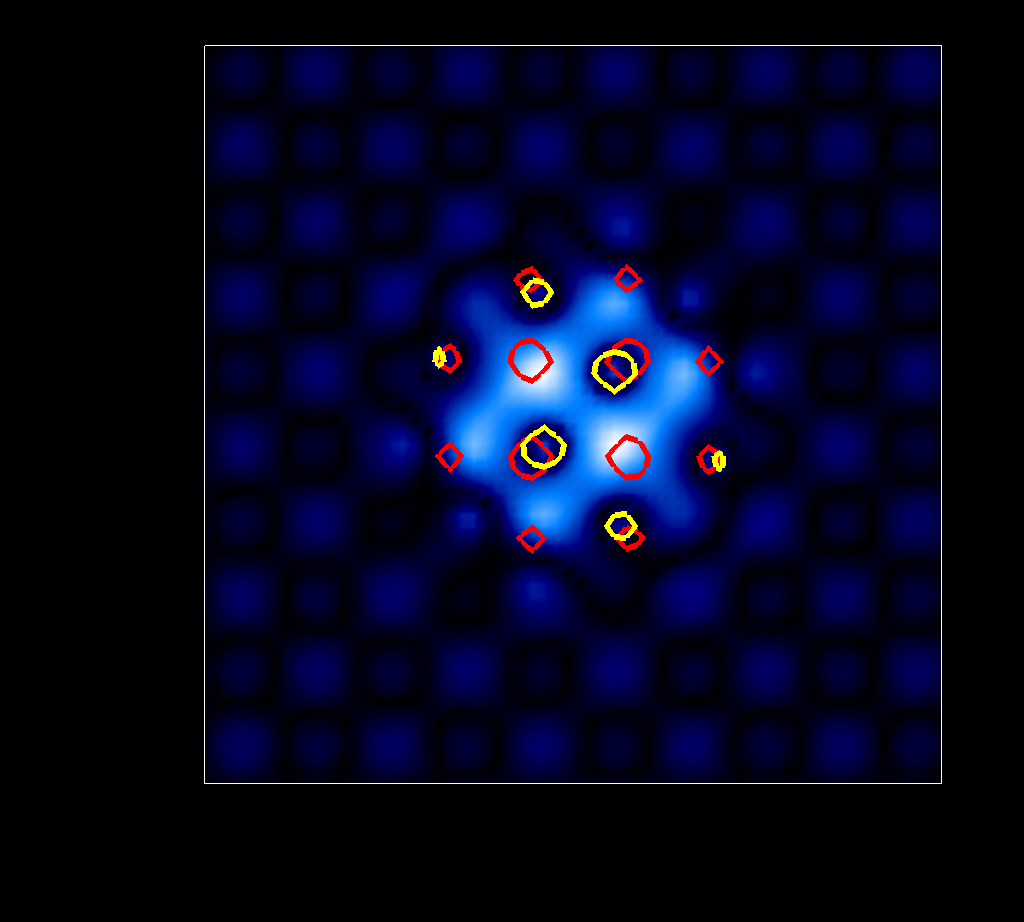}
    \put(-55,110){  $ t_f$}
    \put(-90,88){ \bf \color{white}(b)}
    \caption{Pseudocolor plots, at the plane $z=L/2$, illustrating the magnetic field ${\bf B} = \nabla\times {\bf A}$ in blue at (a) initial $t_i$ and (b) final $t_f$ (imaginary) times, with neutron vortices and proton flux tubes indicated by red and yellow contours, respectively. The vortices and flux tubes are the contour plots of the pseudo-vorticity $\omega_{\rm p}= |\nabla \times (\rho {\bf v})|$, where $\rho$ is the density and ${\bf v}$ is the velocity of the superfluid or superconductor. Both neutron and proton
    subsystems rotate with $\Omega=2.5$; and the external magnetic field is $B_{\rm ext}=0.8$.}
    \label{fig:agle_rho_p_rotation_Bext_iii}
\end{figure}

\textbf{Case(iii):} Both the neutron and proton condensates rotate with ${\bf \Omega}=\Omega\hat{z}$; and the proton condensate is subjected to an external magnetic field  ${\bf B}_{\rm ext}=B\hat{z}$. The equilibrium state of the neutron condensate resembles that of Case (ii), with vortices penetrating the system beyond the neutron critical angular velocity $\Omega_c^n$. The proton condensate manifests a London field ${\rm B}_L$ as in Case (ii). In Figs.~\ref{fig:agle_rho_p_rotation_Bext_iii}(a)-(b) we present blue-scale plots of the magnetic field ${\bf B} = \nabla\times {\bf A}$, in the plane $z=L/2$ and at initial $t_i$ and final $t_f$ representative (imaginary) times; these plots show neutron vortices and proton flux tubes in red and yellow contours respectively. We visualize these vortices and flux tubes via plots of the pseudo-vorticity $\omega_{\rm p} = |\nabla \times (\rho {\bf v})|$, where $\rho$ is the density and ${\bf v}$ is the velocity field of the superfluid neutrons or superconducting protons. The velocity field is calculated using the following
 \begin{eqnarray}
     {\bf v}_{\rm n,p} = \frac{\hbar}{\rho_{\rm n,p}} \frac{\psi^*_{\rm n,p}\nabla\psi_{\rm n,p}-\psi_{\rm n,p}\nabla\psi^*_{\rm n,p}}{2i}\,,
 \end{eqnarray}
 where $(\rm n,p)$ refers to neutron and proton Cooper pairs, respectively.

At the initial (imaginary) time $t_i$ we include proton flux tubes (yellow) in the initial condition [Fig.~\ref{fig:agle_rho_p_rotation_Bext_iii}(a)]; the magnetic field is concentrated near the boundary of the simulation box. In the equilibrium state, neutron vortices (in red) enter the condensate and organize themselves in the manner depicted in Fig.~\ref{fig:agle_rho_p_rotation_Bext_iii}(b). Even though $\gamma = 0$, i.e., there is no direct interaction between the two components, the neutron-superfluid vortices and proton-superconductor flux tubes come close together, as we show in Fig.~\ref{fig:agle_rho_p_rotation_Bext_iii}(b). This effective attraction follows from the coupling induced by the Poisson equation for gravitational potential [Eq.(~\ref{eq:grav_pot})]. 

It is also important to note that the sizes of vortices in Fig.~\ref{fig:agle_rho_p_rotation_Bext_iii} are different. This can be explained using the pseudo-vorticity $\omega_{\rm p}=\nabla \times (\rho {\bf v})$, which can be rewritten as
\begin{equation}
    \omega_{\rm p} = \rho \nabla \times {\bf v}+ (\nabla\rho)\times {\bf v}\,.
    \label{eq:pseudo_vorticity}
\end{equation}
For a uniform density distribution, such as in harmonic confinement, the second term in Eq.~\eqref{eq:pseudo_vorticity} is very small because $\nabla\rho$ is negligible, resulting in vortices of similar sizes throughout the condensate. However, for the self-gravitating case, the density decreases as we move away from the center. This creates a negative density gradient towards the edge, causing $\omega_{\rm p}$ to become small and resulting in smaller vortex sizes at the boundary.
   
\begin{figure*}[!htb]
    \centering
    \includegraphics[scale=0.13]{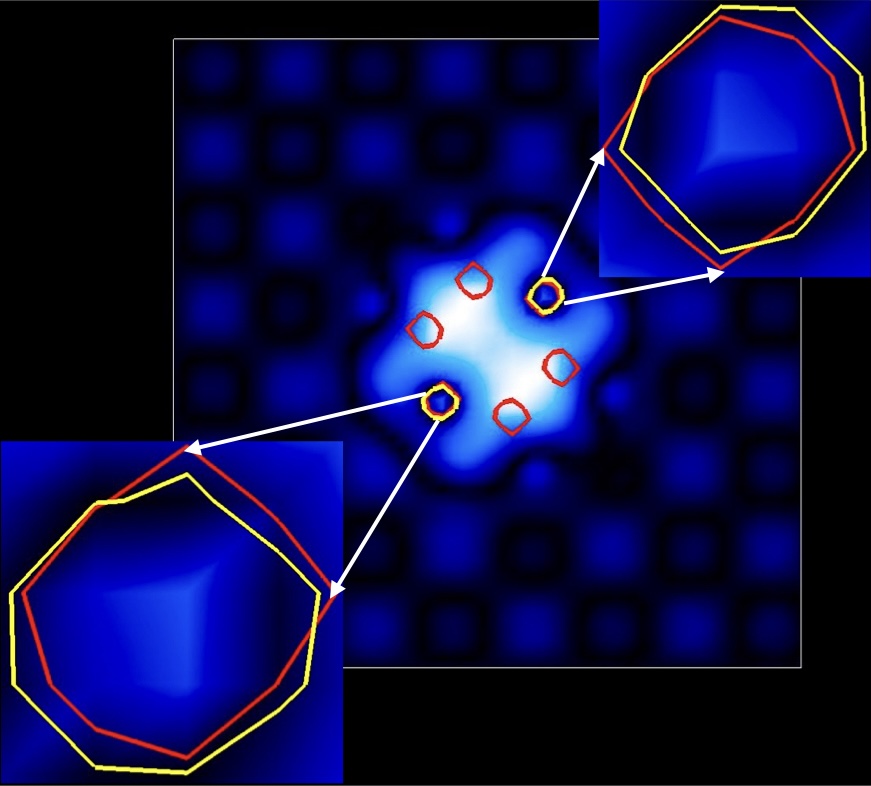}
    \put(-110,90){\color{white} \bf (a)}
    \includegraphics[scale=0.11]{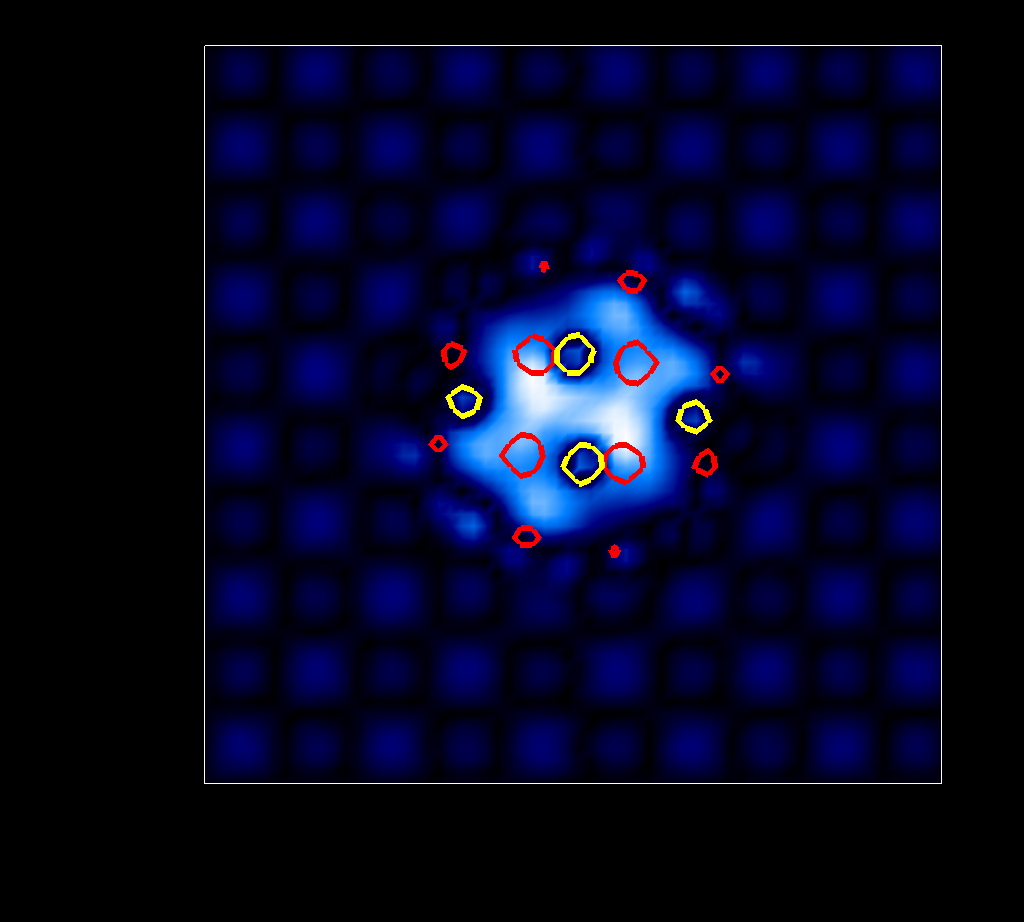}
    \put(-110,90){\color{white} \bf (b)}
    \includegraphics[scale=0.13]{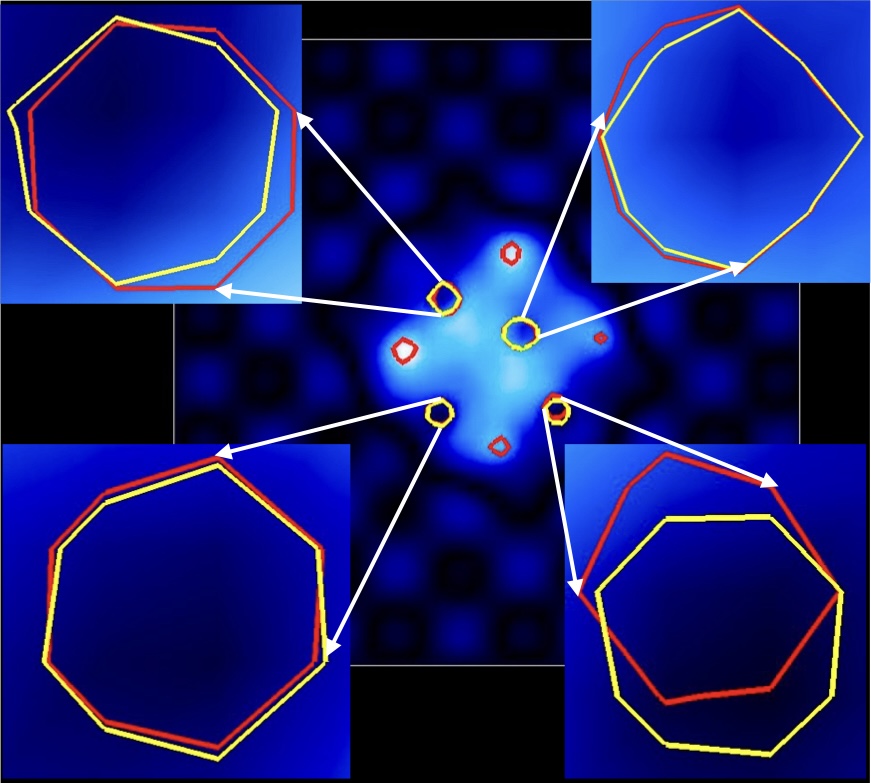}
    \put(-110,90){\color{white} \bf (c)}
    \includegraphics[scale=0.2]{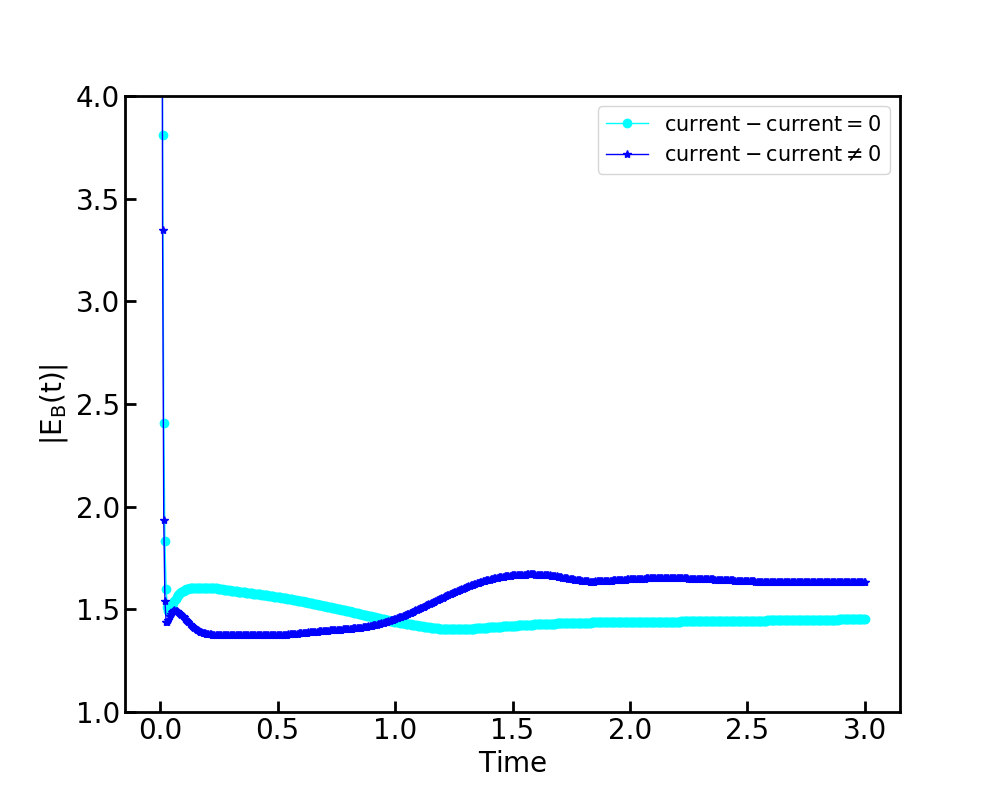}
    \put(-115,88){\bf (d)}
    \caption{Pseudocolor plots, at plane $z=L/2$, illustrating the magnetic field ${\bf B} = \nabla\times {\bf A}$ in blue [at equilibrium, i.e., the final imaginary time in our DNS], with neutron vortices and proton flux tubes shown via red and yellow contours, respectively, for {\bf (a)} gravitational and attractive ($g'<0$) density-density interactions, {\bf (b)} gravitational and repulsive ($g'>0$) density-density interactions, {\bf (c)} gravitational, attractive density-density, and also current-current interactions. {\bf (d)} Imaginary-time series plots of the magnetic-field energy $\rm E_{B}=|{\bf B}_{\rm ext}-\nabla\times {\bf A}|^2$ for zero current-current interaction ($\gamma=0$) in cyan and nonzero current-current interaction ($\gamma\neq0$) in blue. Both neutron and proton subsystems rotate with $\Omega=2.5$; and the external magnetic field is $B_{\rm ext}=0.8$. The vortices and flux tubes are the contour plots of the pseudo-vorticity $\omega_{\rm p}= \nabla \times (\rho {\bf v})$, where $\rho$ is the density and ${\bf v}$ is the velocity of the superfluid or superconductor.}
    \label{fig:agle_rho_p_rotation_Bext}
\end{figure*}

\subsection{Imaginary-time study: $\gamma\neq 0$, $V_{\theta}=0$, $\Theta=0$}
\label{sec:imag_time_int_no_Vtheta_p}
We examine the interacting case of $\gamma\neq 0$ 
[in addition to the gravitational-potential-induced coupling in Case(iii)]. 
Consider first neutron-proton density-density interactions; their effective strength follows from the first term in the interaction Lagrangian~(\ref{eq:interaction_lag}):
\begin{equation}
    g'=g_{np} |\psi_n|^2 |\psi_p|^2\,.
    \label{eq:gprime}
\end{equation}
For attractive (repulsive) couplings $g' < 0$ ($g' > 0$), the density minima of neutron-superfluid vortices and proton-superconductor flux tubes align (the maximum of the neutron-superfluid density aligns with the proton flux tubes). 
[A similar analysis has been conducted in Ref.~\cite{Drummond_2017} without a gravitational interaction.] Given that the gravitational interaction is inherently attractive, the attractive case ($g' < 0$) only promotes this alignment: 
If we start with the same initial condition as in Fig.~\ref{fig:agle_rho_p_rotation_Bext_iii}(a), we obtain the final equilibrium state, shown in Fig.~\ref{fig:agle_rho_p_rotation_Bext}(a), with overlapping density minima of the neutron-superfluid vortices in red and of proton-superconductor flux tubes in yellow (as in Fig.~\ref{fig:agle_rho_p_rotation_Bext_iii}(b), with only gravitation-induced interactions). In contrast, if $g' > 0$, the interplay between the repulsive density-density interaction and the attractive gravitational interaction
is such that cores do not overlap, as evident in Fig. \ref{fig:agle_rho_p_rotation_Bext}(b).

We now incorporate the current-current interaction in the Lagrangian (\ref{eq:interaction_lag}). This leads to the entrainment of the proton-superconductor current because of the term $\frac{\gamma q}{c^2\epsilon_0} {\bf J}_n|\psi_p|^2$ in the vector potential~(\ref{eq:vector_pot}). This introduces a combination of gravitational, negative density-density ($g'<0$), and positive current-current interactions, whose effects we investigate by starting with the same initial condition as in Fig.~\ref{fig:agle_rho_p_rotation_Bext_iii}(a). As imaginary time progresses, neutron-superfluid vortices (in red) and proton-superconductor flux tubes (in yellow) tend to merge and the minima of neutron and proton densities overlap because of the gravitational and negative density-density interaction (see Fig.~\ref{fig:agle_rho_p_rotation_Bext}(c)).
This entrainment causes neutron-superfluid vortices to drag proton-superconductor flux tubes and induces a magnetic field inside these vortices. This is clearly visible as intense blue spots in Fig.~\ref{fig:agle_rho_p_rotation_Bext}(c), inside vortices that do not overlap with proton-superconductor flux tubes. To illustrate the generation of this entrainment-induced magnetic field, we plot the magnetic-field energy $\rm E_B(t) = |{\bf B}_{\rm ext}-\nabla \times {\bf A}|^2$ versus imaginary time $t$ in Fig.~\ref{fig:agle_rho_p_rotation_Bext}(d) both with (blue curve) and without (cyan curve) current-current interaction. In the latter case $\rm E_B(t)$ is higher than in the former at large $t$.

\subsection{Angle between ${\bf B}_{\rm ext}$ and ${\bf \Omega}$ }
\label{sec:imag_time_non_zero_angle}
So far we have examined cases with aligned  ${\bf B}_{\rm ext}$ and ${\bf \Omega}$, i.e., $\Theta=0$.
We turn now to $\Theta > 0$, which is the case in most pulsars.

\subsubsection{{\bf {No interactions:}} $\gamma=0$, $V_{\theta}=0$, and $\Theta=30^o$}
\label{sec:imag_time_non_zero_angle_no_int_no_Vtheta}

As in Case(i) Subsection~\ref{sec:imag_time_no_int_no_Vtheta_p} we study a neutron condensate that rotates [${\bf \Omega} = \Omega \hat{z}$] and a non-rotating proton condensate in an external magnetic field ${\bf B}_{\rm ext}=B (\hat{z}\cos\Theta+\hat{y}\sin\Theta)$.
We examine the case with no direct interactions, i.e., $\gamma=0$, no crust potential, i.e., $V_{\theta}=0$, and the representative values $\Theta=30^o$ and  $\Omega=2.5 > \Omega_c^n$, so vortices enter the neutron condensate, and the proton condensate is stabilized with an Abrikosov lattice. 

\begin{figure}[!htb]
    \centering
    \includegraphics[scale=0.075]{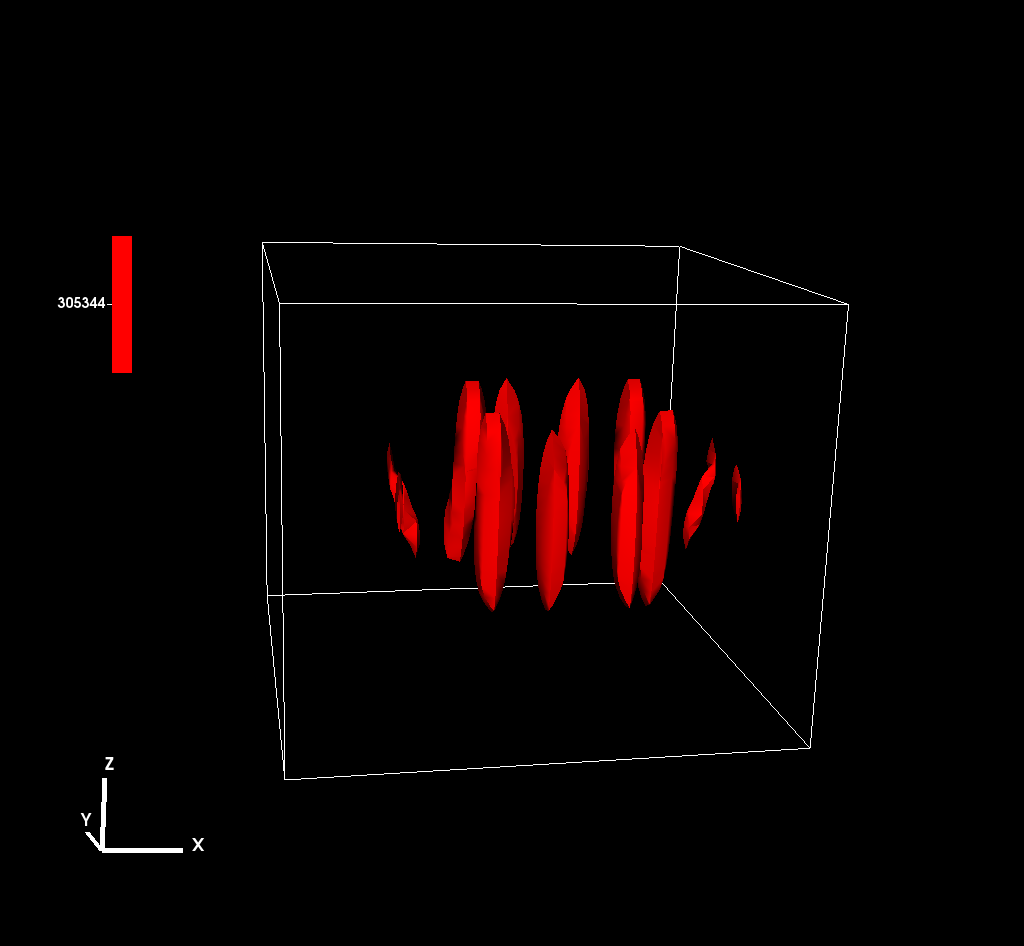}
    \put(-70,60){\color{white} \bf (a)}
    \includegraphics[scale=0.075]{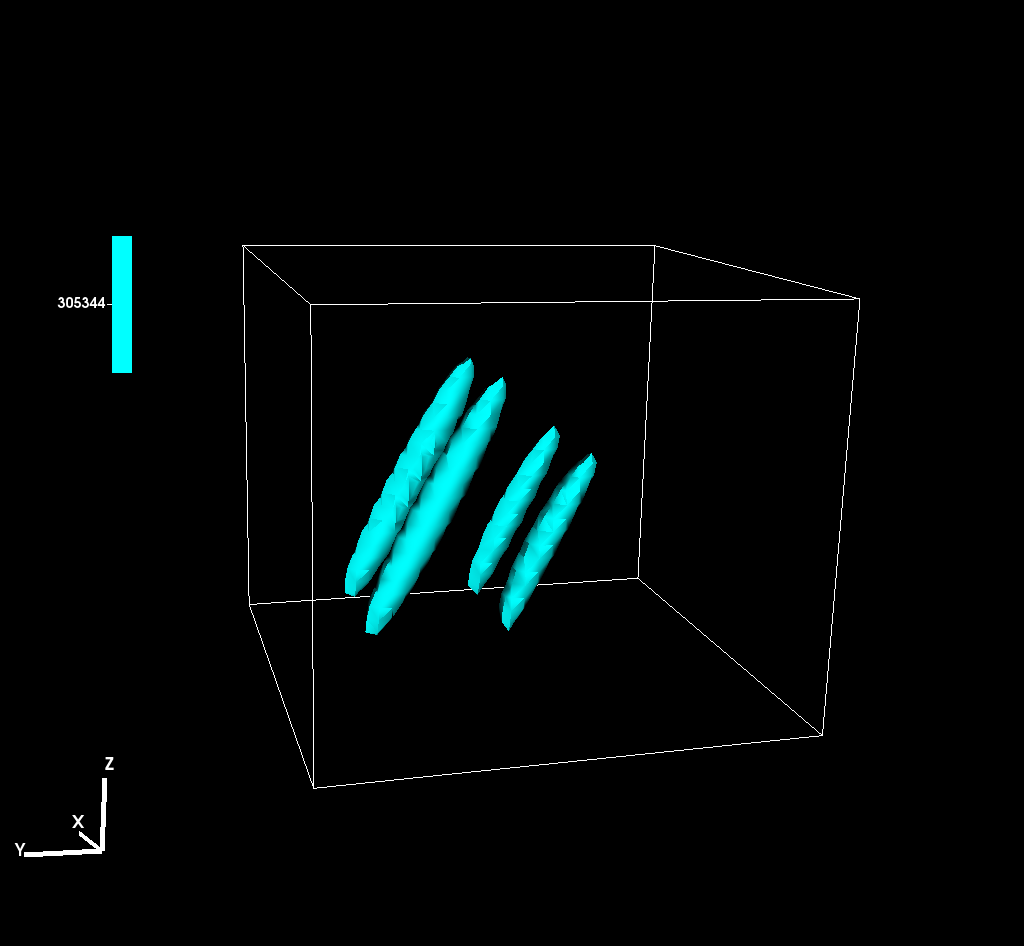}
    \put(-70,60){\color{white} \bf (b)}   
    \includegraphics[scale=0.075]{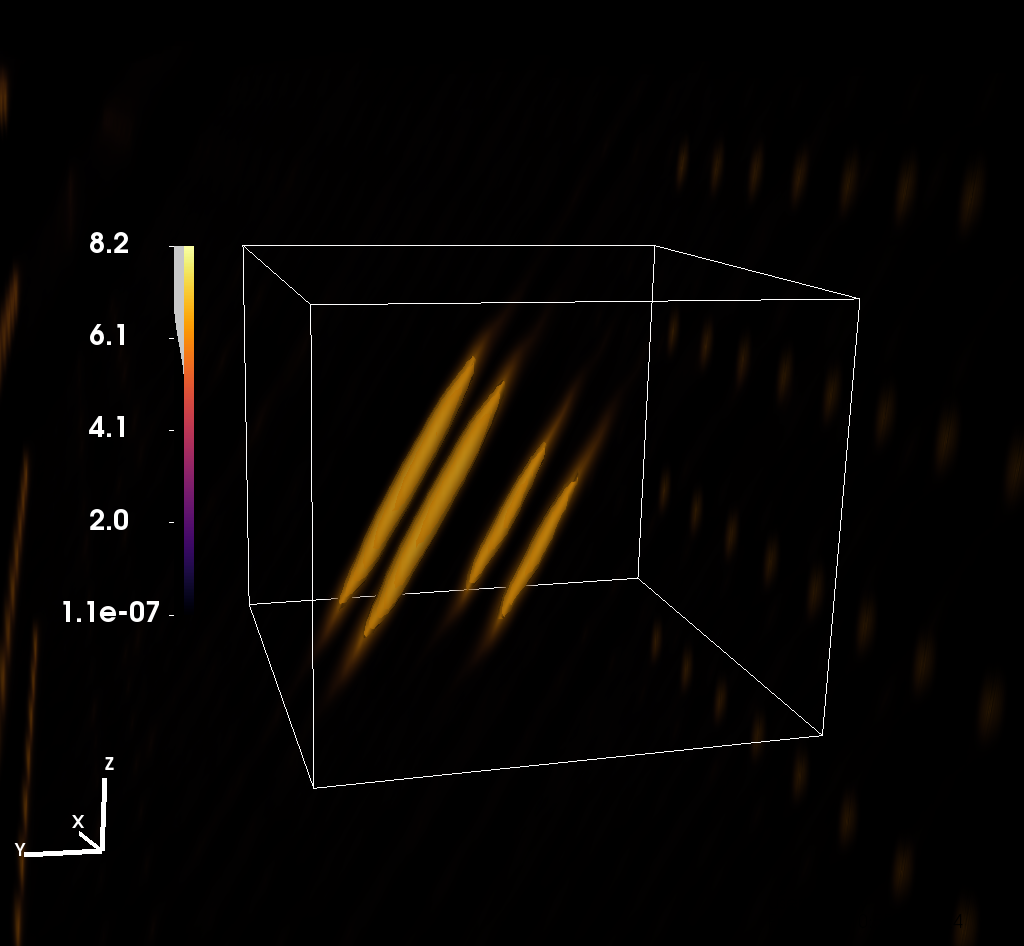}
    \put(-70,60){\color{white} \bf (c)}  
    
    \caption{One-level contour plots of $(\nabla \times (\rho v))^2$ for (${\bf a}$) neutron-superfluid vortices, and (${\bf b}$)  proton-superconductors flux tubes at the final imaginary time. (${\bf c}$) Volume plot (final imaginary time) of the magnetic field ${\bf B} = \nabla\times {\bf A}$. In these plots, the neutron condensate rotates with angular velocity ${\bf \Omega} = \Omega \hat{z}$, with $\Omega=4.0$, and a non-rotating proton condensate; there is an external magnetic field $B_{\rm ext}=0.8$ that makes an angle $\Theta=30^\circ$ with the $z$-axis. Both species interact only through the gravitational potential.}
    \label{fig:agle_3D_no_p_rotation}
\end{figure}

In Fig.~\ref{fig:agle_3D_no_p_rotation}(a), we show 
a one-level red contour plot of the neutron-superfluid vortices at the final imaginary time; these are aligned along the $z$-axis. By contrast, the proton-superconductor flux tubes, illustrated by cyan contour plots in Fig.~\ref{fig:agle_3D_no_p_rotation}(b), form an Abrikosov lattice, have their axes tilted at a fixed angle of $\Theta=30^\circ$ relative to the $z$-axis; the magnetic field manifests itself solely within these flux tubes [Fig.~\ref{fig:agle_3D_no_p_rotation}(c)]. 

\begin{figure}[!htb]
    \centering
    \includegraphics[scale=0.07]{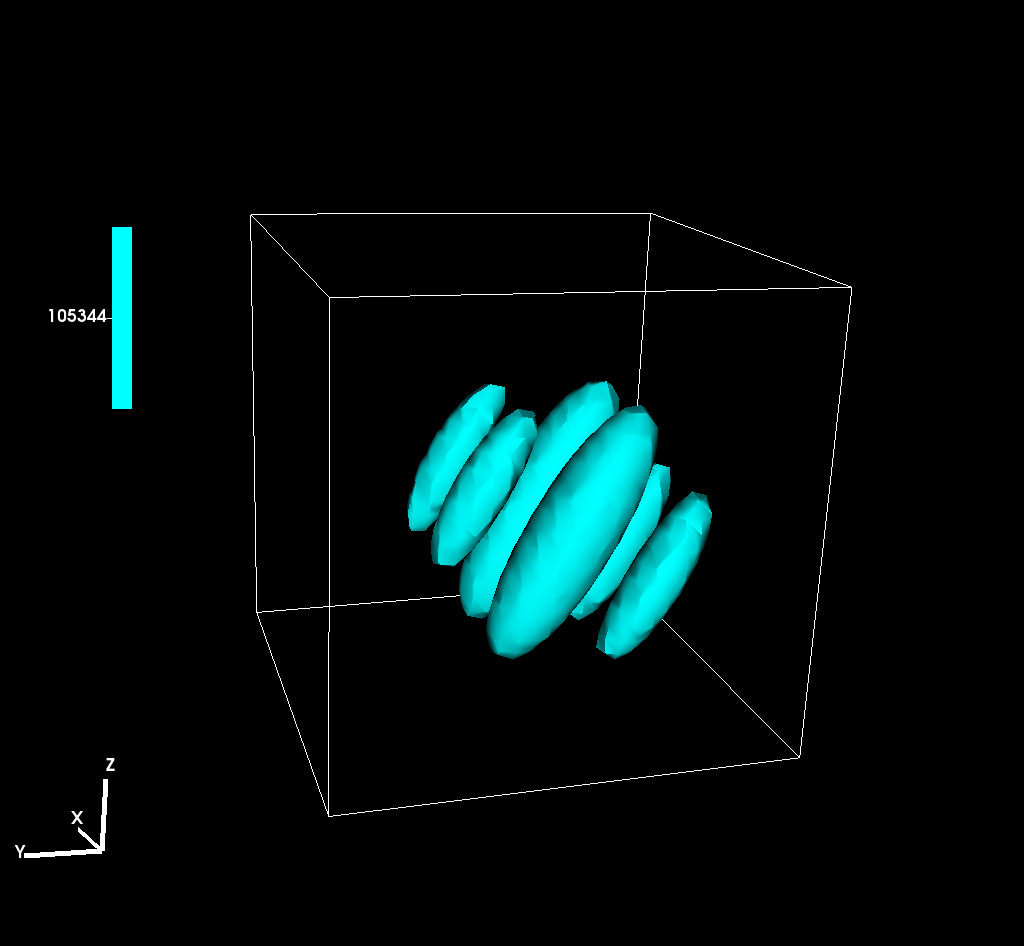}
    \put(-70,60){\color{white} \bf (a)}
    \includegraphics[scale=0.07]{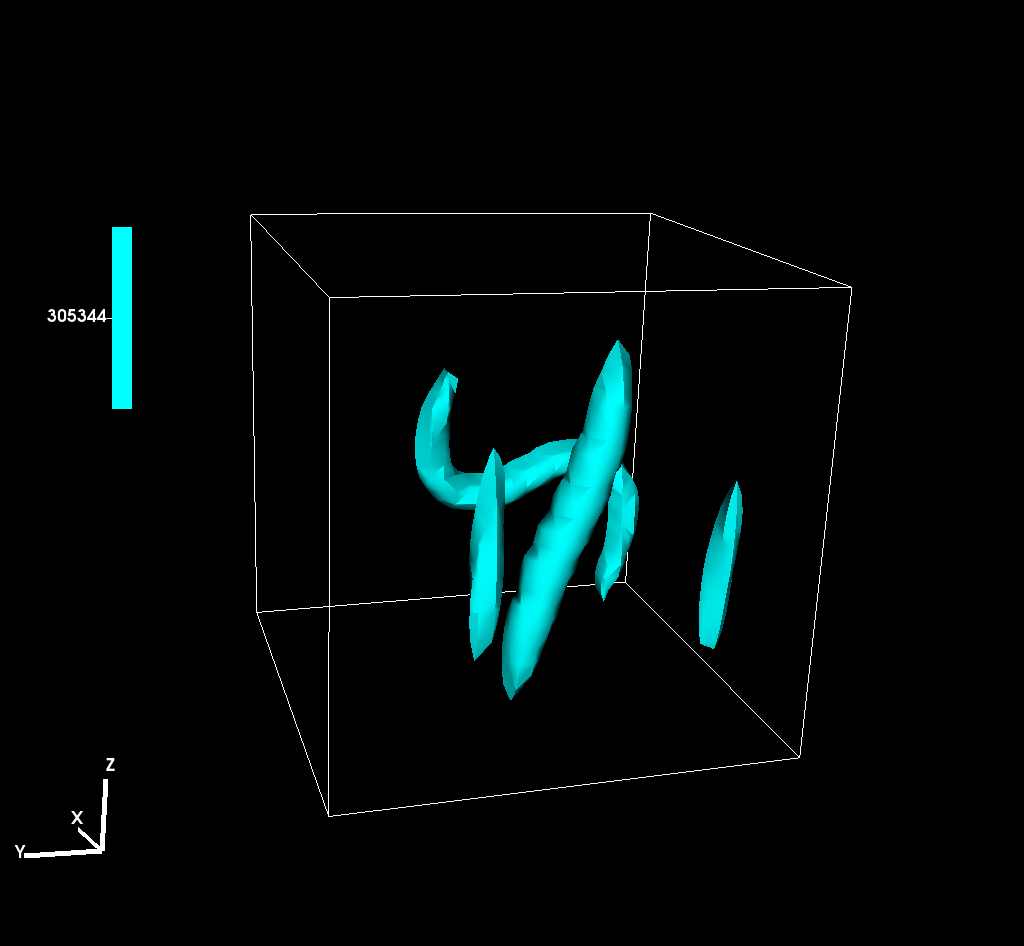}
    \put(-70,60){\color{white} \bf (b)}   
    \includegraphics[scale=0.07]{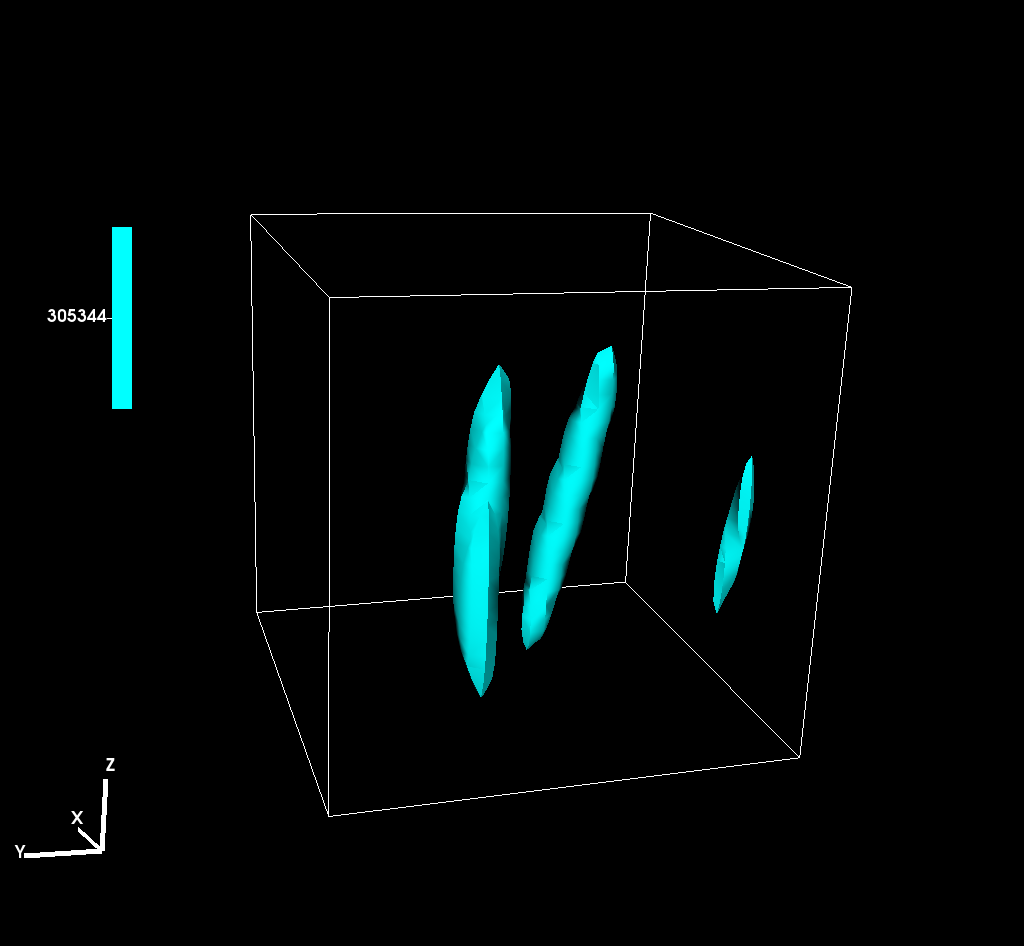}
    \put(-70,60){\color{white} \bf (c)}

    \includegraphics[scale=0.07]{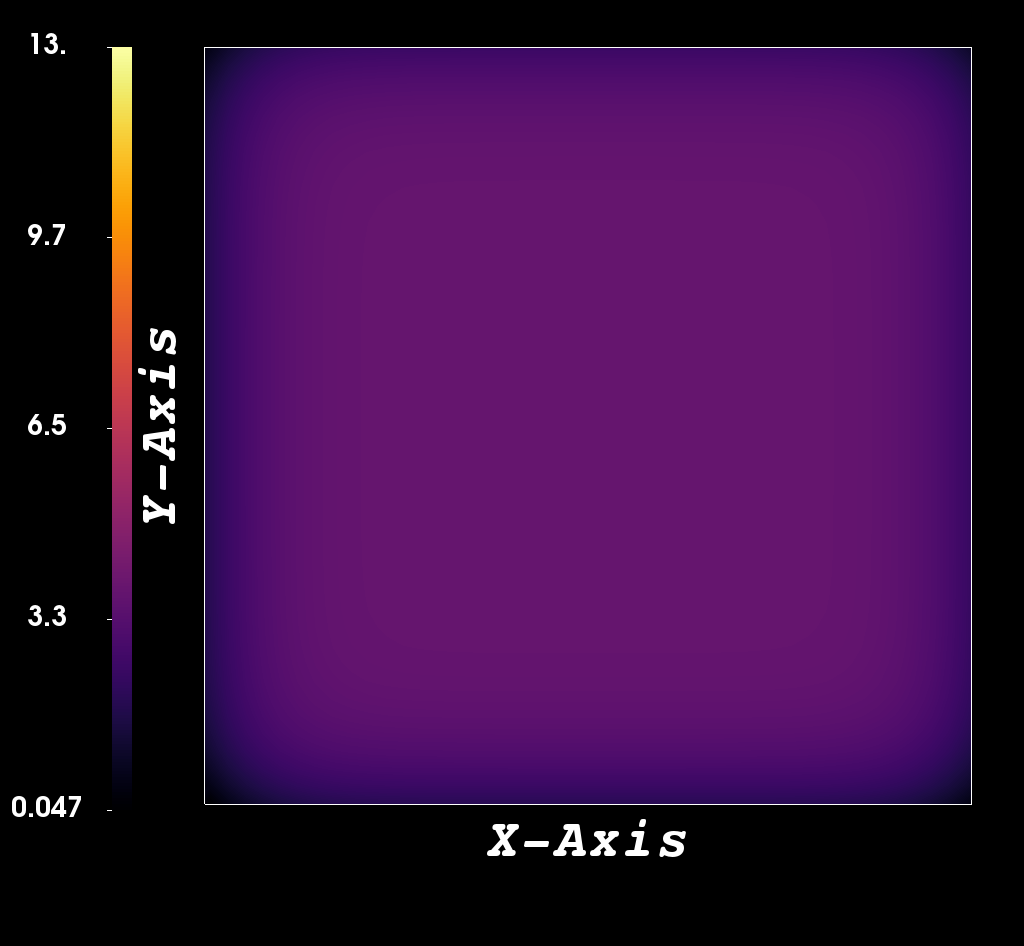}
    \put(-70,60){\color{white} \bf (d)}
    \includegraphics[scale=0.07]{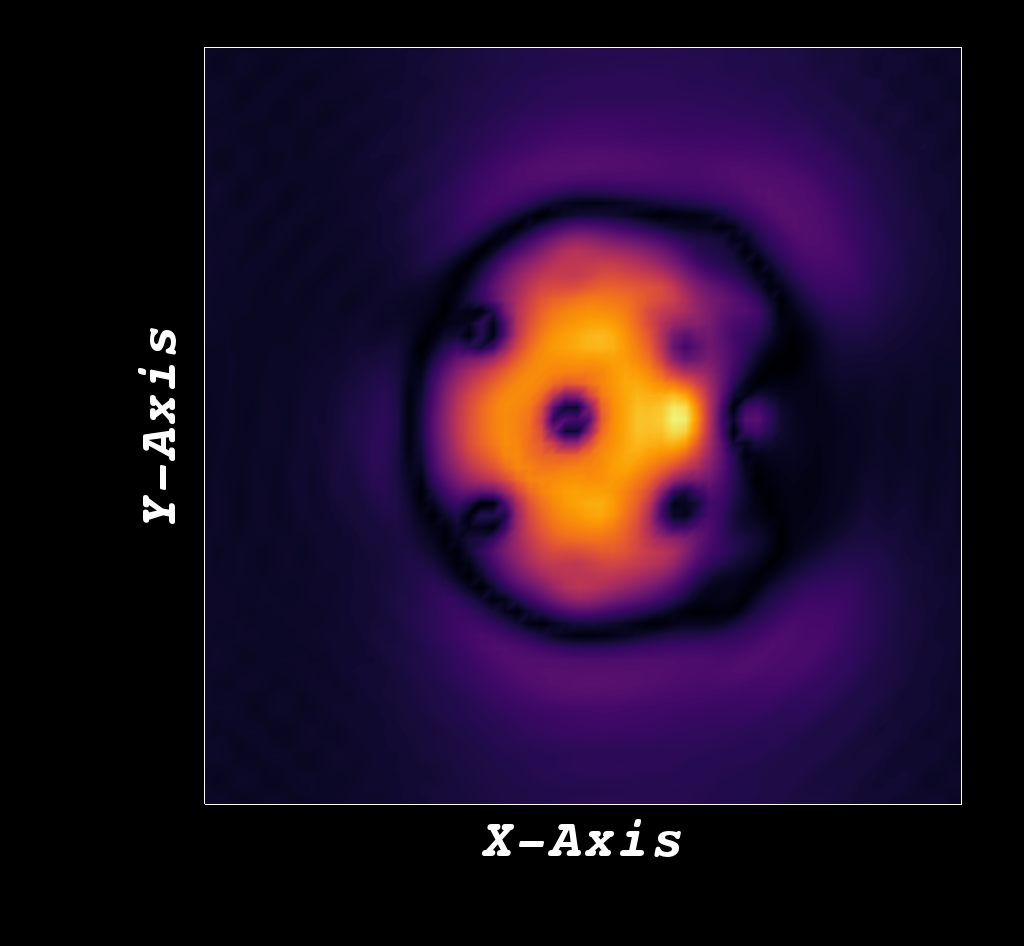}
    \put(-70,60){\color{white} \bf (e)}
    \includegraphics[scale=0.07]{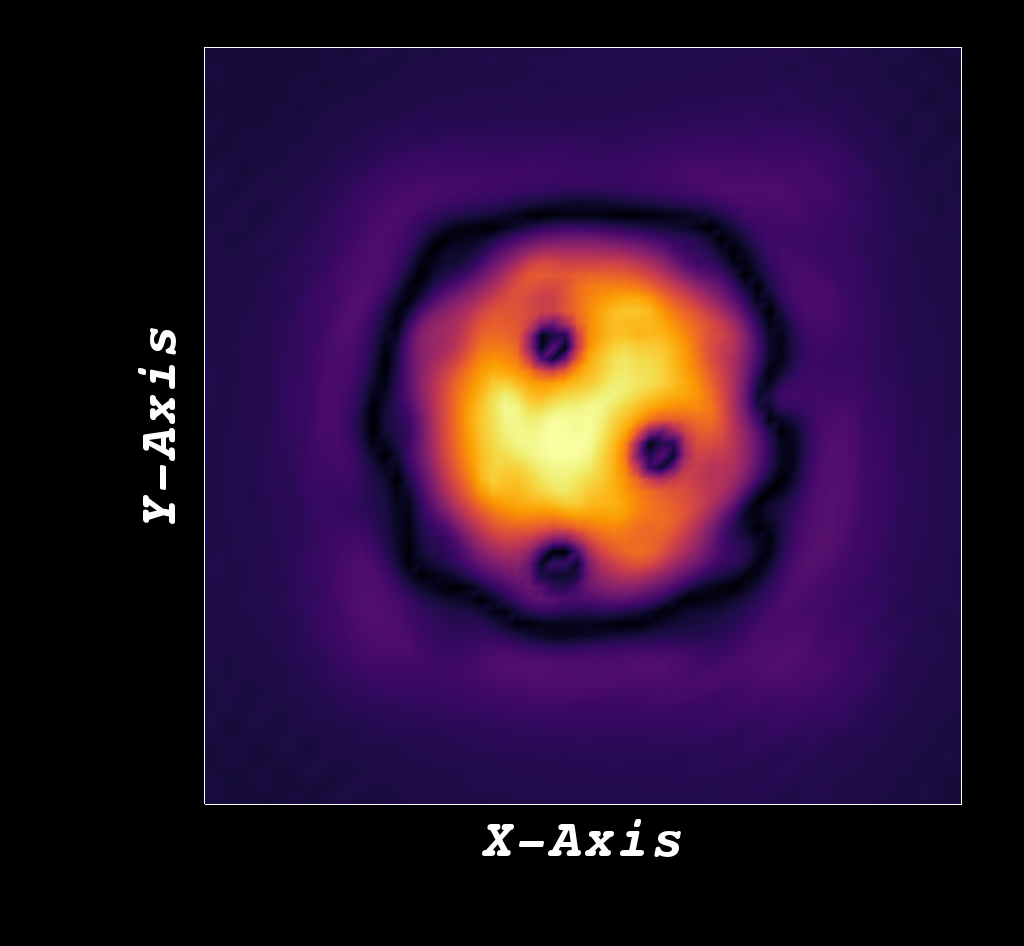}
    \put(-70,60){\color{white} \bf (f)}

    \includegraphics[scale=0.07]{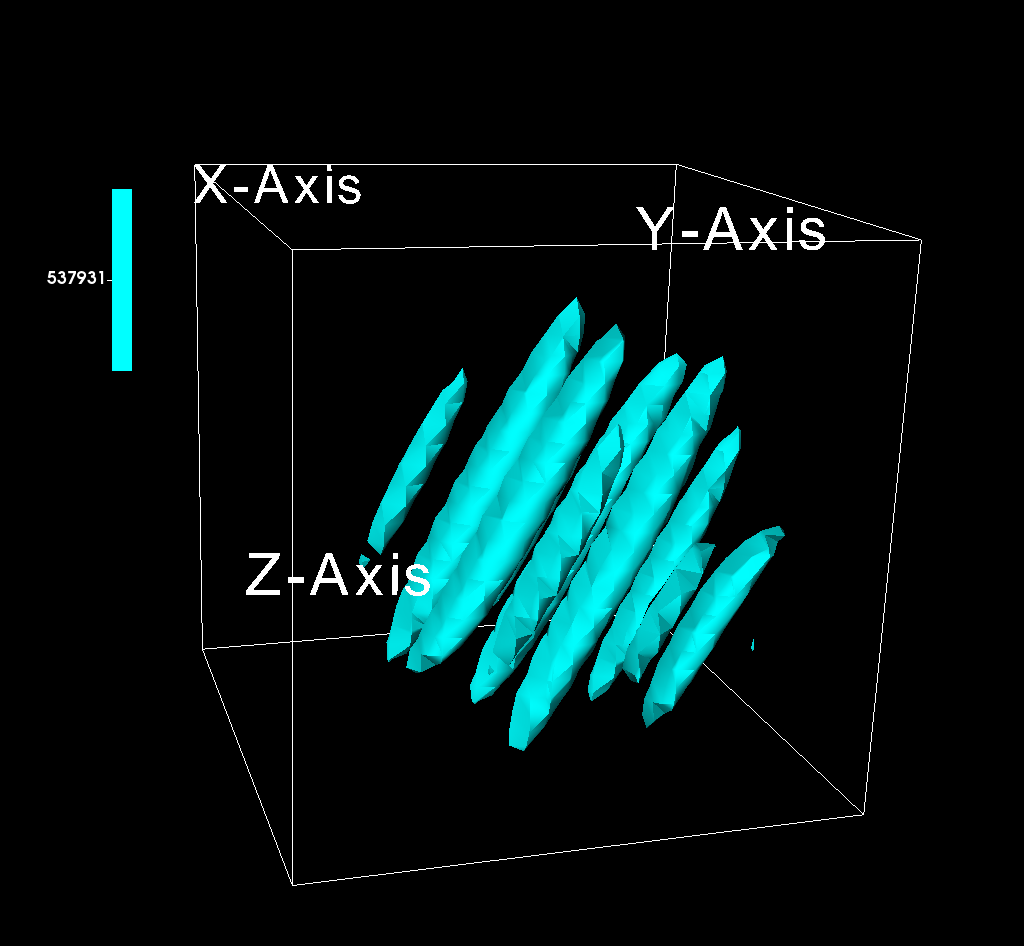}
    \put(-70,60){\color{white} \bf (g)}
    \includegraphics[scale=0.07]{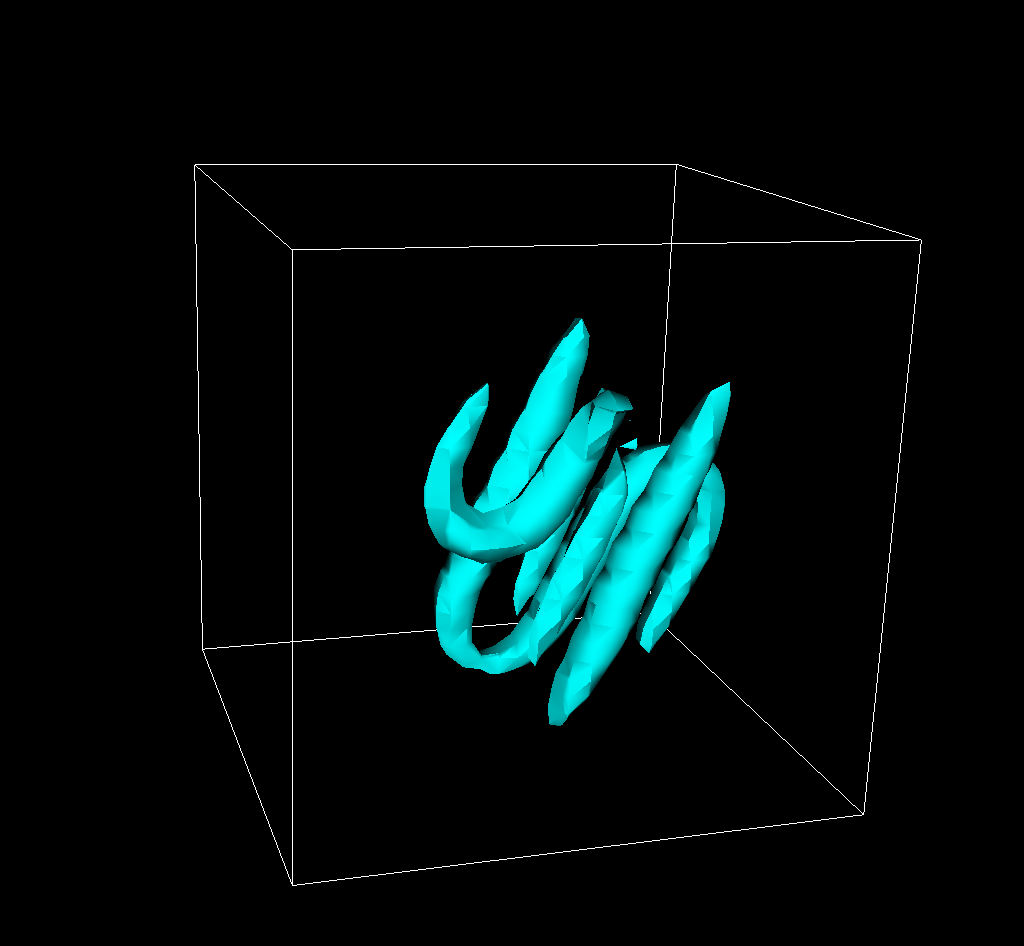}
    \put(-70,60){\color{white} \bf (h)}   
    \includegraphics[scale=0.07]{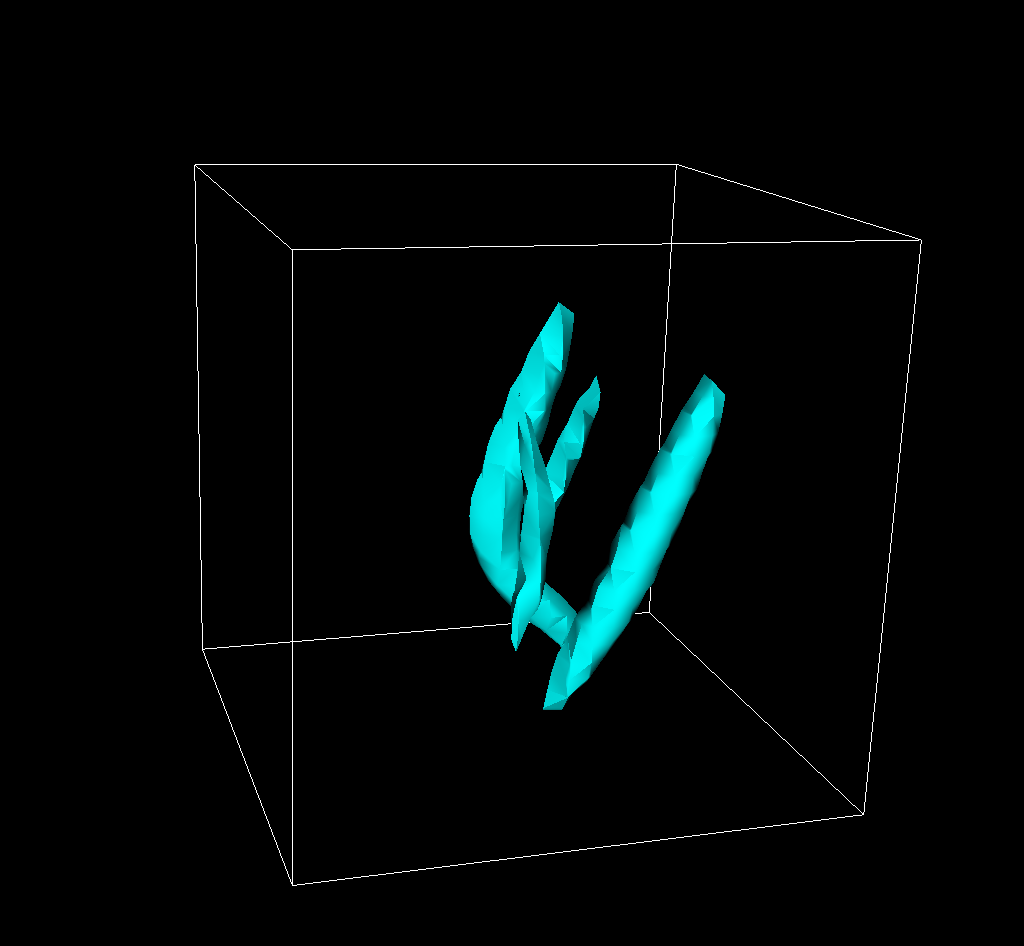}
    \put(-70,60){\color{white} \bf (i)}
    
    \caption{ {\bf (a)}-{\bf (c)} One-level contour plots of $(\nabla \times (\rho v))^2$ for proton flux tubes and {\bf (d)}-{\bf (f)} pseudocolor plots of the magnetic field {\bf B} at the midplane ($z=L/2$) at three representative imaginary times. Both neutron and proton subsystems rotate with the angular velocity ${\bf \Omega} = 4.0 \hat{z}$; and ${\bf B}_{\rm ext}=0$. {\bf (g)}-{\bf (i)} One-level contour plots of $(\nabla \times (\rho v))^2$ for proton flux tubes with ${\bf \Omega} = 4.0 \hat{z}$; and ${\bf B}_{\rm ext}=0.8$.  At the initial imaginary time in {\bf (a)}, the proton flux tubes make an angle $\Theta=30^{\circ}$ with the $z$-axis. Both species interact only through the gravitational potential.}
    \label{fig:agle_3D_p_rotation}
\end{figure}

We now consider the counterpart of Case(ii) Subsection~\ref{sec:imag_time_no_int_no_Vtheta_p}: both neutron and proton condensates rotate with an angular velocity ${\bf \Omega}=\Omega\hat{\bf z}$ and there is no external magnetic field [${\bf B}_{\rm ext}=0$].  Vortices, oriented along the $z$-axis, thread the neutron condensate. At small values of the imaginary time $t$, the proton-superconductor flux tubes are oriented at an angle $\Theta=30^o$ [Fig.~\ref{fig:agle_3D_p_rotation}(a)]. As $t$ increases, these flux tubes try to align themselves along the rotation axis $z$ [ Fig.~\ref{fig:agle_3D_p_rotation}(b)], and they do so ultimately [Fig.~\ref{fig:agle_3D_p_rotation}(c)]. This alignment is facilitated by the absence of an external magnetic field, which tends to counteract this reorientation. We also observe the generation of a London magnetic field ${\rm B}_L$ inside the proton condensate [Fig.~\ref{fig:agle_3D_p_rotation}(d)-(f)].

Finally we consider the counterpart of Case(iii) Subsection~\ref{sec:imag_time_no_int_no_Vtheta_p}: Both neutron and proton condensates rotate [${\bf \Omega}=\Omega\hat{\bf z}$] and there is an external uniform magnetic field ${\bf B}_{\rm ext}$, as in
a pulsar, but with interactions solely through the gravitational potential. At small values of the imaginary time $t$ [Fig.~\ref{fig:agle_3D_p_rotation}(g)], the proton-supercondcutor flux tubes are aligned at an angle $\Theta=30^o$ with the $z$-axis, and the magnetic field is concentrated primarily outside the condensate. As $t$ increases, the proton flux tubes try to orient themselves along the rotation axis [Fig.~\ref{fig:agle_3D_p_rotation}(h)], but 
${\bf B}_{\rm ext}$ resists this alignment. Ultimately, these proton-superconductor flux tubes exhibit frustration [Fig.~\ref{fig:agle_3D_p_rotation}(i)] as they try both to align globally with the rotation axis and to adhere to ${\bf B}_{\rm ext}$, which makes an angle $\Theta=30^\circ$ with the z-axis. This frustration is akin to the glassy behavior of flux tubes, studied in Ref.~\cite{Drummond_2017}, without gravity but with a quadratic confining potential.
In the next Subsection we go beyond the study of Ref.~\cite{Drummond_2017} by incorporating the full Maxwell equations that lead to entrainment. 

\subsubsection{\bf Non-zero interactions: $\gamma \neq 0$, $V_{\theta}= 0$, and $\Theta=30^o$}
\label{sec:imag_time_non_zero_angle_int_no_Vtheta}
\begin{figure}[!htb]
    \centering
    \includegraphics[scale=0.08]{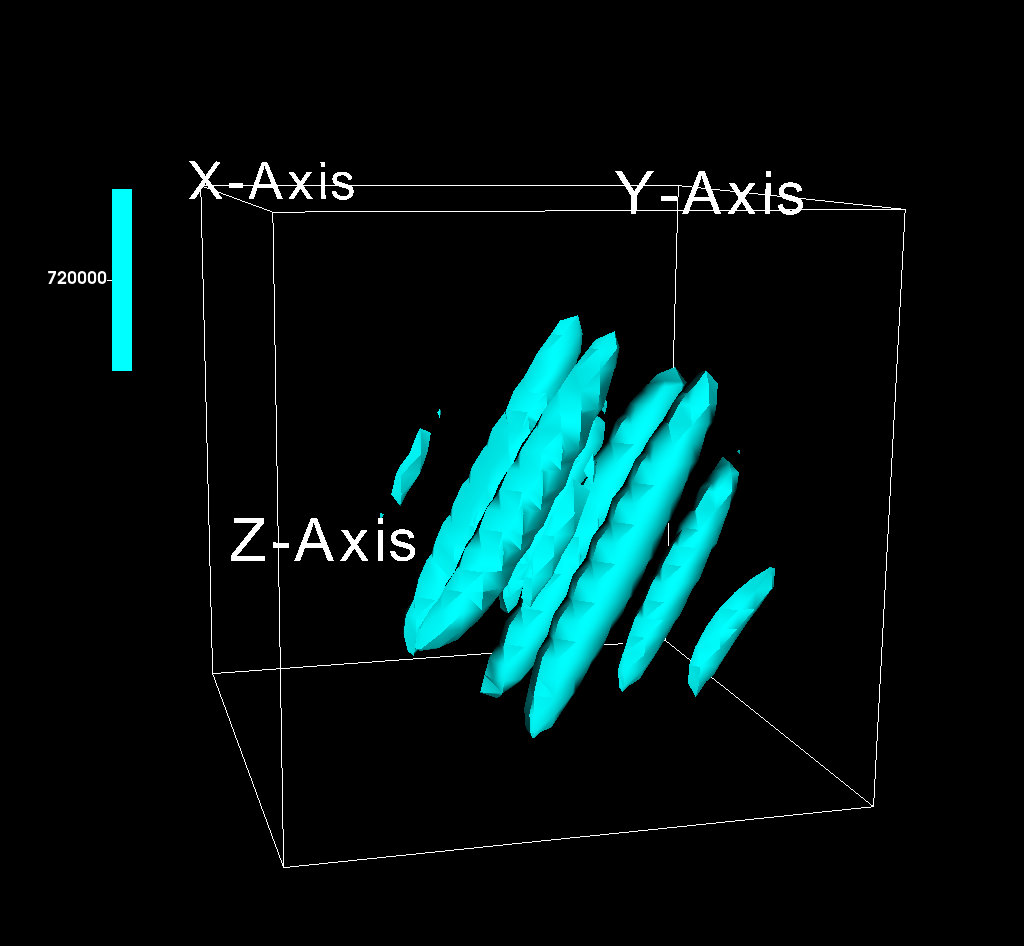}
    \put(-70,65){\color{white} \bf (a)}
    \includegraphics[scale=0.08]{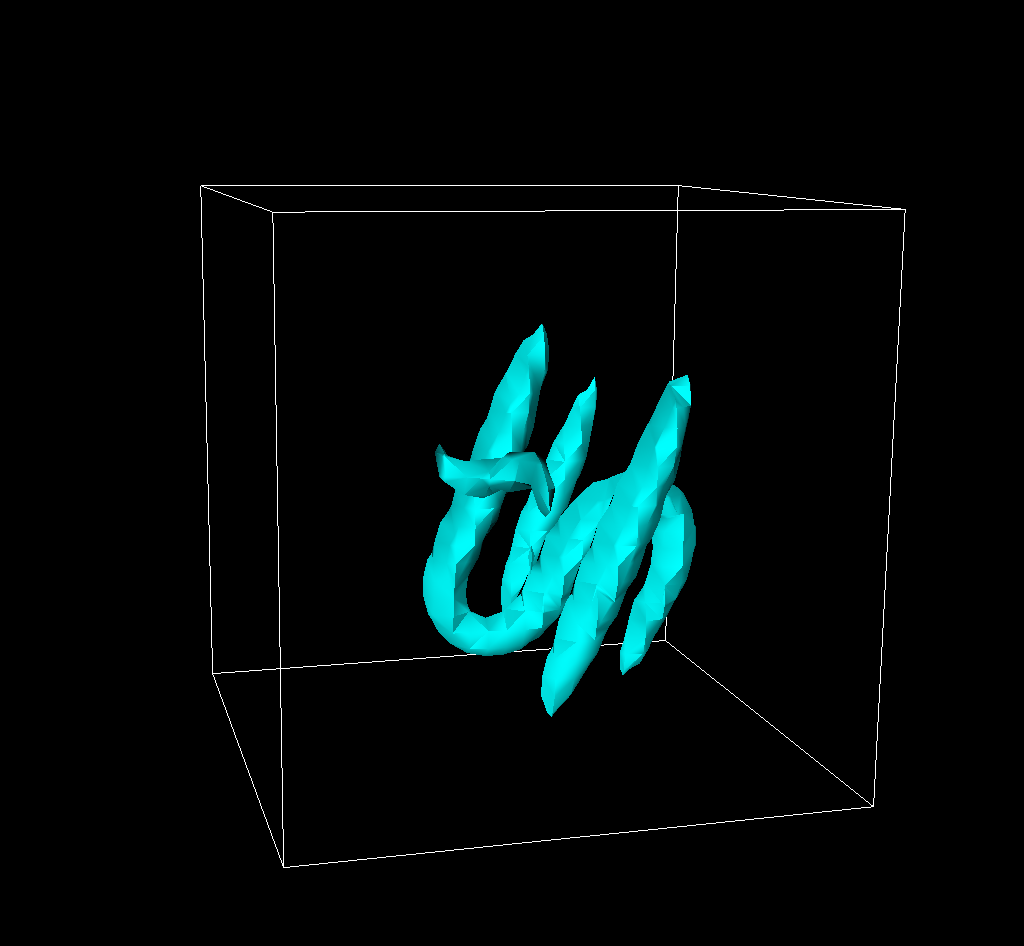}
    \put(-70,65){\color{white} \bf (b)}    
    \includegraphics[scale=0.08]{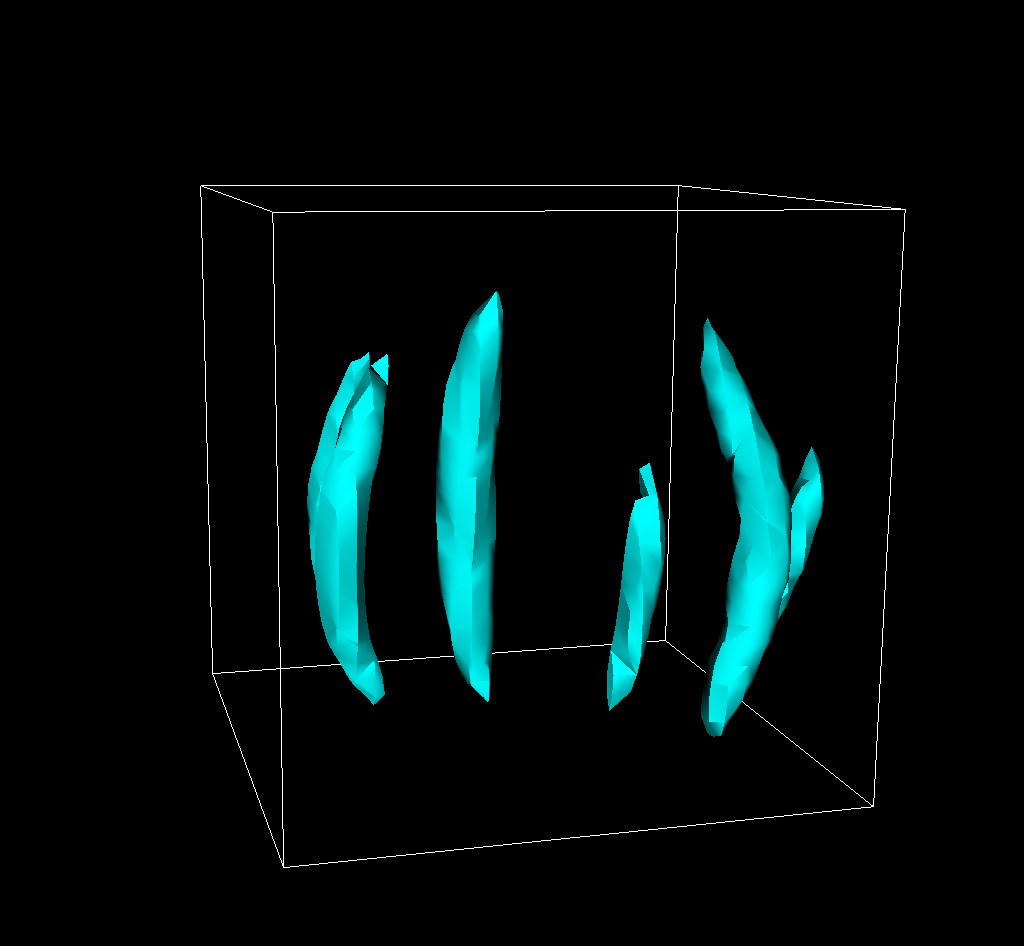}
    \put(-70,65){\color{white} \bf (c)}
   
    \caption{One-level contour plots of $(\nabla \times (\rho v))^2$ for proton flux tubes at three representative imaginary times in {\bf (a)}, {\bf (b)}, and {\bf (c)}. Both neutron and proton subsystems rotate with the angular velocity ${\bf \Omega} = \Omega \hat{z}$, where $\Omega=4.0$; and ${B}_{\rm ext}=0.8$, which makes an angle $\Theta=30^{\circ}$ with the $z$-axis. Furthermore, we have gravitational and attractive density-density ($g'<0$) interactions (first term in Eq.~(\ref{eq:interaction_lag})) between neutron and proton Cooper pairs.}
    \label{fig:agle_3D_p_rotation_Bex}
    \label{fig:agle_3D_p_rotation_Bext_G_NDD}
\end{figure}

Consider now the counterpart of Subsection~\ref{sec:imag_time_int_no_Vtheta_p}, i.e.,
direct interaction ($\gamma\neq0$) between neutron and proton condensates. We first examine attractive density-density interactions [Eq.~(\ref{eq:gprime})], so $g' < 0$.
We show the evolution (in imaginary-time $t$) of the proton-superconductor flux tubes and the magnetic field in Figs.~\ref{fig:agle_3D_p_rotation_Bext_G_NDD}(a)- (c). At the initial time  $t=0$, the proton flux tubes are oriented at an angle $\Theta=30^o$ with respect to the $z$-axis [Fig.~\ref{fig:agle_3D_p_rotation_Bext_G_NDD}(a)]. As $t$ increases, these flux tubes exhibit a rapid realignment with the rotation axis because of the combined gravitational and attractive density-density interactions [Figs.~\ref{fig:agle_3D_p_rotation_Bext_G_NDD}(b)-(c)].

\begin{figure}[!htb]
    \centering
    \includegraphics[scale=0.36]{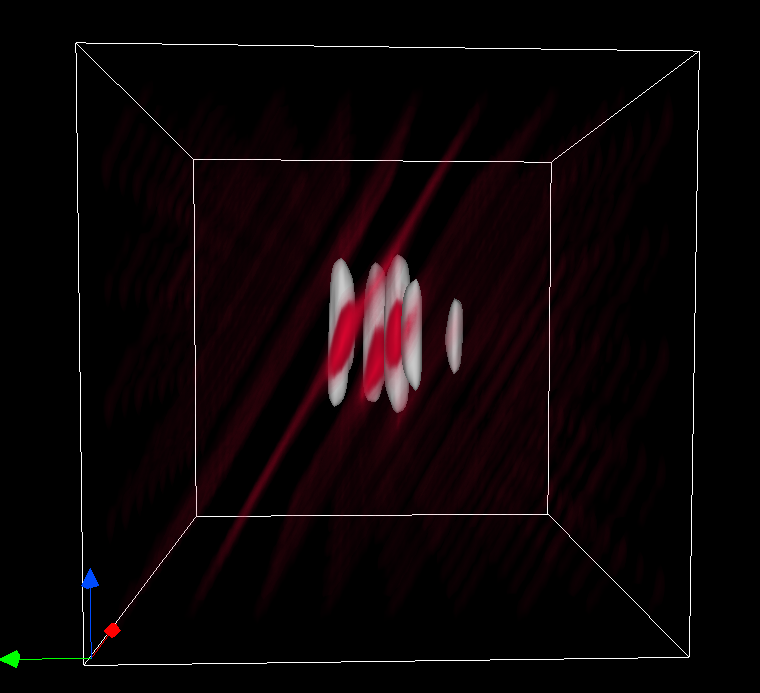}

    \caption{One-level contour plots of $(\nabla \times (\rho v))^2$ for neutron-superfluid vortices, at the final imaginary time with a superimposed volume plot of the magnetic field ${\bf B}$ at the final time. Both neutron and proton subsystems rotate with the angular velocity ${\bf \Omega} = \Omega \hat{z}$, where $\Omega=4.0$; and ${B}_{\rm ext}=0.8$, which makes an angle $\Theta=30^{\circ}$ with the $z$-axis.  Gravitational and other interaction terms are included  [Eq.~(\ref{eq:interaction_lag}))].}
    \label{fig:agle_3D_p_rotation_Bex}
    \label{fig:agle_3D_p_rotation_Bext_G_NDD_CC}
\end{figure}

To obtain entrainment, we must introduce the current-current interactions [the second term in Eq.~(\ref{eq:interaction_lag})]. The proton flux tubes evolve in imaginary time $t$ as in the previous case with attractive density-density interactions. However, the current-current interaction, $\frac{\gamma q}{c^2\epsilon_0} {\bf J}_n|\psi_p|^2$ in the vector potential~(\ref{eq:vector_pot}), leads to an entrained proton-superconductor current. This entrainment results in an induced magnetic field inside the neutron-superfluid vortices [gray and red isosurfaces, respectively, in Fig.\ref{fig:agle_3D_p_rotation_Bext_G_NDD_CC}]. 

\subsection{Real-time evolution}
\label{sec:real_time}
\begin{figure*}[!htb]
\includegraphics[scale=0.27]{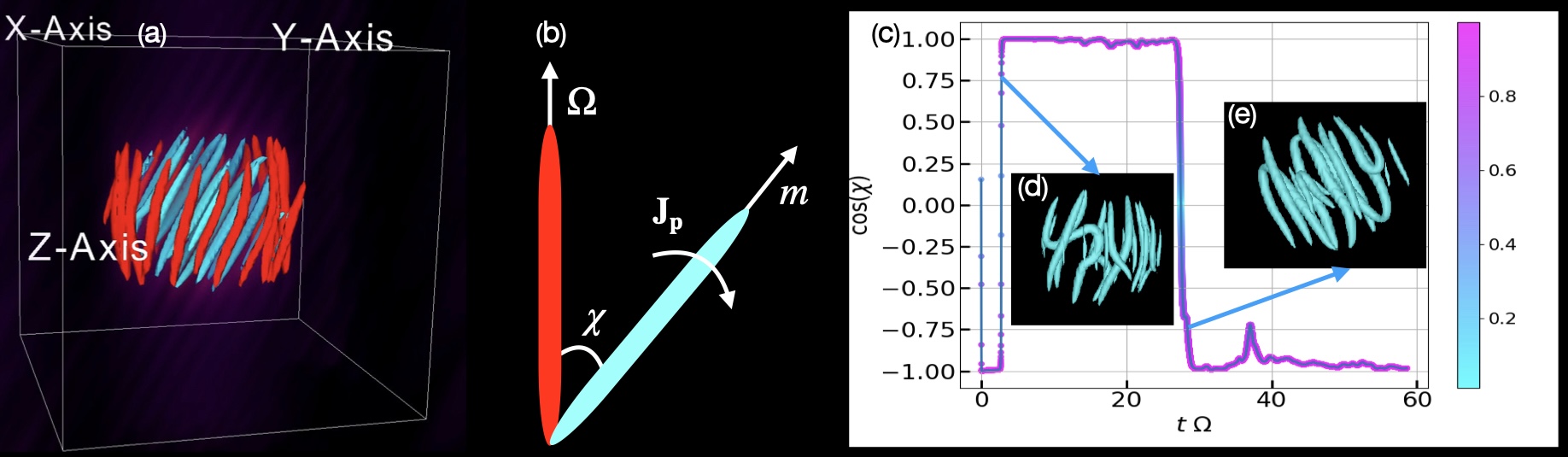}

\caption{Real-time evolution: {\bf (a)} One-level contour plot of $(\nabla \times (\rho v))^2$ for neutron vortices (in red) and proton flux tubes (in cyan) at the initial time. {\bf (b)} Schematic diagram showing the angle $\chi$ between the rotation axis and the magnetic moment [Eq.~(\ref{eq:mag_mom})]. {\bf (c)} The evolution of the angle $\chi$ with time. Both neutron and proton subsystems rotate with angular velocity ${\bf \Omega} = \Omega \hat{z}$, where $\Omega=4.0$; and ${B}_{\rm ext}=0.8$, which makes an angle $\Theta=30^{\circ}$ with the $z$-axis. Insets (d) and (e) show illustrative proton flux-tube configurations before and after the reversal.}
\label{fig:agle_3D_magnetic_moment}
\end{figure*}

We now delve into the real-time equations~(\ref{eq:GPE_neutron_ndim})-(\ref{eq:scalar_pot_ndim}).
We first follow the dynamics of the alignment of proton-superconductor flux tubes with the rotation axis [Fig.~\ref{fig:agle_3D_p_rotation_Bext_G_NDD}]. We begin with the configuration of vortices (cyan) and flux tubes (red)
shown in Fig.\ref{fig:agle_3D_magnetic_moment}(a), which we obtain as the equilibrium state of the imaginary-time versions of Eqs.(\ref{eq:GPE_neutron_ndim})-(\ref{eq:vector_pot_ndim}). The axes of rotation and the magnetic moment ${\bf m}$ make an angle $\chi$ [see the 
schematic diagram in Fig.\ref{fig:agle_3D_magnetic_moment}(b)]; we define them as follows:
\begin{eqnarray}
    {\bf m}&=&\frac{1}{2}\int {\bf r} \times {\bf J}_p dV \,;\\
    \cos(\chi) &=& \frac{{\bf \Omega}\cdot {\bf m}}{|{\bf \Omega}||{\bf m}|}\,.
    \label{eq:mag_mom}
\end{eqnarray}
The angle $\chi$ depends on time $t$; for considerable lengths of time it shows minor fluctuations, but, occasionally, it changes dramatically, as we show in Fig.\ref{fig:agle_3D_magnetic_moment}(c) via a plot of $\cos(\chi)$ versus $t$. The sudden change in $\cos(\chi)$, from positive to negative values, indicates that the magnetic moment undergoes reversals, which are familiar 
in many other hydrodynamical~\cite{mishra2015dynamics,shukla2016statistical}, dynamo~\cite{berhanu2007magnetic,petrelis2008chaotic,gissinger2010morphology}, geomagnetic~\cite{valet2016deciphering,laj2021geomagnetic,alberti2023unveiling}, and astrophysical~\cite{dobbs2016magnetic,han2018pulsar,charbonneau2014solar} settings.
Measurements for several pulsars~\cite{Malov_2011} indicate that $0^o\leq \chi \leq 90^o$; e.g., the pulsar (PSR B1055-52) is an aligned rotator, whereas another (PSR B1702-19) is an orthogonal rotator; the time scales of these observations are such that, for any given pulsar, $\chi$ has a steady, time-independent value.  

In our minimal model the crust potential $V_{\theta}$ is a function of a single polar angle $\theta$ whose dynamics is given by Eq.~(\ref{eq:crust_pot_eq}). To study the interplay of crust, neutron-superfluid, and proton-superconductors, we use Eqs.~(\ref{eq:GPE_neutron_ndim}-(\ref{eq:scalar_pot_ndim}) along with Eq.~(\ref{eq:crust_pot_eq}) and initial conditions
that we obtain from the equilibrium states of imaginary-time studies of the previous Sections, with the same angular velocity $\Omega=\frac{d\theta}{dt}$ for the crust, the neutron condensate, and the proton condensate. For purposes of illustration, we consider $\Theta=0$, i.e., the angle between the rotation axis and the magnetic field is zero. The neutrons and protons interact only through the gravitational potential via the Poisson equation ($\gamma = 0$). 
 
\begin{figure*}[!htb]
    \centering
    \includegraphics[scale=0.26]{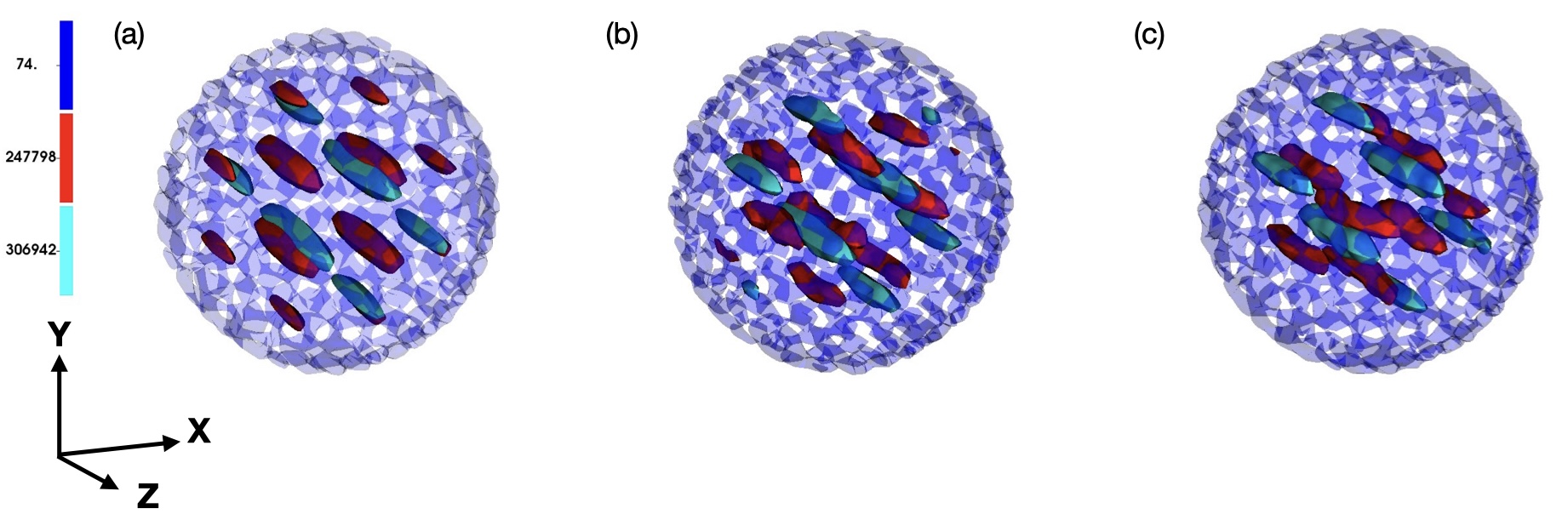}
    
    \caption{One-level contour plots of the crust potential together with the neutron vortices (in red) and proton flux tubes (in cyan) at three different times in {\bf (a)}, {\bf (b)}, and {\bf (c)} obtained by using the real-time GPPE (Eq.~(\ref{eq:GPE_neutron_ndim})) and  RTGLE (Eq.~(\ref{eq:GLE_proton_ndim})). Both neutron and proton subsystems rotate with an angular velocity ${\bf \Omega} = \Omega \hat{z}$, where $\Omega=4.0$; and ${B}_{\rm ext}=4.0$, which is along the $z$-axis.}
    \label{fig:gpe_3D_npv_crust}
\end{figure*}

In Figs.~\ref{fig:gpe_3D_npv_crust}(a)-(c), we show isosurface plots of the crust potential  in blue, isosurfaces of neutron-superfluid vortices in red, and proton-superconductor flux tubes in cyan at three representative times. At the initial time step (Fig.\ref{fig:gpe_3D_npv_crust}(a)), the system features $12$ neutron vortices and $6$ proton flux tubes. Given the parameters of our simulation, the proton-superconductor flux tubes are more effectively pinned by the crust potential than
neutron-superfluid vortices, in part because of the anchoring of the flux tubes to the strong external magnetic field ${\bf B}_{\rm ext}$. Over the duration of the simulation presented in Figs.~\ref{fig:gpe_3D_npv_crust}(a)-(c), the number of neutron-superfluid vortices reduces by a factor of $2$ but the number of proton flux tubes remains unchanged. Both superfluid vortices and proton-superconductor flux tubes undergo differential rotation because of the friction coefficient $\alpha$ in equation~(\ref{eq:crust_pot_eq})) for the polar angle. 

The angular momentum of the crust is given as ${\rm J}_c = I_c d\theta/dt$, with $I_c$ is the moment of inertia of the crust [see the Appendix~\ref{sec:ang_mom_apen} for details]. The temporal evolution of ${\rm J}_c$ is complicated because of the subtle interplay between the friction, which slows down the crust, and the angular momentum in the neutron-superfluid vortices. When such a vortex is ejected from the pulsar, its angular momentum is transferred to the crust.
Some neutron vortices linger close to proton-superconductor flux tubes because of the  Poisson-equation-induced gravitational attraction between them. This also affects the time-dependence of ${\rm J}_c$. Finally, we have an effective stick-slip dynamics for ${\rm J}_c$ that displays glitches whose statistics has properties that are akin to those seen in several pulsars and which yield pulsar glitches~\cite{Radhakrishnan_1969,reichley1969observed,manchester_2017,AK_verma_2022}. We observe small quasi-oscillatory structures in the time series of ${\rm J}_c$. This occurs because we use the periodic version of coordinates ($x_p$,$y_p$) defined in Eq.~\eqref{eq:x_y_periodic}. It is also important to note that, in our model, the crust potential rotates with the superfluid when vortices are pinned; however, when a vortex becomes unpinned, the crust experiences a sudden decrease in its angular momentum, followed by an increase in the angular momentum as the unpinned vortex moves from the condensate to the crust,
thus transferring its angular momentum.

\begin{figure*}[!htb]
    \centering
    \includegraphics[scale=0.25]{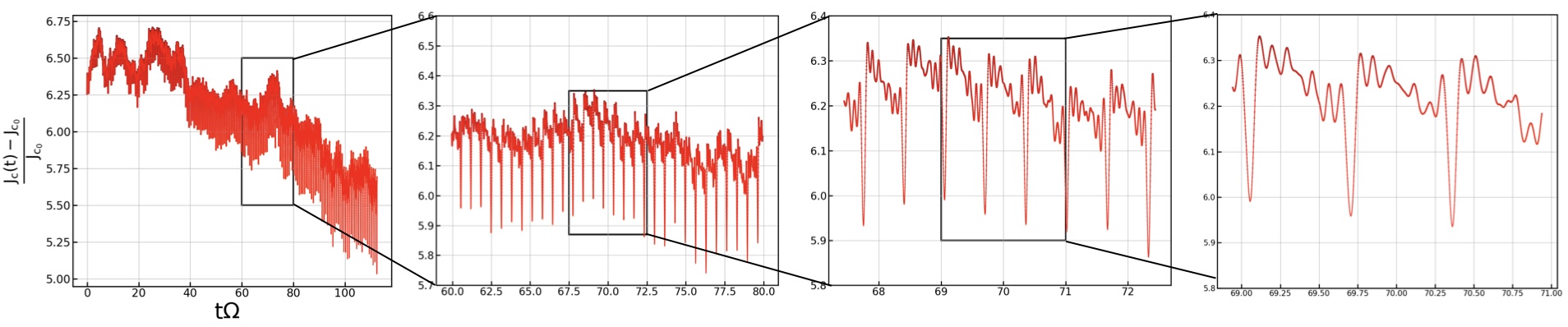}
    \put(-450,88){ \bf (a)}
    \put(-340,88){ \bf (b)}
    \put(-220,88){ \bf (c)}
    \put(-100,88){ \bf (d)}

    \includegraphics[scale=0.255]{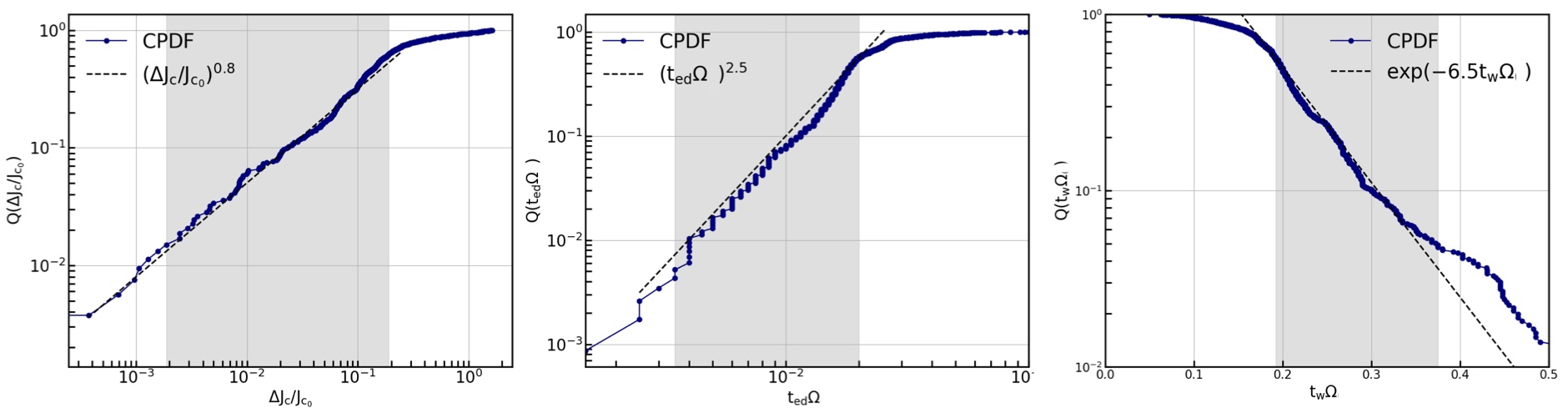}
    \put(-410,110){ \bf (e)}
    \put(-250,110){ \bf (f)}
    \put(-90,110){ \bf (g)}
    \caption{{\bf (a)} Time series of the crust angular momentum $\rm (J_c-J_{c_0})/J_{c_0}$. {\bf (b)}, {\bf (c)}, and {\bf (d)} are the zoomed versions of the rectangular regions shown in the preceding plots. Log-Log plots of ${\bf (e)}$ the CPDF $\rm Q(\Delta J_c/J_{c_0})$ of the event size and ${\bf (f)}$ the CPDF $\rm Q(\rm t_{ ed}\Omega)$ of the event duration. ${\bf (g)}$ semilog plot of the CPDF $\rm Q(\rm t_{ w}\Omega)$ of the waiting time. $\rm J_{c_0}$ and $\rm \Omega$ are the initial angular momentum and initial angular velocity of the crust, respectively.}
    \label{fig:gpe_3D_crust_angmom}
\end{figure*}
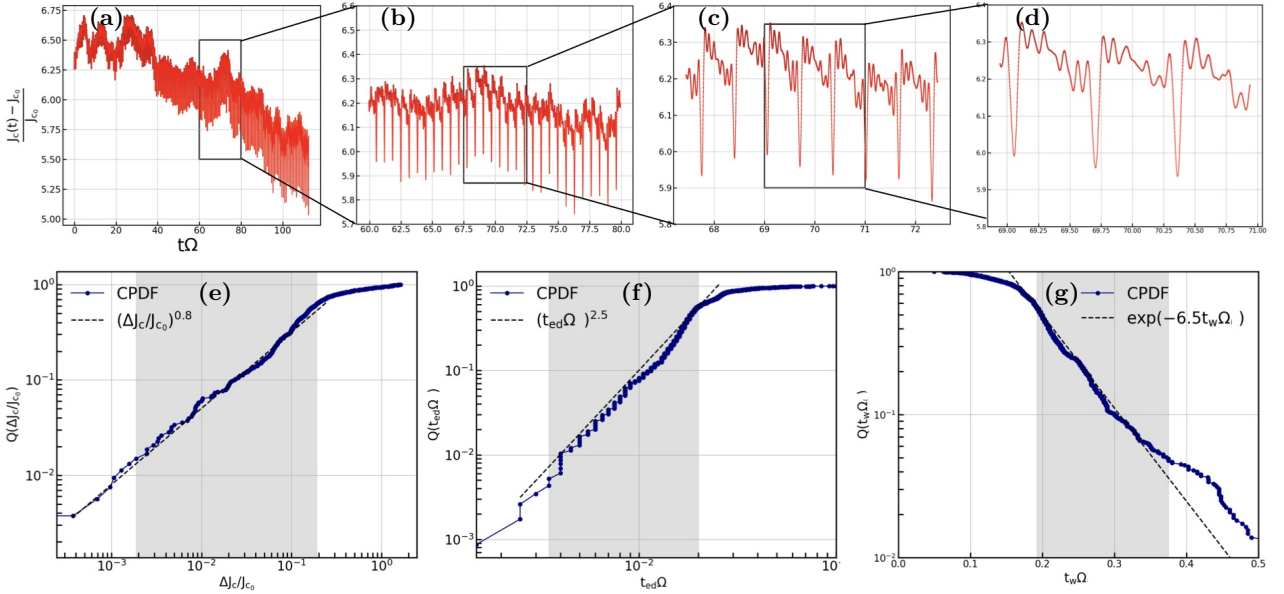

We now examine the analogues of pulsar glitches in our model, by following the methods developed in Ref.~\cite{AK_verma_2022}. In particular, we present the time series of the angular momentum $ ({\rm J}_c-{\rm J}_{c_0})/{\rm J}_{c_0}$ of the crust in Fig.\ref{fig:gpe_3D_crust_angmom}(a). This time series of $ ({\rm J}_c-{\rm J}_{c_0})/{\rm J}_{c_0}$ exhibits characteristic features that are associated with Self-Organized Criticality (SOC)~\cite{Bak_1987,Jensen_1998,Donald_Turcotte_1999,Melatos_2008}, which we have explored, in the context of gravitationally collapsed boson stars, in our earlier work~\cite{AK_verma_2022}. Figures~\ref{fig:gpe_3D_crust_angmom}(b)-(d) present expanded views of specific segments [indicated by black boxes] of the time series in Fig.\ref{fig:gpe_3D_crust_angmom}(a). From the time dependence of ${\rm J}_c$, we observe that the crust can either lose angular momentum to the superfluid or can gain angular momentum from it, because of the stick-slip dynamics mentioned above. 

To characterize SOC, we quantify the statistics ${\rm J}_c$ as follows: We measure (a) the event size $\Delta  {\rm J}_c$, which is the difference between successive minima and maxima in ${\rm J}_c$, (b) the event-duration time $t_{ed}$, which is the time difference between successive minima and maxima of $ {\rm J}_c(t)$, and (c) the waiting time $t_{w}$, which is the time between successive maxima in $ {\rm J}_c(t)$. We then obtain cumulative probability distribution functions (CPDFs) of $\Delta  {\rm J}_c$, $t_{ed}$,
and $t_{w}$; as in Ref.~\cite{AK_verma_2022}, the former two CPDFs exhibit power-law tails, whereas the last has an exponential tail. In Fig.~\ref{fig:gpe_3D_crust_angmom}(e), we plot the CPDF $Q(\Delta {\rm J}_c/{\rm J}_{c_0})$; it scales as $Q(\Delta {\rm J}_c/{\rm J}_{c_0})\sim (\Delta  {\rm J}_c/{\rm J}_{c_0})^{\beta}$, within the gray-shaded region. Therefore, the corresponding probability distribution function (PDF) scales as $P(\Delta {\rm J}_c/{\rm J}_{c_0})\sim (\Delta {\rm J}_c/{\rm J}_{c_0})^{\beta-1}$; for our representative run, we obtain the scaling exponent $\beta = 0.86\pm 0.15$, by using local-slope analysis. The CPDF of $t_{ed}$ shows the power law $Q(t_{ed}\Omega)\sim (t_{ed}\Omega)^{\gamma_t}$ in the gray-shaded region of Fig.~\ref{fig:gpe_3D_crust_angmom}(f), with an exponent $\gamma_t=2.5\pm 0.2$ for our run. The CPDF of $t_w$ shows the exponential form $Q(t_{w}\Omega)\sim \exp(-6.5 t_{w}\Omega)$ [Fig.~\ref{fig:gpe_3D_crust_angmom}(f)]. The qualitative forms of these CPDFs is similar to those seen in experiments, as has been noted in Ref.~\cite{AK_verma_2022}, which uses the GPPE system without the proton superconductor and the Maxwell equations that we include. The values of the exponents $\beta$ and $\gamma$ lie close to those that have been observed for certain pulsars [e.g., PSR J 1825-0935 has glitch-size-PDF exponent $\simeq 0.36 $ ]~\cite{Melatos_2008}. In Ref.~\cite{AK_verma_2022}, it has been noted that range of glitching sizes depends on $\Omega$. In addition, we find that these sizes also depend on $B_{\rm ext}$. 

We have noted above that Poisson-equation-induced interaction between the neutron superfluid and the proton superconductor makes superfluid vortices approach superconductor flux tubes. As the crust decelerates, the neutron vortices leave the condensate abruptly. The associated jumps in $ {\rm J}_c(t)$ are somewhat sharper in time but smaller in amplitude than those in the GPPE model of Ref.~\cite{AK_verma_2022}. Consequently, our values of the exponents $\beta$ and $\gamma$ are about $10\%$ larger than those in Ref.~\cite{AK_verma_2022}, but still comfortably in the observational range~\cite{Melatos_2008} $-0.13\lesssim -(\beta-1)\lesssim 2.4$. The inclusion of current-current and density-density interactions [Eqs.~(\ref{eq:interaction_lag})] may reduce the sizes of glitches, by slowing down the ejection of vortices from the condensate. Furthermore, we expect that the current-entrainment term in Eq.~(\ref{eq:vector_pot_ndim}), which induces a magnetic field inside neutron vortices, could reduce glitch sizes.

\section{Conclusions}
\label{sec:conclusions}
We have developed a theoretical framework for studying the coupled motion of neutron-superfluid vortices and proton-flux tubes in a gravitationally collapsed condensate. In this framework we have employed (a) a 3D Gross-Pitaevskii-Poisson-Equation (GPPE) for neutron Cooper pairs, (b) the Real-Time-Ginzburg-Landau equation (RTGLE) for proton Cooper pairs, (c) the Maxwell equation for the vector potential ${\bf A}$, and (d) Newtonian gravity and interactions, both direct and induced by the Poisson equation, between the neutron and proton subsystems. For a pulsar we have included, in addition, a crust potential as in Ref.~\cite{AK_verma_2022}. The recent studies in Refs.~\cite{Drummond_2017,Drummond_2017_b,Drummond_2023_c} use the Gross-Pitaevskii-Equation and Ginzburg-Landau equation (imaginary time) together with a \textit{static Ansatz} for ${\bf A}$ in a \textit{harmonic trap}. We have gone well beyond these earlier studies by including Newtonian gravity in the GPPE and RTGLE together with \textit{ the complete Maxwell equations} for ${\bf A}$. To the best of our knowledge, this has not been attempted hitherto in the context of pulsars.

Our imaginary-time studies of the GPPE~(\ref{eq:GPE_neutron_ndim}) and the RTGLE~(\ref{eq:GLE_proton_ndim}) reveals that, even in the absence of any direct interaction ($\gamma=0$), the neutron-superfluid vortices and proton-superconductor flux tubes interact gravitationally through the Poisson equation~\eqref{eq:vector_pot_ndim}.
By including the Maxwell equations, we demonstrate, for the first time, that neutron-superfluid vortices display an induced magnetization whose magnitude is proportional to $\frac{\gamma q}{c^2\epsilon_0} {\bf J}_n|\psi_p|^2$. This magnetization plays a crucial role in the expulsion of vortices from the pulsar. 
The angle $\Theta$ is an important control parameter in our model.
For example, if $\Theta = 30^{\circ}$, proton flux tubes gradually endeavor to align themselves with the rotation axis over time [Figs.\ref{fig:agle_3D_p_rotation}(a)-(c)], but they also exhibit a tendency to adhere to the external magnetic field; 
this competition leads to frustration in the proton-superconductor flux tubes, which are no longer straight but become distorted [Figs.~\ref{fig:agle_3D_p_rotation}(b) and (h)].

The real-time dynamics of the GPPE~\eqref{eq:GPE_neutron_ndim}, RTGLE~\eqref{eq:GLE_proton_ndim}, and the Maxwell equations~\eqref{eq:vector_pot_ndim} can be applied qualitatively to pulsars. We must, of course, incorporate a pulsar-crust potential $V_{\theta}$, described, at the simplest level, by the polar angle $\theta$ in Eq.~(\ref{eq:crust_pot_eq}). This provides a minimal model for studying pulsar glitches. Our investigation reveals that the proton-superconductor flux tubes remain anchored to the crust by the external magnetic field ${\bf B}_{\rm ext}$, while neutron-superfluid vortices leave the condensate and give rise to the glitching phenomenon, the complicated time evolution of the crust angular momentum ${\rm J}_c(t)$, which displays signatures of self-organised criticality (SOC). Although pulsar glitches have been obtained recently in the GPPE model~\cite{AK_verma_2022}, they have not been studied in the presence of proton-superconductor flux tubes, whose dynamics affects pulsar glitches and the time series of ${\rm J}_c(t)$ significantly [as we can see by comparing Fig.~\ref{fig:gpe_3D_crust_angmom} with Fig. 4(b) in Ref.~\cite{AK_verma_2022}].

The SOC that we obtain in our model for pulsars, which generalises the earlier work from our group~\cite{AK_verma_2022}, is akin to what has been obtained in some pulsars; e.g., in the pulsar PSR J 1825-0935, the glitch-size exponent $\beta \simeq 0.36$. Given the simplicity of our model, this is indeed gratifying. It is important to note that neutrons in the outer core of a neutron star are strongly interacting and are sometimes argued to scatter through $p$-wave interactions~\cite{Haskell2018}. However, the Gross-Pitaevskii (GP) model of neutron Cooper pairs is applicable for weakly interacting neutrons and considers only $s$-wave interactions. Our interest lies in the dynamics of neutron vortices, which can be modelled using the s-wave interacting GP equation. This approach has previously been applied in modelling the outer core of neutron stars~\cite{Drummond_2017,Drummond_2017_b,Drummond_2023_c}. In our simulations, some ratios match closely the values found in typical pulsars: In the outer core of a neutron star, the proton to neutron number density ratio is $\frac{n_p}{n_n}\simeq0.05$; we consider the range $0.5 \leq \frac{n_p}{n_n} \leq 1$. Furthermore, we choose the ratio of the neutron and proton coherence lengths $\frac{\xi_n}{\xi_p}=2$, which agrees with the value found in neutron stars~\cite{Toby_2022}. The London parameter $\kappa$ for type II superconducting proton Copper pairs is chosen to be greater than $\frac{1}{\sqrt{2}}$, so that we have an Abrikosov phase. However, it is important to note that numerical studies cannot achieve spatial and temporal scales and resolutions that are in the ranges of direct relevance to pulsars. In particular, the radius of a typical neutron star is $\simeq 10$ km; by contrast, the core sizes of neutron-superfluid vortices are $\simeq 10^{-15}$m; the ratio of the speed of light to that of sound $\frac{c}{c_s}\simeq 10^6$, which is a challenge for any simulation. The number of neutron-superfluid vortices that thread a pulsar is estimated to be $N_v = 10^{16}$; this is far in excess of what can be simulated on even the world's biggest computers. The number of vortices in our model is given as $N_v = \frac{L}{2}\sqrt{\frac{\beta}{\alpha}}$, where $L=2\pi$ is the length of the simulation box and $\alpha$ and $\beta$ are given in Table~\ref{tab:paramters}. For the values we use in our simulations, namely, $\alpha=0.2$ and $\beta = 5$, we have $N_v\sim 15-20$ vortices. We can increase $N_v$ by increasing either the size of our simulation or the ratio $\beta/\alpha$ (or both); but these are limited severely by computational facilities.

\section*{Acknowledgments}
We thank the Indo-French Centre for Applied Mathematics (IFCAM), the Science and Engineering Research Board (SERB), and the National Supercomputing Mission (NSM), India for support, and the Supercomputer Education and Research Centre (IISc) for computational resources.

\appendix
\section{}
\label{sec:ang_mom_apen}
The total angular momentum of the system is ${\rm J} = {\rm J}_z+{\rm J}_c$, with ${\rm J}_c$, the crust angular momentum, and ${\rm J}_z$, the angular momentum of the system without the crust, which are, respectively,
\begin{eqnarray}
    {\rm J}_c &=& I_c \frac{d\theta}{dt}\qquad {\rm{and}}\nonumber\\
    {\rm J}_z &=& \int d^3  {\bf x} \ \psi_n^* (\hat{\bf e}_z\times {\bf r})\cdot (-i\hbar \nabla)\psi_n \nonumber\\
    &+& \int d^3 {\bf x}\ \psi_p^* (\hat{\bf e}_z\times {\bf 
 r})\cdot (-i\hbar \nabla)\psi_p\,,
\end{eqnarray}
where $I_c$ is the moment of inertia of the crust. In the absence of friction [$\alpha=0$ in Eq.~(\eqref{eq:crust_pot_eq})], the total angular momentum is conserved in an infinite system. In the spatially periodic cubical domain that we consider, this conservation is only approximate because this domain does not have strict rotational invariance [see Ref.~\cite{AK_verma_2022} for a detailed discussion].

%\bibliography{reference.bib}


\end{document}